\documentclass[lettersize,journal]{IEEEtran}
\usepackage{amssymb,amsfonts}
\usepackage{algorithmic}
\usepackage{graphicx}
\usepackage{textcomp}
\usepackage{xcolor}
\usepackage{amsmath}
\usepackage{bm}
\usepackage{empheq}
\usepackage{caption}
\usepackage{subcaption}
\usepackage{xurl}
\usepackage[normalem]{ulem}
\usepackage{cancel}
\usepackage[font=small,labelfont=bf]{caption}
\usepackage{makecell}
\usepackage{amssymb}
\usepackage{rotating}
\usepackage{booktabs}
\usepackage{longtable}
\usepackage{multirow}
\usepackage{tabularx}
\usepackage{adjustbox}
\def\BibTeX{{\rm B\kern-.05em{\sc i\kern-.025em b}\kern-.08em
    T\kern-.1667em\lower.7ex\hbox{E}\kern-.125emX}}
\usepackage{balance}
\begin{document}

\title{Towards SISO Bistatic Sensing for ISAC}
\author{Zhongqin Wang, \IEEEmembership{Member, IEEE},
		J. Andrew Zhang, \IEEEmembership{Senior Member, IEEE},\\
		Kai Wu, \IEEEmembership{Member, IEEE},
        Min Xu, \IEEEmembership{Member, IEEE},
		Y. Jay Guo, \IEEEmembership{Fellow, IEEE}

\IEEEcompsocitemizethanks{
\IEEEcompsocthanksitem Zhongqin Wang, J. Andrew Zhang (Corresponding Author), Kai Wu, and Min Xu are with the School of Electrical and Data Engineering, University of Technology Sydney, Sydney 2007, Australia. E-mail:\{zhongqin.wang, andrew.zhang, kai.wu, min.xu\}@uts.edu.au
\IEEEcompsocthanksitem Y. Jay Guo is with the Global Big Data Technologies Centre, University of Technology Sydney, Sydney 2007, Australia. E-mail: jay.guo@uts.edu.au
}}
\maketitle

\begin{abstract}
Integrated Sensing and Communication (ISAC) is a key enabler for next-generation wireless systems. However, real-world deployment is often limited to low-cost, single-antenna transceivers. In such bistatic Single-Input Single-Output (SISO) setup, clock asynchrony introduces random phase offsets in Channel State Information (CSI), which cannot be mitigated using conventional multi-antenna methods. This work proposes WiDFS 3.0, a lightweight bistatic SISO sensing framework that enables accurate delay and Doppler estimation from distorted CSI by effectively suppressing Doppler mirroring ambiguity. It operates with only a single antenna at both the transmitter and receiver, making it suitable for low-complexity deployments. We propose a self-referencing cross-correlation (SRCC) method for SISO random phase removal and employ delay-domain beamforming to resolve Doppler ambiguity. The resulting unambiguous delay-Doppler-time features enable robust sensing with compact neural networks. Extensive experiments show that WiDFS 3.0 achieves accurate parameter estimation, with performance comparable to or even surpassing that of prior multi-antenna methods, especially in delay estimation. Validated under single- and multi-target scenarios, the extracted ambiguity-resolved features show strong sensing accuracy and generalization. For example, when deployed on the embedded-friendly MobileViT-XXS with only 1.3M parameters, WiDFS 3.0 consistently outperforms conventional features such as CSI amplitude, mirrored Doppler, and multi-receiver aggregated Doppler.
\end{abstract}


\section{Introduction}
\IEEEPARstart{I}{ntegrated} Sensing and Communication (ISAC) is rapidly emerging as a key enabler for next-generation wireless systems by embedding sensing capabilities into existing communication infrastructure \cite{zhang2021overview, zhang2022integration, liu2023integrated, lu2024integrated}. ISAC enables a wide range of applications, such as environment monitoring \cite{wu2025isac, wang2025water}, human-computer interaction \cite{he2024forward, miao2025wi}, and healthcare \cite{feng2021lte}. As ISAC moves toward real-world deployment, bistatic Single-Input Single-Output (SISO) architectures, where the transmitter and receiver are separately deployed with only one antenna each, have attracted increasing attention due to their low hardware complexity. Such setups are common in low-power IoT nodes, smart sensors, and embedded edge platforms, while still supporting reliable communication functions. They are particularly well suited for applications such as in-home gesture control, behavior monitoring in elder-care settings, and occupancy detection for energy-efficient buildings. These make SISO-based bistatic ISAC a practical and scalable solution for ubiquitous sensing.

Most existing CSI-based sensing work in ISAC focuses on Single-Input Multiple-Output (SIMO) or Multiple-Input Multiple-Output (MIMO) architectures, where the receiver is typically equipped with multiple antennas. This is primarily because, in bistatic setups where the transmitter and receiver are not synchronized, the CSI is distorted by timing offset (TO) and carrier frequency offset (CFO), resulting in time-varying random phase shifts. To address this issue, prior works have proposed many compensation techniques such as Cross-Antenna Cross-Correlation (CACC) \cite{qian2018widar2, wang2023single} and Cross-Antenna Signal Ratio (CASR) \cite{feng2021lte, ni2023uplink, 10678871}. CACC performs by computing the conjugate multiplication of CSI from two Rx antennas, while CASR achieves by dividing CSI across the two antennas. These methods have enabled a range of sensing applications, including target tracking \cite{chen2023development, pegoraro2024jump, 10737138}, activity recognition \cite{zhang2021widar3, meneghello2022sharp, zheng2024pushing}, and even water-level monitoring \cite{ni2025deep}. Recent advances in deep learning have given rise to a variety of data-driven sensing frameworks. Models leveraging attention mechanisms \cite{luo2024vision}, domain adaptation \cite{10930823}, and spatiotemporal feature extraction \cite{11026879} have significantly improved performance across diverse ISAC tasks. 

Meanwhile, communication standardization is also progressing rapidly. The IEEE 802.11bf amendment introduces native sensing capabilities into Wi-Fi for CSI-based sensing \cite{du2024overview}. This initiative represents a key milestone toward enabling commercial devices to support wireless sensing, facilitating the transition of ISAC technologies from research prototypes to real-world deployment. While existing research has established a strong foundation under multi-antenna settings, further efforts are still required to enable robust sensing in practical, low-complexity configurations such as bistatic SISO systems.

However, towards SISO bistatic sensing for ISAC, several challenges remain to be addressed:

\textit{1) CSI random phase removal under SISO constraints.}  
In multi-antenna systems, the TO and CFO caused by clock asynchrony are identical across Rx antennas, enabling efficient methods such as CACC and CASR to fully eliminate phase offsets. However, these methods are not applicable in SISO configurations. The research on SISO-based sensing remains limited. Several methods \cite{meneghello2022sharp, 11079818} typically assume strong line-of-sight (LoS) conditions, which may not hold in practical deployments, or involve high computational costs, making them unsuitable for embedded platforms. Therefore, it is essential to clean raw CSI without relying on multi-antenna diversity or ideal propagation assumptions.

\textit{2) Doppler mirror ambiguity.} In ISAC systems, the limited bandwidth of communication signals, for example in LTE-based systems where it ranges from 1.4 MHz to 20 MHz, results in coarse range resolution. As a result, Doppler features play a crucial role in capturing target motion for accurate sensing. However, many existing systems suffer from Doppler mirror ambiguity, where the energy is nearly symmetric across positive and negative Doppler frequencies, making it difficult to determine the true direction of motion. This commonly arises in CACC-based systems, which inherently induces symmetric signal components. While CASR avoids this symmetry, it is primarily designed for single-target scenarios and may introduce nonlinear distortions in feature extraction. The Doppler mirroring is often overlooked in prior work.

\textit{3) Learning robust representations from wireless signals.} Unlike image and language tasks where abundant, high-quality datasets enable end-to-end learning directly from raw inputs, wireless sensing faces unique challenges. Due to multipath propagation, environmental variability, hardware diversity, and differing frequency bands, the received signals vary significantly across settings. Collecting diverse and large-scale labeled datasets is costly and challenging. While some works \cite{11036155, 11026879} employ simulation data for training, their effectiveness in real-world deployment remains uncertain. As a result, directly feeding raw CSI into deep networks often leads to poor generalization. Moreover, common data augmentation techniques used in vision such as rotation or cropping lack theoretical validation in the context of wireless signals. Therefore, designing interpretable signal features and augmentation strategies is critical for enhancing model generalization.

In this work, we propose \textit{WiDFS 3.0}, a real-time bistatic sensing system for SISO-based ISAC tasks. It features a separately deployed transmitter and receiver, each with only a single antenna, providing a low-cost ISAC solution. WiDFS 3.0 enables precise estimation of delay, Doppler, and ambiguity-resolved micro-Doppler under the impact of clock asynchrony. Based on the extracted high-quality features, lightweight neural networks deployable on embedded platforms can achieve strong sensing performance. WiDFS 3.0 also supports extension to multi-antenna setups for angular estimation and is compatible with different operating frequencies and bandwidths.

At first, to mitigate the impact of TO and CFO, we construct an energy-adjusted reference CSI from the original CSI itself, called \textit{Self-Referencing Cross-Correlation (SRCC)}. Specifically, we apply an inverse fast Fourier transform (IFFT) to transform the CSI into the delay domain. A Gaussian window is then applied to strengthen the dominant path (a superposition of multiple paths) while suppressing weaker multipath components. The windowed result is transformed back to the frequency domain via Fourier transform (FFT) to reconstruct a new CSI. On this basis, cross-correlation between the raw and reference CSI is performed to remove TO and CFO while preserving the linear structure of delay and Doppler features, facilitating subsequent feature extraction.

Second, we adopt a beamforming-based approach to suppress Doppler mirroring. We first aggregate multiple consecutive CSI measurements to form a coherent processing interval (CPI) \cite{zhang2021overview}. Within each CPI, we apply Capon beamforming \cite{mailloux2017phased} to estimate weights for each delay bin, suppress dynamic by-product noise, and extract beamformed Doppler signatures to construct a 2D delay-Doppler map. By concatenating the map across multiple CPIs, we construct an unambiguous delay-Doppler-time tensor for downstream sensing tasks.

Third, we use embedded-friendly lightweight neural networks like MobileViT-XXS \cite{mehta2022mobilevit} by inputting our unambiguous features to achieve accurate and robust sensing. Due to the coarse range resolution in communication systems, we compress the delay dimension to extract Doppler-time features for fine-grained sensing in single-target scenarios. In contrast, for multi-target tasks, preserving the delay dimension enables spatial discrimination among targets. We also introduce a Doppler-guided data augmentation strategy grounded in physical principles, providing meaningful variations during training.

Our main contributions are highlighted as follows:

\textit{1)} We propose a low-complexity SISO SRCC based on energy-adjusted CSI for random phase removal in bistatic systems, without relying on ideal propagation assumptions.

\textit{2)} We develop a lightweight delay-domain beamforming approach to suppress dynamic by-product noise and resolve Doppler ambiguity during feature extraction.

\textit{3)} We introduce a Doppler-guided data augmentation strategy grounded in physical motion principles, improving model robustness and generalization.

\textit{4)} We conduct extensive experiments validating the efficiency and effectiveness of WiDFS 3.0. 

\begin{itemize}
\item On an edge platform (Raspberry Pi 4B), our unambiguous delay and Doppler feature extraction runs in just 8.5 milliseconds, demonstrating real-time capability. 

\item Despite using a single antenna at both ends, WiDFS 3.0 achieves a 2.05 m median range error (20 MHz, without smoothing), outperforming prior multi-antenna systems by 0.5--2 m, while offering Doppler mirror suppression performance comparable to multi-antenna approaches.

\item The Doppler ambiguity-resolved features enable accurate and generalizable sensing. A compact model like MobileViT-XXS (1.3M parameters) achieves an F1 score of 0.928--0.938 on the single-target Widar 3.0 dataset \cite{zhang2021widar3}, significantly outperforming the classical BVP baseline (0.849--0.859). On the multi-target WiMANS dataset \cite{huang2024wimans}, it achieves an F1 score of 0.659 for behavior recognition and a people counting accuracy of 0.629, surpassing the commonly-used CSI amplitude baseline.
\end{itemize}

\section{Related Work}
This section reviews representative methods for random phase removal, feature extraction, and data-driven learning.

\subsection{CSI Random Phase Removal}
Many techniques \cite{wu2024sensing} have been proposed to eliminate the CSI random phase offsets in bistatic sensing, typically categorized by their use of spatial (antenna), spectral (subcarrier), or temporal (Doppler) dimensions. One widely used approach is CACC \cite{qian2018widar2}, which requires two Rx antennas at least. By leveraging the fact that clock-induced phase distortions are identical across antennas, CACC can removes TO and CFO while preserving the linear relationships among delay, Doppler, and Angle of Arrival (AoA). However, the conjugate operation in CACC introduces symmetry in signal representations. Specifically, Doppler spectra often exhibit nearly identical energy at positive and negative frequencies, making it difficult to infer true motion direction. To mitigate this, WiDFS \cite{wang2023single} proposes Differential CACC (DCACC), which applies linear transformations across different antenna pairs to suppress mirror ambiguity, but it still requires three Rx antennas. Other multi-antenna works \cite{dong2024signal} exploit subspace methods to extract dominant AoA and build reference signals. Another popular method is CASR \cite{feng2021lte, ni2023uplink, 10678871}, which computes the CSI ratio between two Rx antennas to cancel out random phase offsets along with automatic gain control (AGC) variations. CASR is effective in Doppler estimation for single-target scenarios. However, its nonlinear formulation complicates accurate estimation of delay and AoA.

Recently, several compensation methods have started exploring SISO configurations by constructing reference signals in the spectral domain across subcarriers. SHARP \cite{meneghello2022sharp} employs compressed sensing to recover a reference signal, but its performance depends on a strong LoS path, and its computational cost limits real-time applicability. The work \cite{kai_wifisensing_icc23} still relies on the presence of a LoS path to extract relative TO and CFO for CSI compensation. And cross-frequency cross-correlation (CFCC) \cite{11079818} also requires a strong LoS condition to function effectively. In addition, linear regression-based methods \cite{Navid2019} aim to estimate relative TO and CFO, though their effectiveness degrades in multipath environments due to nonlinear phase variations. In contrast, our SISO SRCC constructs an energy-adjusted CSI, sharing the same TO and CFO as the original CSI, without relying on LoS assumptions.

\subsection{Multi-Dimension Signal Feature Extraction}
ISAC systems often extract motion-related features across multiple domains, such as Doppler (temporal), delay (spectral), and AoA (spatial). Many existing works extract only Doppler features to perform tasks such as human tracking and behavior sensing. For instance, PITrack \cite{niu2022rethinking} estimates target trajectories using Doppler shifts observed across multiple receivers. However, such methods are prone to accumulated trajectory drift over time. Other methods \cite{zhang2023autoloc} focus solely on AoA estimation across multiple Rx antennas to localize the target, but tend to lose robustness in the presence of low spatial diversity.

More advanced approaches adopt joint estimation of Doppler, delay, and AoA, requiring only a receiver equipped with multiple Rx antennas. For example, Widar2.0 \cite{qian2018widar2} and mD-Track \cite{xie2019md} apply maximum likelihood estimation to jointly infer these parameters. While such optimization-based frameworks can achieve high accuracy, they typically incur high computational overhead and are sensitive to initialization. Recent works have explored alternatives such as mirrored-MUSIC \cite{ni2021uplink} to reduce the complexity of multidimensional estimation, or tensor decomposition techniques \cite{du2024tensor}. WiDFS \cite{wang2023single} leverages DCACC to suppress both random phase distortions and Doppler ambiguity, and then uses MUSIC-based high-resolution Doppler estimation as a basis to infer delay and AoA for precise target localization. WiDFS 2.0 \cite{10737138} further enhances this pipeline by identifying multiple latent Doppler components from a tracked target, followed by delay and AoA estimation for each component. This method improves tracking accuracy. In this work, we focus on the SISO bistatic setup, where only Doppler and delay dimensions are available. We introduce a lightweight delay-domain beamforming method to achieve unambiguous delay and Doppler estimation.

\subsection{Data-driven Sensing Techniques}
Data-driven techniques are playing an increasingly important role in ISAC applications. Inspired by the success of end-to-end learning in vision and natural language processing, many studies \cite{zheng2024pushing, yang2023slnet} directly feed raw or minimally processed CSI such as amplitude, real, or imaginary components into deep neural networks. These approaches focus on network design and aim to learn feature representations implicitly. While they often achieve high sensing accuracy in specific conditions, generalization across diverse environments remains a key challenge. To improve robustness, many work focuses on first removing random phase distortions and then extracting Doppler spectrograms for network input. For instance, SHARP \cite{meneghello2022sharp} mitigates TO and CFO, applies short-time Fourier transform (STFT) to generate Doppler features, and uses a custom lightweight CNN for behavior recognition. Widar 3.0 \cite{zhang2021widar3} aligns multi-receiver Doppler profiles to form the BVP feature, which is then processed by a sequence modeling network.

Recently, large language models (LLMs) have also been explored for CSI understanding, leveraging their powerful capacity to model complex spatiotemporal relationships in multipath channels. Some works \cite{11026879, 11036155} train LLMs using simulated CSI to overcome the scarcity of labeled real-world data. While promising, the practical effectiveness of these LLM-based methods remains to be fully validated. In this work, we extract physically grounded delay-Doppler-time features under SISO bistatic settings, where Doppler mirror ambiguity is explicitly suppressed. These interpretable features not only support lightweight neural network inference on embedded platforms, but also provide a strong foundation for generalizable and data-efficient sensing.

\section{System Overview}
\begin{figure*}
\centering
\includegraphics[width=\textwidth]{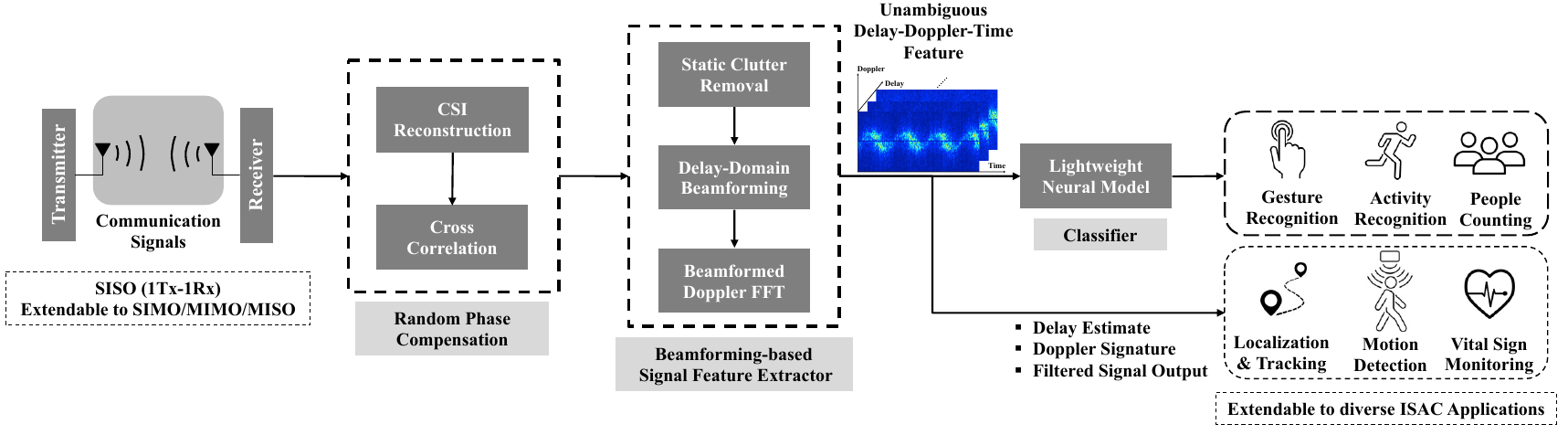}
\caption{System overview of WiDFS 3.0: A SISO bistatic sensing framework for ISAC.}
\label{fig:workflow}
\vspace{-1em}
\end{figure*}

In this work, we propose a lightweight SISO bistatic sensing scheme \textit{WiDFS 3.0} for ISAC applications. The system is designed for real-world deployment with separately deployed transmitter and receiver, each equipped with only a single antenna, offering a low-cost and low-complexity solution. Despite the SISO constraint and clock asynchrony, WiDFS 3.0 achieves accurate delay, Doppler, and ambiguity-resolved micro-Doppler estimation. Moreover, the framework is extensible to Single-Input Multiple-Output (SIMO), Multiple-Input Multiple-Output (MIMO), and Multiple-Input Single-Output (MISO) configurations to further exploit Angle of Departure (AoD) and AoA information for enhanced sensing. The overall workflow of WiDFS 3.0 is illustrated in Fig. \ref{fig:workflow}, which consists of the following key modules:

\textit{1) Phase Compensation via CSI Reconstruction:} To mitigate the effects of TO and CFO caused by clock asynchrony, we construct an energy-adjusted reference CSI along the subcarrier dimension. Sharing the same TO and CFO as the original CSI, this reference enables effective phase cancellation through conjugate multiplication while preserving both temporal and spectral linear structures.

\textit{2) Beamforming-based Signal Feature Extractor:} We adopt a delay-domain beamforming approach to suppress mirror ambiguity and dynamic noise in Doppler estimation. By aggregating across multiple CPIs, we obtain an unambiguous delay-Doppler-time tensor for downstream sensing tasks.

\textit{3) Lightweight Neural Network Classifier:} The unambiguous features are fed into off-the-shelf compact neural networks, such as MobileViT-XXS, which enable accurate activity recognition and are well-suited for embedded deployment. In addition, a Doppler-guided data augmentation strategy is introduced to enhance model generalization.

\section{Phase Compensation via CSI Reconstruction}
In this section, we present a CSI model in a SISO bistatic setup and reconstruct CSI values for random phase removal.

\subsection{SISO Bistatic CSI Model}
In SISO bistatic systems, the transmitter and receiver are spatially separated and operate without a shared clock. Each side is equipped with a single antenna. Let $\mathit{CSI}_{i,j}$ denote the CSI measured at the $i$-th subcarrier and the $j$-th OFDM symbol (CSI sampling time). Within the short-duration CPI (around 100 milliseconds), the transmit power, automatic gain control settings, and the amplitude of each signal are typically assumed to be constant \cite{zhang2021overview}. The CSI measurements are $\mathbf{csi} \in \mathbb{C}^{N \times M}$, where $N$ is the number of subcarriers and $M$ is the number of OFDM symbols. The CSI is modeled as:
\begin{equation}
\mathit{CSI}_{i,j}= \underbrace{{{e}^{-\bm{J}{\left( 2\pi {f}_{i} \tau_j^{\text{TO}} + \phi_j^{\text{CFO}}\right)}}} {{e}^{-\bm{J}\phi^h}}}_{\text{Imperfect Signal Processing}} \left( H^S_{i}+H^X_{i,j}\right),
\label{equation1}
\end{equation}
where
\begin{equation}
\left\{
\begin{aligned}
&H^S_{i} = \sum_{l_1}{\rho _{i}^{S}\left[ l_1 \right]}{{e}^{-\bm{J}{2\pi {f}_{i} {\tau^{S}}\left[ l_1 \right]}}} \\
&H^X_{i,j}=\sum_{l_2} {\rho _{i}^{X}\left[ l_2 \right]}{e}^{-\bm{J}2\pi \left( f_i\tau^X\lbrack l_2 \rbrack  +  f^D\lbrack l_2 \rbrack \left(j-1\right) \Delta t \right)}\\
\end{aligned} 
\right..
\label{equation2}
\end{equation}
These variables are explained as follows:

1) $\tau_j^{\text{TO}}$ and $\phi_j^{\text{CFO}}$ denote the random TO and CFO, caused by clock asynchrony between the transmitter and receiver. 

2) $\phi^h$ denotes a hardware-induced phase offset, initialized by the phase-locked loop at power-up. Due to the randomness of the initial phase, it may vary across different power cycles.

3) $H^S_i$ represents the channel frequency response (CFR) associated with static paths. These include both the direct line-of-sight (LOS) or non-line-of-sight (NLOS) path between the transmitter and receiver, as well as other reflections from static environmental objects such as walls, floors, and furniture. Each static path is characterized by an attenuation $\rho_i^S[l_1]$ and a propagation delay $\tau^S[l_1]$. $f_i$ is the subcarrier frequency.

4) $H^X_{i,j}$ is the CFR contributed by reflections from moving objects. Each dynamic path has an attenuation $\rho_i^X[l_2]$, a delay $\tau^X[l_2]$, and a Doppler frequency shift (DFS) $f^D[l_2]$ caused by object motion. $\Delta t$ is the OFDM symbol interval.

\subsection{CSI Reconstruction via Delay-Domain Windowing}
For the $j$-th OFDM symbol, we reconstruct the CSI by utilizing the frequency domain across subcarriers. The goal is to enhance the power of the dominant propagation path while suppressing weaker paths in the reconstructed CSI. Here, the dominant path typically represents a composite of closely-spaced strong reflections due to the limited delay resolution. To achieve that, we first transform the original CSI from the frequency domain to the time domain using IFFT. The time-domain Channel Impulse Response (CIR) is given by:
\begin{equation}
h_j\left(\tau\right) = \mathcal{F}^{-1} \{ \mathit{CSI}_{i,j} \} = \sum_{i=0}^{N-1} \mathit{CSI}_{i,j} e^{\bm{j}2\pi f_i \tau},
\label{equation3}
\end{equation}
where $\tau$ denotes a discrete delay bin.

A Gaussian window function $\mathcal{G}\left(\tau\right)$ is then applied, centered at the peak energy bin $\tau^{\text{peak}}_j$, so the windowed CIR is
\begin{equation}
{h}^{'}_j\left(\tau\right) = \mathcal{G}\left(\tau - \tau^{\text{peak}}_j\right) \cdot h_j\left(\tau\right),
\label{equation4}
\end{equation}
where
\begin{equation}
\mathcal{G}\left(\tau - \hat\tau_j\right) = e^{-\left( \frac{\tau - \tau^{\text{peak}}_j}{2\sigma} \right)^2}.
\label{equation5}
\end{equation}
The parameter $\sigma$ controls the width of the Gaussian window. 

After that, we transform the windowed CIR back to the frequency domain using FFT to obtain the reconstructed CSI. This process is expressed as:
\begin{equation}
{\mathcal{CSI}}_{i,j} = \mathcal{F} \{ {h}^{'}_j\left(\tau\right) \} = \sum_{\tau=0}^{N-1} {h}^{'}_j\left(\tau\right) e^{-\bm{j} 2\pi f_i \tau}.
\label{equation6}
\end{equation}

Compared to the original CSI, the reconstructed version offers an energy-adjusted representation in which the dominant path is enhanced while the power of secondary paths is effectively suppressed. Note that, overly small $\sigma$ may introduce spectral distortion in the frequency domain due to severe filtering, which may introduce additional noise in CSI reconstruction. Its impact will be analysed in the following.

\subsection{Random Phase Removal via Cross-Correlation}
To removal the impact of TO and CFO, we perform cross correlation between the original and reconstructed CSI:
\begin{equation}
\begin{aligned}
\Delta \mathit{CSI}_{i,j} & = \mathit{CSI}_{i,j}  {\overline{\mathcal{CSI}}_{i,j}} \\
& = \left( H^S_{i} + H^X_{i,j} \right) \left( \overline{\mathcal{H}}^S_{i}+\overline{\mathcal{H}}^X_{i,j} \right) \\
&= \underbrace{H^S_{i}\overline{\mathcal{H}}^S_{i}}_{\text{static}} + \underbrace{H^S_{i}\overline{\mathcal{H}}^X_{i,j} + \overline{\mathcal{H}}^S_{i}H^X_{i,j}}_{\text{dynamic}} + \underbrace{H^X_{i,j}\overline{\mathcal{H}}^X_{i,j}}_{\text{by-product}}.
\end{aligned}
\label{equation7}
\end{equation}
This random phase removal process is referred to as \textit{Self-Referencing Cross-Correlation (SRCC)}. It can simultaneously mitigate the impacts of clock asynchronism and hardware mismatch. When extended to multi-antenna systems, our SRCC can eliminate the need for complex antenna-specific correction in AoA estimation. It also preserves the linear relationship of signal features such as Doppler and delay.

In addition, Eq. \eqref{equation7} contains four cross-correlation terms. \textit{(1)} $H^S_{i}\overline{\mathcal{H}}^S_{i}$ represents the static component, capturing the signal contribution between the fixed transmitter and receiver, as well as reflections from static objects in the environment. \textit{(2)} $H^S_{i}\overline{\mathcal{H}}^X_{i,j}$ and $ \overline{\mathcal{H}}^S_{i}H^X_{i,j}$ are cross-terms that encode dynamic variations of interest. However, they are complex conjugates of each other, exhibiting symmetric structures in delay and Doppler domains. Specifically, the DFS may result in nearly equal energy at positive and negative frequencies, making it difficult to determine the true direction of movement. In our SRCC, the windowing function can enhance the energy of $H^S_i$ (typically containing dominant paths) while suppressing $H^X_{i,j}$ (which includes weaker dynamic components). As a result, the windowed terms often satisfy $ \|H^S_i \overline{\mathcal{H}}^X_{i,j}\|<\|\overline{\mathcal{H}}^S_i H^X_{i,j} \|$, which helps mitigate mirror ambiguity to some extent. However, narrower windows may amplify phase noise. Therefore, it is necessary to apply an appropriate Gaussian window and further exploit the delay domain to effectively suppress mirror ambiguity. \textit{(3)} $ H^X_{i,j}\overline{\mathcal{H}}^X_{i,j}$ is a by-product term that generally acts as noise and should be suppressed. Most prior works \cite{wang2023single, 10737138, li2017indotrack} simply neglect this term, assuming its energy is negligible compared with other components. However, this assumption does not always hold in practice, especially under weak LoS or NLoS conditions where dynamic multipath reflections may retain significant energy. In such cases, the by-product term can generate considerable interference, leading to degraded sensing performance.

\subsection{Impact of Windowing on CSI Reconstruction Accuracy}
To quantify the impact of delay-domain windowing in CSI reconstruction, we analysis the phase variance of the reconstructed CSI across subcarriers using the Cram\'{e}r-Rao Lower Bound (CRLB). Assuming additive Gaussian noise and unbiased estimation, the CRLB for estimating the phase $\phi_{i,j}$ of the reconstructed CSI at the $i$-th subcarrier and the $j$-th OFDM symbol is inversely proportional to the energy of the windowed CIR (see Appendix \ref{appendix:crlb} for derivation):
\begin{equation}
\text{Var}\left({\phi}_{i,j}\right) \geq \frac{\eta^2}{\left\| \mathcal{G}\left(\tau - \tau_j^{\text{peak}}\right) \cdot h_j\left(\tau\right) \right\|^2},
\label{equation8}
\end{equation}
where $\eta^2$ is noise power and $\| \cdot \|^2$ is to compute signal power. 

A wider window can reduce the phase noise in the reconstructed CSI by preserving more signal energy, while it introduces additional multipath components and increases ambiguity in the cross-correlation results. In contrast, a narrower window suppresses side lobes and reduces such ambiguity, but at the cost of higher phase noise due to reduced signal energy. We evaluate the practical impact of this trade-off in Sec. VIII.

\section{Beamforming-based Signal Feature Extractor}
While deep learning networks are capable of learning rich representations, the generated features are often difficult to interpret. In this work, we adopt a signal-model-driven approach to extract interpretable features. Specifically, we employ a beamforming-based method to generate a delay-Doppler-time feature that captures motion dynamics under SISO bistatic setups, while maintaining low computational complexity.

\subsection{Dynamic Component Separation}
Within each CPI window, we compute the average of the cross-correlation CSI $\Delta H_{i,j}$ along the time dimension at each subcarrier. The average is denoted as $\mathcal{U}_i$, given by:
\begin{equation}
\mathcal{U}_i = \frac{1}{M} \sum_{j=1}^{M} \Delta \mathit{CSI}_{i,j} \approx H^S_{i}\overline{\mathcal{H}}^S_{i},
\label{equation9}
\end{equation}

We then subtract the static term from $\Delta \mathit{CSI}_{i,j}$ to obtain the dynamic component:
\begin{equation}
\begin{aligned}
\mathcal{V}_{i,j} &= \Delta \mathit{CSI}_{i,j} - \mathcal{U}_i \approx H^S_{i}\overline{\mathcal{H}}^X_{i,j} + \overline{\mathcal{H}}^S_{i}H^X_{i,j} + H^X_{i,j}\overline{\mathcal{H}}^X_{i,j}.
\end{aligned}
\label{equation10}
\end{equation}
This operation can suppress static clutter and highlight motion-induced changes. To enhance motion sensitivity, we normalize the dynamic component $\mathcal{V}_{i,j}$ by the static term $\mathcal{U}_i$ to get
\begin{equation}
\mathcal{W}_{i,j}=\frac{\mathcal{V}_{i,j}}{\mathcal{U}_i}\approx  \frac{H^X_{i,j}}{H^S_{i}} + \frac{\overline{\mathcal{H}}^X_{i,j}}{\overline{\mathcal{H}}^S_{i}}  + \frac{H^X_{i,j}}{H^S_{i}} \frac{\overline{\mathcal{H}}^X_{i,j}}{\overline{\mathcal{H}}^S_{i}}.
\label{equation11}
\end{equation}
According to \cite{wang2023single, 10678871}, this ratio can amplify motion variations relative to the static background, effectively increasing the signal-to-noise ratio (SNR) for dynamic components. 

\subsection{Delay-Domain Beamforming with By-Product Suppression}
We first apply the MVDR (Minimum Variance Distortionless Response) algorithm \cite{mailloux2017phased} to estimate the delay profile, yielding a set of beamforming weights for each delay bin.

\textit{1) Observation Matrix.} As shown in Eq.~\eqref{equation11}, $\mathcal{W}_{i,j}$ contains two conjugate-mirrored dynamic components, i.e., ${H^X_{i,j}}/{H^S_{i}}$ and ${\overline{\mathcal{H}}^X_{i,j}}/{\overline{\mathcal{H}}^S_{i}}$, whereas the by-product term does not exhibit such symmetry. To preserve this conjugate relationship in the beamforming process, we construct the observation matrix by concatenating both the original $\mathcal{W}_{i,j}$ and its conjugate counterpart $\overline{\mathcal{W}}_{i,j}$. This joint representation enables the MVDR beamformer to fully exploit the phase correlation between the mirrored dynamic components, ensuring coherent alignment of target-related phases while attenuating uncorrelated interference, including the by-product term. The observation matrix $\bm{\Lambda} \in \mathbb{C}^{N \times 2M}$ is formulated as:
\begin{equation}
\bm{\Lambda}=
\begin{bmatrix}
\mathcal{W}_{1,1} &\cdots & \mathcal{W}_{1,M} & \overline{\mathcal{W}}_{1,1} &\cdots &  \overline{\mathcal{W}}_{1,M}\\
\mathcal{W}_{2,1} &\cdots & \mathcal{W}_{2,M} &  \overline{\mathcal{W}}_{2,1} &\cdots &  \overline{\mathcal{W}}_{2,M}\\
\vdots & \ddots  & \vdots & \vdots & \ddots  & \vdots\\
\mathcal{W}_{N,1}  &\cdots & \mathcal{W}_{N,M} &  \overline{\mathcal{W}}_{N,1}  &\cdots &  \overline{\mathcal{W}}_{N,M}
\end{bmatrix}_{N \times 2M}.
\label{equation12}
\end{equation}

\textit{2) Steering Matrix.} A 2D steering matrix $\bm{A} \in \mathbb{C}^{N \times L}$ is then constructed, where each steering vector $\bm{\mathsf{a}}\left(\Delta \tau_l\right)$ corresponds to a delay value $\Delta \tau_l$. Here, $\Delta \tau = \tau^X - \tau^S$ denotes the relative delay between a dynamic path and a static path. The steering matrix is expressed as:
\begin{equation}
\bm{A} = 
\left[
\bm{\mathsf{a}}\left( \Delta \tau_1 \right), 
\bm{\mathsf{a}}\left( \Delta \tau_2 \right), 
\cdots, 
\bm{\mathsf{a}}\left( \Delta \tau_L \right)
\right]_{N \times L},
\label{equation13}
\end{equation}
where 
\begin{equation}
\bm{\mathsf{a}}\left( \Delta \tau \right) =
\left[
e^{-j 2\pi f_1 \Delta \tau}, 
e^{-j 2\pi f_2 \Delta \tau}, 
\cdots, 
e^{-j 2\pi f_N \Delta \tau}
\right]^\mathsf{T}.
\label{equation14}
\end{equation}

Since the observation matrix already incorporates both the original and conjugate dynamic components, the resulting delay-domain response becomes conjugate-symmetric about zero delay. This symmetry makes the delay profile on one side of the axis fully redundant with the other, allowing the beamforming search to be restricted to a single side (either $\tau \ge 0$ or $\tau \le 0$) without loss of information. Here, we set the delay search range to $\tau \ge 0$.

\textit{3) Covariance Matrix.} To obtain a more robust estimation of the covariance matrix, we apply forward-backward smoothing,
\begin{equation}
\widetilde{\bm{R}}_{\bm{\Lambda}} = \bm{R}_{\bm{\Lambda}} + \bm{J} \bm{R}_{\bm{\Lambda}} \bm{J}+ \epsilon \bm{I},
\label{equation15}
\end{equation}
where $\bm{R}_{\bm{\Lambda}} = \bm{\Lambda} \bm{\Lambda}^\mathsf{H}$ is the covariance matrix, $\bm{J}$ is the exchange matrix of size $N \times N$, obtained by flipping the identity matrix from left to right, and $\epsilon \bm{I}$ is a small regularization term added to improve stability. The forward-backward smoothing can improve the effective rank of the matrix and strengthen the separation between the signal and noise subspaces \cite{li2003robust}. 

\textit{4) Beamforming Weights.} The beamforming weight vector $\bm{\mathsf{w}}\left( \Delta \tau \right) \in \mathbb{C}^{N \times 1}$ for the delay bin $\Delta \tau$ can be computed as:
\begin{equation}
\bm{\mathsf{w}}\left( \Delta \tau \right) = \frac{\widetilde{\bm{R}}_{\bm{\Lambda}}^{-1} \bm{\mathsf{a}}\left( \Delta \tau \right)}{\bm{\mathsf{a}}^\mathsf{H}\left( \Delta \tau \right) \widetilde{\bm{R}}_{\bm{\Lambda}}^{-1} \bm{\mathsf{a}}\left( \Delta \tau \right)},
\label{equation16}
\end{equation}
which can suppress noise and interference from all other delays while preserving the amplitude and phase of the signal at the target delay $\Delta \tau$. The corresponding beamforming weights are then applied separately to the original and conjugate parts of the observation matrix to get
\begin{equation}
\left\{
\begin{aligned}
&{\bm{X}}^{\text{orig}}\left(\Delta \tau\right) =  \bm{\mathsf{w}}^\mathsf{H}\left(\Delta \tau\right)\, \bm{\Lambda}^{\text{orig}} \\
&{\bm{X}}^{\text{conj}}\left(\Delta \tau\right) =  \bm{\mathsf{w}}^\mathsf{H}\left(\Delta \tau\right)\, \bm{\Lambda}^{\text{conj}}
\end{aligned} 
\right.,
\label{equation17}
\end{equation}
where $\bm{\Lambda}^{\text{orig}} \in \mathbb{C}^{N \times M}$ and $\bm{\Lambda}^{\text{conj}}=\overline{\bm{\Lambda}}^{\text{orig}} \in \mathbb{C}^{N \times M}$ are the first and last $M$ columns of the matrix $\bm{\Lambda} \in \mathbb{C}^{N \times 2M}$. The resulting outputs, ${\bm{X}}^{\text{orig}}\left(\Delta \tau\right) \in \mathbb{C}^{M}$ and ${\bm{X}}^{\text{conj}}\left(\Delta \tau\right) \in \mathbb{C}^{M}$, coherently enhance target-related signals due to their preserved conjugate symmetry, whereas the by-product term, lacking such symmetry, suffers phase misalignment between its two components and is thus suppressed.

\textit{5) Beamformed Signal.} The final beamformed signal for Doppler processing is obtained by  summing the two outputs:
\begin{equation}
{\bm{X}}\left(\Delta \tau\right) = {\bm{X}}^{\text{orig}}\left(\Delta \tau\right) + {\bm{X}}^{\text{conj}}\left(\Delta \tau\right).
\label{equation18}
\end{equation}
Each element corresponding to the $j$-th OFDM symbol can be expressed as:
\begin{equation}
{X}_j\left(\Delta \tau\right) = \sum_{l} {\rho\left[ l \right]}{e}^{-\bm{J} \left( 2\pi f^D\lbrack l \rbrack \left(j-1\right) \Delta t + \phi \left[ l \right] \right)} + \mathcal{N}_j
\label{equation19}
\end{equation}
where there may contain multiple targets at the delay bin, and $\mathcal{N}_j$ represents residual noise.

\subsection{Beamformed Doppler FFT}
To reveal fine-grained motion patterns, such as limb swings, quick gestures, or subtle posture transitions, we perform a Doppler analysis on the beamformed signal at each delay bin $\Delta \tau$. We apply the FFT along the temporal dimension (i.e., across OFDM symbols) within a CPI as follows:
\begin{equation}
\begin{aligned}
{Y}\left(f^D \mid \Delta \tau\right) &= \mathcal{F} \left\{ {{X}}_j\left(\Delta \tau\right) \right\} \\
&= \sum_{j=0}^{M} {{X}}_j\left(\Delta \tau\right) e^{-\bm{j}2\pi f^D \left(j-1\right) \Delta t}.
\end{aligned}
\label{equation20}
\end{equation}

Repeating this process across delay bins yields a 2D delay-Doppler spectrum, where delay offers coarse spatial separation and Doppler captures motion velocity and periodicity. This joint spectrum enables fine-grained  motion characterization.

\subsection{Spatiotemporal Feature Representation across CPIs}
To capture continuous motion patterns, we extend the delay-Doppler spectrum over multiple CPIs. Specifically, a delay-Doppler frame is computed per CPI and stacked along the temporal axis to form a 3D tensor:
\begin{equation}
\bm{\mathcal{S}} \in \mathbb{R}^{L_{\text{Delay}} \times L_{\text{Doppler}} \times L_{\text{CPI}}},
\label{equation21}
\end{equation} 
where each element $\mathcal{S}\left(\Delta \tau, f^D, \ell\right)$ in $\bm{\mathcal{S}}$ denotes the delay-Doppler-time feature at the delay bin $\Delta \tau$, DFS $f^D$, and CPI frame index $\ell$. Here, $L_{\text{delay}}$ is the number of delay bins, $L_{\text{doppler}} $ is the number of Doppler bins, and $L_{\text{CPI}}$ is the number of consecutive CPIs. Such a structured feature enables downstream neural models to recognize complex motion patterns.

\subsection{Algorithm Complexity Analysis}
We analyze the computational complexity of the proposed delay-Doppler-time feature extraction pipeline.

\textit{1) CSI Reconstruction:} Each OFDM symbol requires an IFFT of size $N$, yielding $\mathcal{O}\left(N \log N\right)$ per symbol and $\mathcal{O}\left(M N \log N\right)$ per CPI.

\textit{2) Static Component Removal:} Temporal averaging and subtraction over the $N \times M$ matrix leads to $\mathcal{O}(N M)$ complexity.

\textit{3) Delay-domain Beamforming:} For each delay bin, MVDR weight computation involves matrix inversion with cost $\mathcal{O}\left(N^3\right)$, and applying weights adds $\mathcal{O}\left(N M\right)$. For $L_{\text{Delay}}$ bins, the total cost is $\mathcal{O}\left(L_{\text{Delay}} N^3 + L_{\text{Delay}} N M\right)$.

\textit{4) Doppler FFT:} Performing an FFT of size $M$ for each delay bin yields $\mathcal{O}\left(L_{\text{Delay}} M \log M\right)$.

The overall complexity is dominated by MVDR beamforming and can be approximated as $\mathcal{O}\left(L_{\text{Delay}} N^3\right)$. The number of delay bins is typically small due to limited bandwidth, keeping the computational load low. For example, based on Intel 5300 CSI data, our algorithm requires only 8.5 ms per CPI on a Raspberry Pi without applying any code-level optimization (see Section VIII for details).

\section{Lightweight Network as Classifier}
The proposed interpretable and unambiguous features are robust to variations in transceiver position, environment, hardware, and individual differences, which simplifies the learning task for the recognition network. Accordingly, a lightweight neural network suitable for deployment on embedded devices is employed to perform efficient classification.

\subsection{Input Representation}
\textit{1) Single-Target Scenario.} For a single moving target, we compress the feature over the delay dimension. This yields a micro-Doppler signature that preserves Doppler-time characteristics while discarding spatial variation,
\begin{equation}
\mathcal{S}\left(f^D, \ell\right) = \sum_{k=1}^{L_{\text{Delay}}} \mathcal{S}\left(\Delta \tau_k, f^D, \ell\right).
\label{equation22}
\end{equation}
The resulting representation is a 2D Doppler-time map:
\begin{equation}
\bm{\mathcal{S^{'}}} \in \mathbb{R}^{L_{\text{Doppler}} \times L_{\text{CPI}}},
\label{equation23}
\end{equation}
which simplifies the model input and emphasizes motion periodicity and velocity variation over time.

\textit{2) Multiple-Target Scenario.} In the presence of multiple moving targets, directly ignoring the delay dimension may obscure spatially distinct motion patterns. Instead, we retain delay bins as separate spatial channels, forming the 3D tensor $\bm{\mathcal{S}}$ defined in Eq. \eqref{equation21}. This representation allows the model to capture simultaneous motions at different distances and improves discrimination in crowded environments.

\subsection{Exploring Doppler Characteristics for Data Augmentation}
In a bistatic system, the Doppler velocity $v^D$ is determined by the geometry relationship among the transmitter, receiver, and the moving target,
\begin{equation}
\begin{aligned}
v^D=\frac{cf^D}{f_c} = \bm{v} \cdot
\left( 
\frac{\bm{P} - \bm{P}_{\text{Tx}}}{\|\bm{P} - \bm{P}_{\text{Tx}}\|} 
+ 
\frac{\bm{P} - \bm{P}_{\text{Rx}}}{\|\bm{P} - \bm{P}_{\text{Rx}}\|} 
\right),
\end{aligned}
\label{equation24}
\end{equation}
where $\bm{v}=\left(v_{x}, {v}_{y}\right)$ is the 2D velocity vector of the moving target on the x-y plane, $\bm{P}=\left(x, y\right)$ is the corresponding 2D position, and $\bm{P}_{\text{Tx}}=\left(x_{\text{Tx}}, y_{\text{Tx}}\right)$ and $\bm{P}_{\text{Rx}}=\left(x_{\text{Rx}}, y_{\text{Rx}}\right)$ are the 2D coordinates of the transmitter and receiver, respectively. When a global coordinate system is defined with the transmitter placed at the origin, i.e., $\bm{P}_{\text{Tx}} = (0, 0)$, all positions and velocities are expressed relative to this reference point. In the following, we leverage this relationship to propose several physically meaningful data augmentation strategies to enhance the robustness of our classifier. For multi-target scenarios, the same augmentation scheme is applied to the Doppler-time features across all delay bins.

\textit{1) Location.} Changing the target position $\bm{P}$ alters the geometric projection vectors toward the transmitter and receiver, which causes a displacement along the Doppler dimension. For example, consider $\bm{v} = \left(0, 1\right)$ m/s, $\bm{P}_{\text{Tx}} = \left(0, 0\right)$ m, and $\bm{P}_{\text{Rx}} = \left(4, 0\right)$ m. When the target is at $\bm{P} = \left(2, 2\right)$ m, the Doppler velocity is $v^D \approx 1.41$ m/s, while at $\bm{P} = \left(3, 2\right)$ m, it increases slightly to $v^D \approx 1.45$ m/s. To simulate such location-induced variations, we apply Doppler-axis translations and affine transformations (e.g., stretching or scaling).

\textit{2) Orientation.} Changing the direction of the velocity vector $\bm{v}$ while keeping its magnitude and the target position $\bm{P}$ fixed can alter the projection onto the transmitter-receiver axis. For example, with $\bm{P} = \left(2, 2\right)$ m, $\bm{P}_{\text{Tx}} = \left(0, 0\right)$ m, and $\bm{P}_{\text{Rx}} = \left(4, 0\right)$ m, consider two velocity vectors with the same magnitude $\|\bm{v}\| = 1$ m/s: upward motion $\bm{v} = \left(0, 1\right)$ m/s yields a projected Doppler velocity of $v^D \approx 1.41$ m/s, while diagonal motion $\bm{v} = \left(\frac{\sqrt{2}}{2}, \frac{\sqrt{2}}{2}\right)$ m/s results in $v^D \approx 1.00$ m/s. These differences can be emulated via Doppler-axis translation, stretching, or mirroring to simulate reversed motion.

\textit{3) Speed.} Scaling the magnitude of the velocity vector $\bm{v}$ directly influences the Doppler shift, since $v^D \propto \|\bm{v}\|$. Increasing the target's speed results in a broader Doppler spread, while decreasing it compresses the Doppler range. To simulate such effects, we scale the Doppler-time spectrogram along the Doppler dimension accordingly.

\textit{4) Timing Variation.} Varying the starting time of a gesture or activity shifts the entire Doppler-time spectrogram along the temporal axis, without altering the underlying Doppler content. During augmentation, such variation can be emulated by applying temporal shifts along the time axis.

\textit{5) Noise Injection.} In practice, Doppler measurements are affected by abnormal noise and multipath interference. To enhance the model robustness, we simulate these effects by injecting Gaussian noise into the Doppler-time spectrogram.

\subsection{Lightweight Network Selection}
Many real-time ISAC sensing tasks require not only accurate motion recognition but also low computational overhead to enable deployment on resource-constrained edge platforms such as wireless routers and embedded IoT devices. While numerous customized neural architectures \cite{chen2023cross, singh2025slim} have been proposed to enhance recognition performance by increasing model complexity and parameter count, such designs often come at the cost of computational efficiency. In this work, we adopt several representative lightweight backbone networks, including MobileViT \cite{mehta2022mobilevit}, MobileNetV2 \cite{8578572}, SqueezeNet \cite{iandola2016squeezenet}, ShuffleNet \cite{ma2018shufflenet}, and shallow CNN, all of which are well-suited for on-device deployment. For comparison, we also include ResNet18 \cite{he2016deep} and a simple MLP as baselines. Specifically, the MLP consists of two fully connected layers with ReLU activation and dropout, sharing the extracted features for classification. The CNN comprises three convolutional blocks with pooling, followed by flattening and classification heads. It is worth noting that if these lightweight models can achieve strong sensing performance, it is reasonable to expect further improvements with stronger networks.

\subsection{Loss Function}
We adopt different loss functions for single-target and multi-target activity recognition tasks.

\textit{1) Single-Target Scenario.} For single-target activity recognition, we adopt the categorical cross-entropy (CE) loss:
\begin{equation}
\mathcal{L}_{\text{single}} = \text{CE}\left(\hat{\bm{y}}, \bm{y}\right) = -\sum_{j=1}^{C} y_j \log\left(\hat{y}_j\right),
\end{equation}
where $C$ is the number of activity classes, $\bm{y} \in \{0,1\}^C$ is the one-hot ground-truth label, and $\hat{\bm{y}} \in [0,1]^C$ is the predicted probability distribution over activity classes. The model uses a 2D Doppler-Time tensor $\bm{\mathcal{S}^{'}}$ as input.

\textit{2) Multi-Target Scenario.} We formulate multi-target activity recognition as a joint task comprising two components: (1) a $K$-class multi-label classification task for activity recognition, and (2) a $(C+1)$-class classification task for people counting (including the case when no person is present), where $K$ denotes the number of predefined activity categories, and $C$ denotes the maximum number of individuals. Each input sample may contain multiple individuals performing different activities simultaneously. The ground truth consists of a binary vector $\bm{y} \in \{0,1\}^K$ indicating the presence of each activity, and a scalar count label $\bm{c} \in \{0, 1, \dots, C\}$ representing the number of active individuals. The model takes a 3D Delay-Doppler-Time tensor $\bm{\mathcal{S}}$ as input. The total loss is:
\begin{equation}
\mathcal{L}_{\text{multi}} = \frac{1}{K} \sum_{j=1}^{K}\text{BCE}\left(\hat{y}_j, y_j\right) + \lambda_c \text{CE}\left(\hat{\bm{c}}, \bm{c}\right),
\label{equation26}
\end{equation}
where $\text{BCE}\left( \cdot \right)$ denotes binary cross-entropy for each activity label, $\text{CE}\left(\cdot \right)$ is the CE loss for people count, and $\lambda_c$ is a hyperparameter. At inference, a fixed threshold of 0.5 is applied to each $\hat{y}_j$ to determine the presence of activity $j$, while the predicted number of individuals is obtained by selecting the class with the highest confidence in $\hat{\bm{c}}$.

\section{Implementation}
\subsection{CSI Datasets}
\subsubsection{Object Tracking Datasets}
We validate the effectiveness of our SISO feature extraction using WiFi and LTE object tracking datasets, collected in indoor environments under LOS conditions with a transmitter-receiver separation of approximately 2 m. The target moves at a speed of 1 m/s.
\begin{itemize}
\item \textit{WiFi:} The WiFi dataset from WiDFS \cite{wang2023single} and WiDFS2.0 \cite{10737138} is collected using Intel 5300 NICs over a 20 MHz bandwidth at 5 GHz. CSI from 30 subcarriers is sampled at 1 kHz using the Linux 802.11n CSI Tool \cite{halperin2011tool}. A single target walks repeatedly along elliptical, V-shaped, and rectangular trajectories, with ground truth provided by a millimeter-wave radar sensor.

\item \textit{LTE:} The 3.1 GHz LTE dataset \cite{chen2023development} is collected using an NI Massive MIMO testbed serving as the base station (BS) and a USRP device acting as the user equipment (UE). Pilot streaming is implemented via LabVIEW Communications 2.0. The extracted CSI contains 100 active subcarriers with an effective frequency resolution of 180 kHz (i.e., resource block-level spacing) over a 20 MHz bandwidth. A person repeatedly walks back and forth along a linear trajectory, facing the tranceiver side.
\end{itemize}

\begin{figure*}
\centering
\begin{subfigure}[t]{0.325\textwidth}
    \centering
    \includegraphics[width=\textwidth]{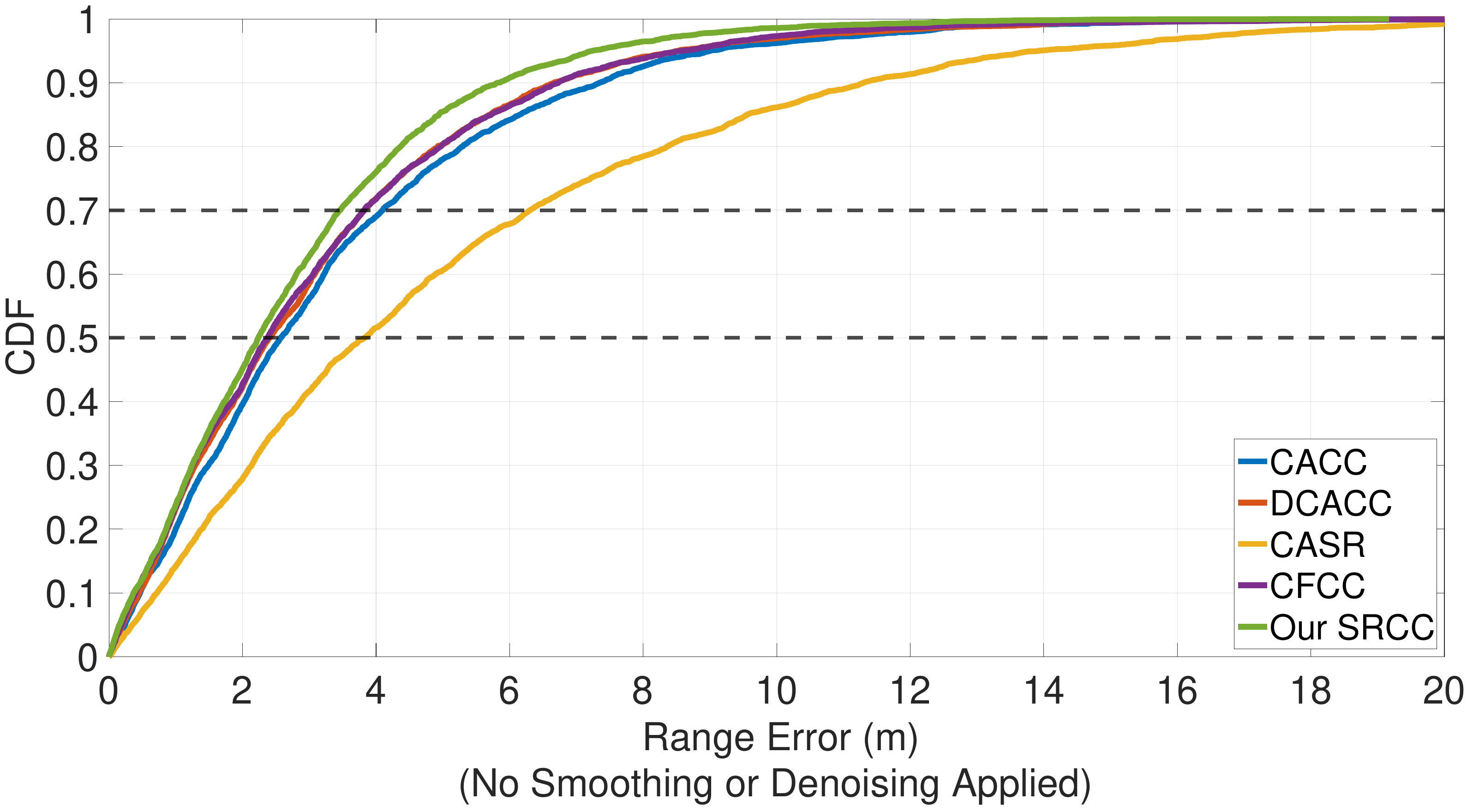}
    \subcaption{Elliptical trajectory}
    \label{Fig9a}
\end{subfigure}
\hfill
\begin{subfigure}[t]{0.325\textwidth}
    \centering
    \includegraphics[width=\textwidth]{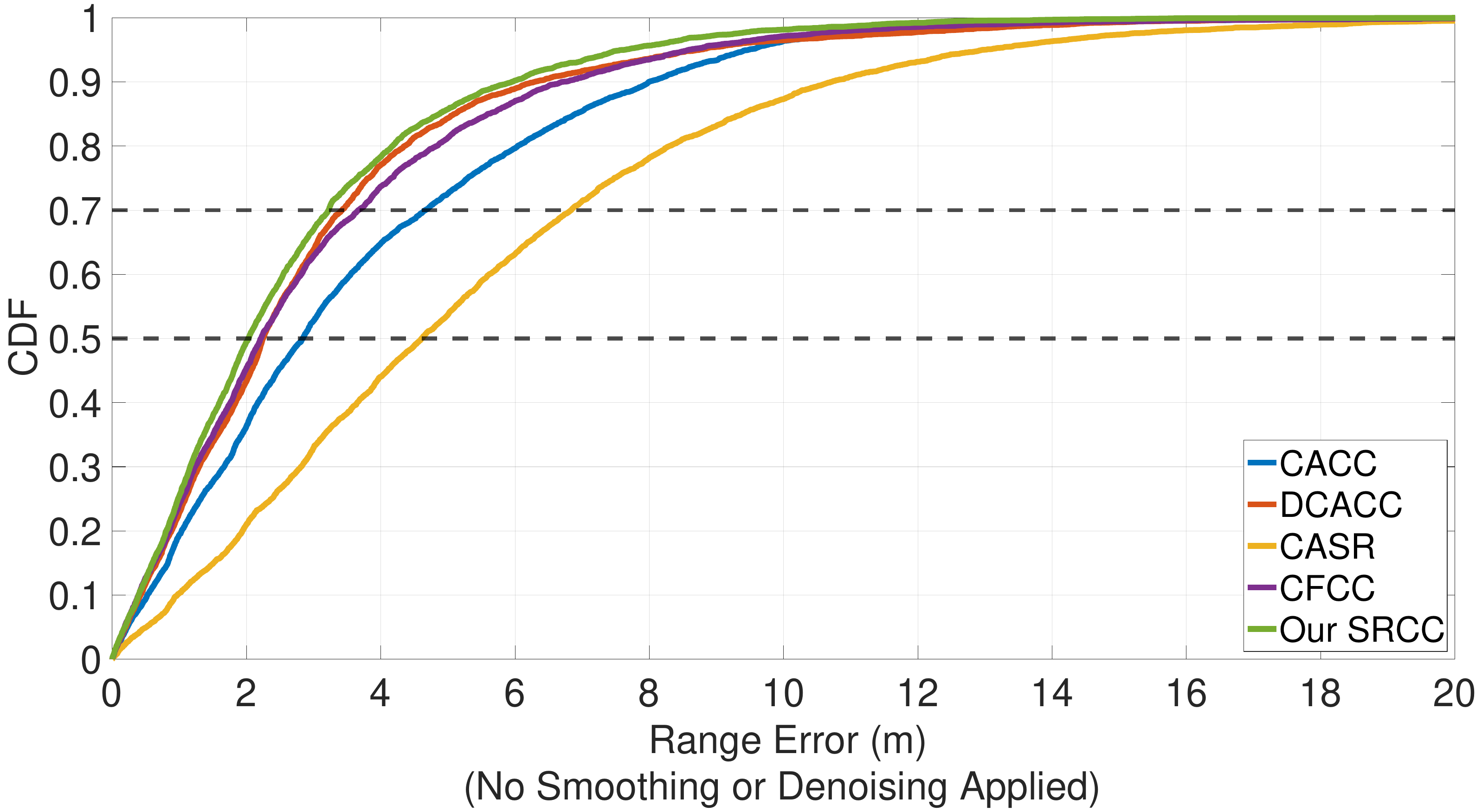}
    \subcaption{Linear trajectory}
    \label{Fig9b}
\end{subfigure}
\hfill
\begin{subfigure}[t]{0.325\textwidth}
    \centering
    \includegraphics[width=\textwidth]{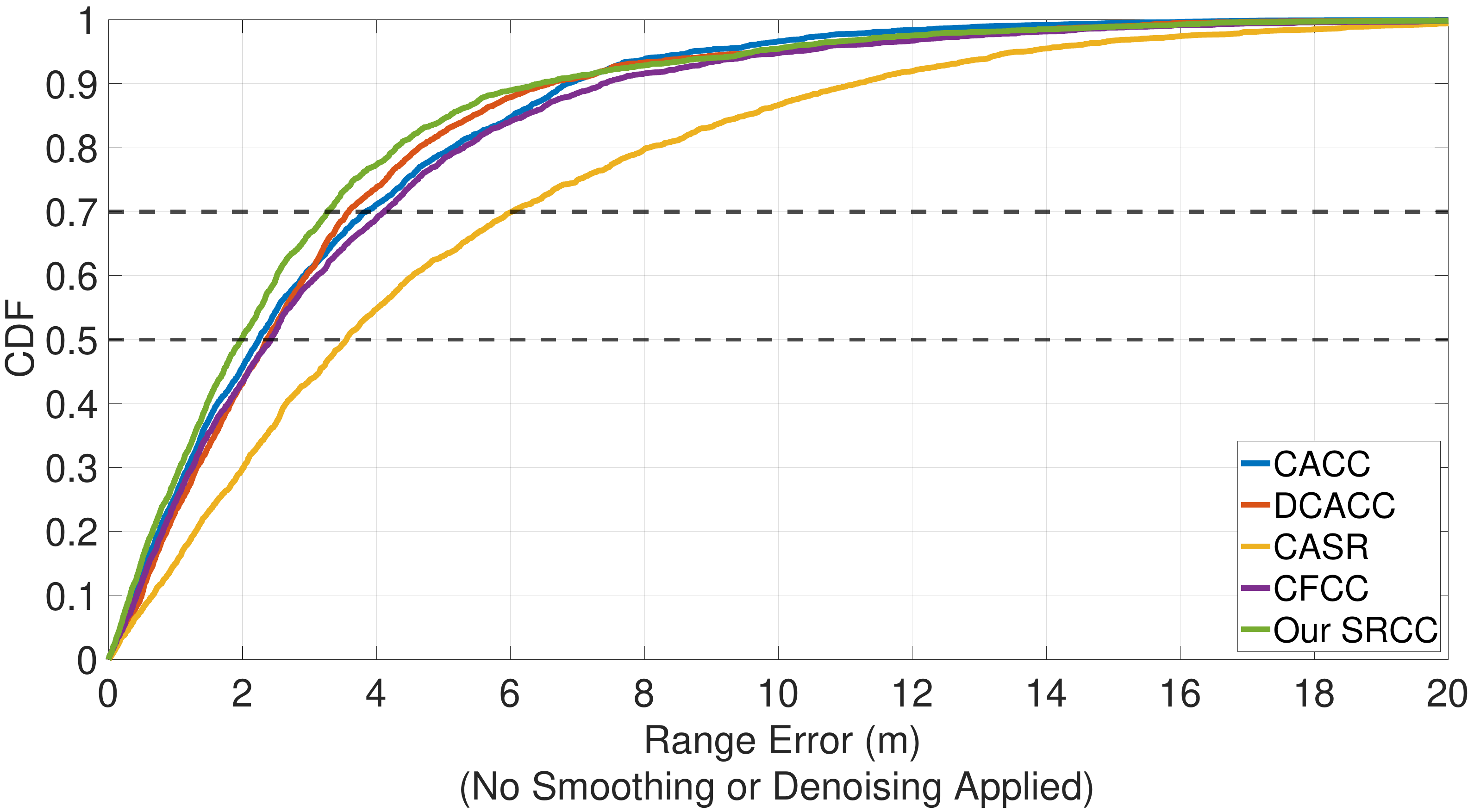}
    \subcaption{Rectangular trajectory}
    \label{Fig9c}
\end{subfigure}
\caption{CDF of range estimation error across three trajectories.}
\label{fig:range_cdf}
\vspace{-1em}
\end{figure*}

\subsubsection{Activity Recognition Datasets}
To evaluate the effectiveness of our SISO features for activity classification, we adopt single-target and multi-target activity recognition datasets, both collected using Intel 5300 NICs.

\begin{itemize}
\item \textit{Single-Target:} The Widar3.0 dataset \cite{zhang2021widar3} is collected at 5 GHz using multiple receivers, each equipped with three antennas. \textit{Dataset 1} contains six human-computer interaction gestures, including Clap (25,039 samples), Draw-O (26,543 samples), Draw-Zigzag (34,046 samples), Push \& Pull (25,799 samples), Slide (29,248 samples), and Sweep (25,049 samples), performed by 16 participants across 5 environments and 5 orientations, totalling over 1,65,724 samples. \textit{Dataset 2} contains numeric drawing gestures (0-9), collected across 5 locations and 5 orientations by 2 users. Each class contains 3,900 samples, leading to a total of 39,000 samples.

\item \textit{Multi-Target:} The WiMANS dataset \cite{huang2024wimans} is a multi-user WiFi CSI dataset captured using a 3Tx-3Rx antenna configuration at both 2.4 GHz and 5 GHz. CSI is sampled at 1 kHz, with each 3-second segment containing 3000 CSI frames in a $3 \times 3 \times 30$ format each. It covers 9 daily activities, including \textit{Nothing, Walking, Rotation, Jumping, Waving, Lying Down, Picking Up, Sitting Down, and Standing Up}, performed by 6 users across 3 environments and 5 locations, totaling 11,286 samples.
\end{itemize}

\subsection{SISO Feature Extraction Parameters}
We use the following parameters for feature extraction:

\begin{itemize}
\item \textit{CPI:} Each CPI has $128$ CSI samples (i.e., OFDM symbols), with a sliding step of $32$ between adjacent CPIs.

\item \textit{CSI Reconstruction:} The number of IFFT bins is  128. A Gaussian window with $\sigma = 64$ is applied.

\item \textit{Delay Beamforming:} MVDR beamforming is applied over a $[0, 32)$ m delay range with 1 m resolution, producing 32 bins. For single- and multi-target activity recognition datasets, the resolution is reduced to 4 m per bin, yielding 8 bins to limit data and computation.

\item \textit{Weighted Doppler FFT:} A Doppler FFT of size 128 is applied to each delay bin, covering a Doppler frequency range of $[-150, 150]$ Hz. For 5 GHz signals, this corresponds to a Doppler velocity range of $[-8, 8]$ m/s.
\end{itemize}

\subsection{Model Training Parameter}
All models are trained for 512 epochs with a batch size of 128 and an initial learning rate of 0.001 using the Adam optimizer. A step scheduler halves the learning rate every 196 epochs. Input features are Z-score normalized, and the hyperparameter $\lambda_c$ in Eq.~\eqref{equation26} is set to 0.2. For the single-target dataset, a 70\%-30\% train-test split is used, while an 80\%-20\% split is adopted for the multi-target dataset. All features are rescaled to a fixed temporal length of 64 via interpolation, downsampling longer and upsampling shorter sequences. In the delay-Doppler-time representation, the delay dimension is treated as the channel dimension for batch training. Models using delay-Doppler-time or Doppler-time features employ the augmentation strategies in Section~VI.B. To address slight class imbalance in \textit{Widar 3.0 Dataset 1}, class weighting is applied to the cross-entropy loss, with weights inversely proportional to class frequencies.

\subsection{Baselines}
\subsubsection{Delay-Doppler Feature Extraction} 
We compare our beamforming-based WiDFS 3.0 with 1Tx-1Rx SRCC \textit{(hereafter referred to as SRCC for simplicity in the following experiments)} with other approaches, all of which mitigate TO and CFO using lightweight signal processing techniques:
\begin{itemize}
\item \textit{CACC (1Tx-2Rx)} \cite{qian2018widar2}: A two-antenna method using conjugate multiplication between Rx antennas.

\item \textit{DCACC (1Tx-3Rx)} \cite{wang2023single}: A three-antenna method that first applies CACC to eliminate random phase offsets, followed by a differential CACC (DCACC) transformation across three Rx antennas to remove Doppler ambiguity.

\item \textit{CASR (1Tx-2Rx)} \cite{feng2021lte, ni2023uplink}: A two-antenna method based on CSI ratio between Rx antennas.

\item \textit{CFCC (1Tx-1Rx)} \cite{11079818}: A single-antenna method based on frequency-domain correlation across subcarriers.
\end{itemize}
All the above baseline methods estimate the target's delay and Doppler velocity using 2D FFT, without smoothing or post-processing, in order to obtain raw measurements.

\subsubsection{Activity Recognition}
We evaluate activity recognition performance using both delay-Doppler-time (multi-target scenario) and Doppler-time (single-target scenario) features. To further benchmark performance, we also include the BVP feature \cite{zhang2021widar3}, derived from a multi-receiver setup. In addition, we assess the effectiveness of compact neural network architectures. For the single-target scenario, we report accuracy, precision, recall, and F1-score. For the multi-target scenario, we use class-wise precision, recall, and F1-score to evaluate per-class performance under more complex conditions. 

\section{Experimental Results}
This section presents a comprehensive evaluation of WiDFS 3.0 in terms of feature quality and classification performance.

\begin{figure*}
\centering
\begin{minipage}[t]{0.245\linewidth}
\vspace{0pt} 
    \centering
    \begin{subfigure}{\linewidth}
        \includegraphics[width=0.975\linewidth]{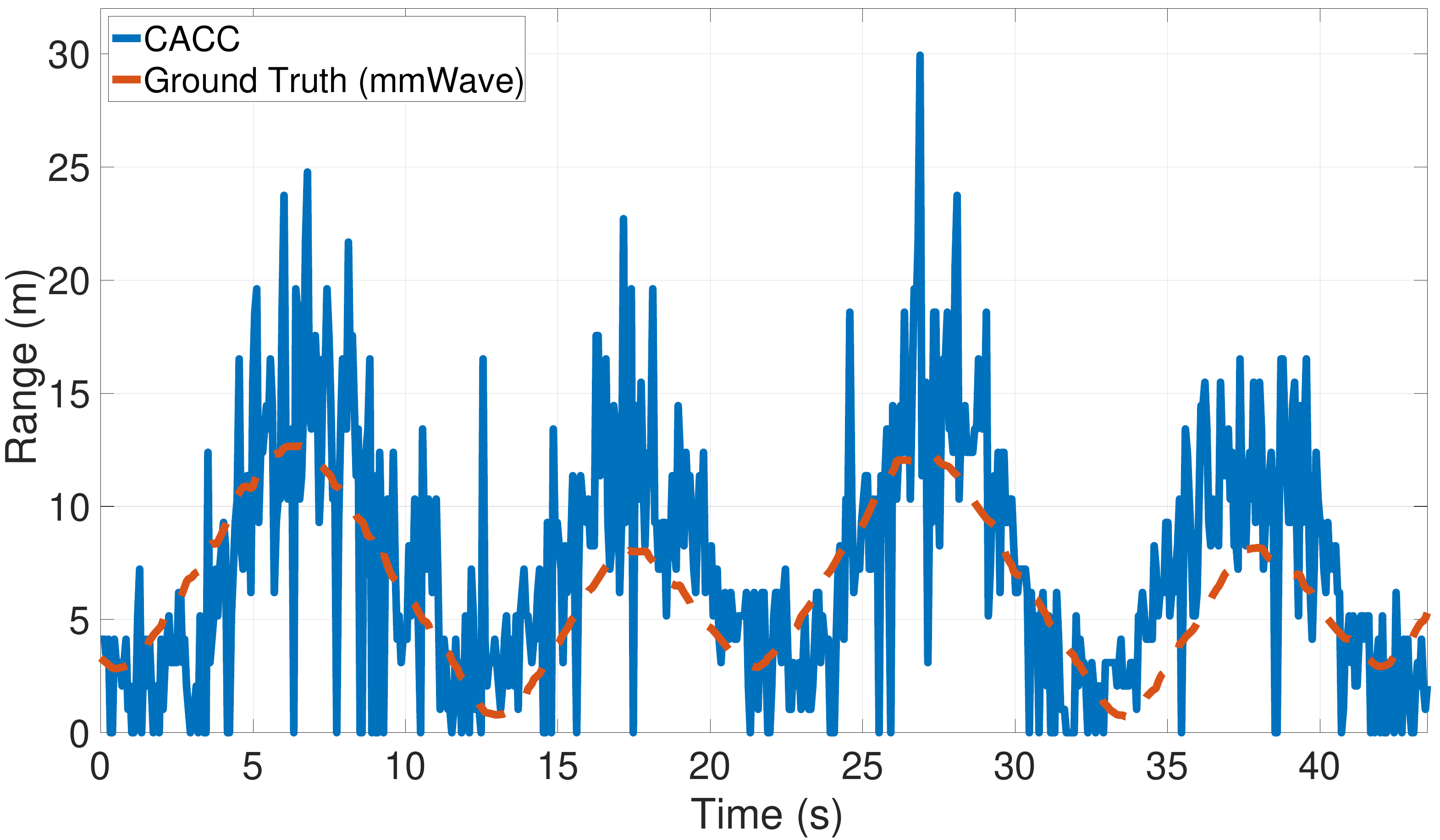}
        \subcaption{CACC (1Tx-2Rx)}
    \end{subfigure}\\
\begin{subfigure}{\linewidth}
        \includegraphics[width=0.975\linewidth]{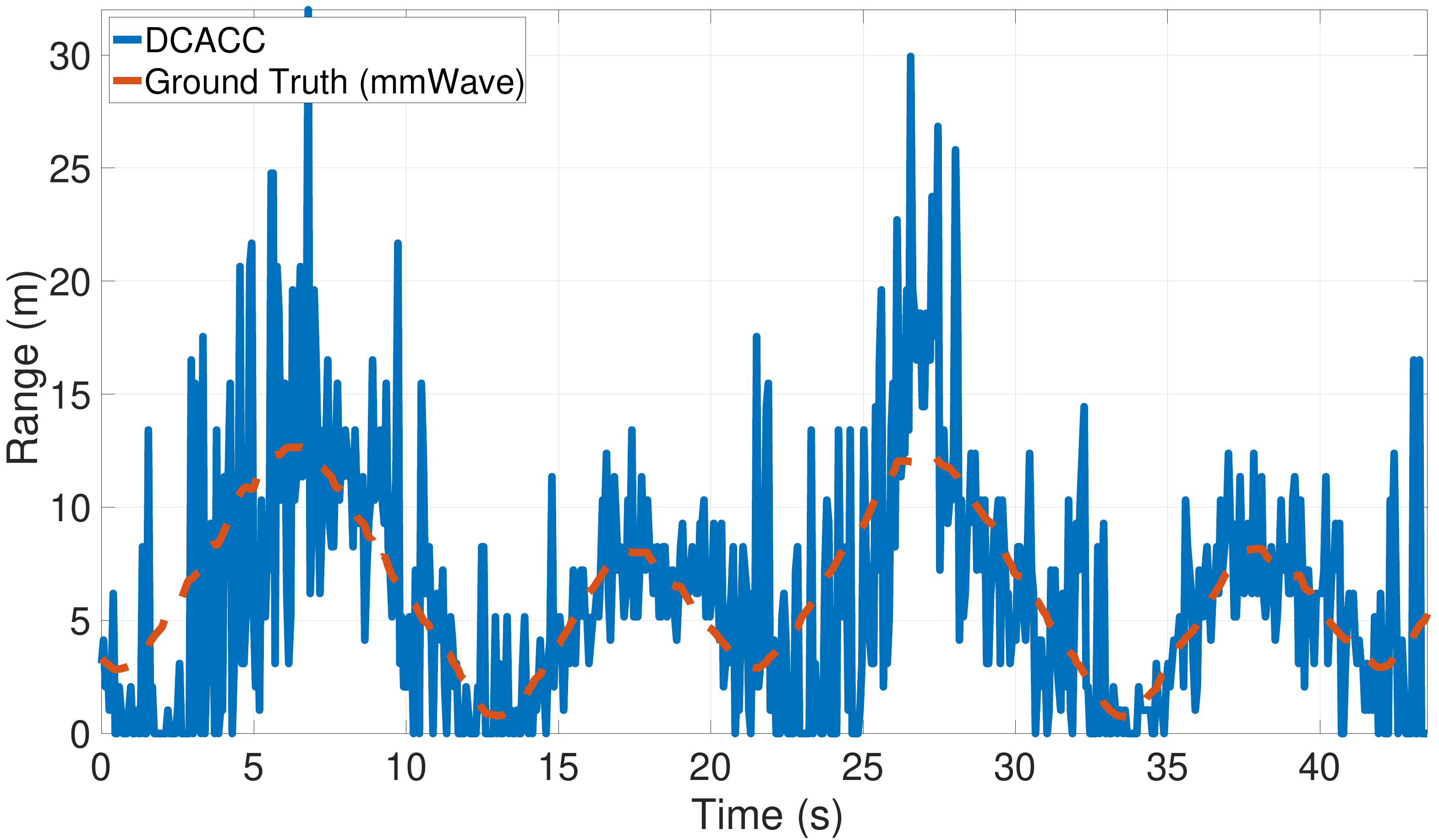}
        \subcaption{DCACC (1Tx-3Rx)}
    \end{subfigure}\\
    \begin{subfigure}{\linewidth}
        \includegraphics[width=0.975\linewidth]{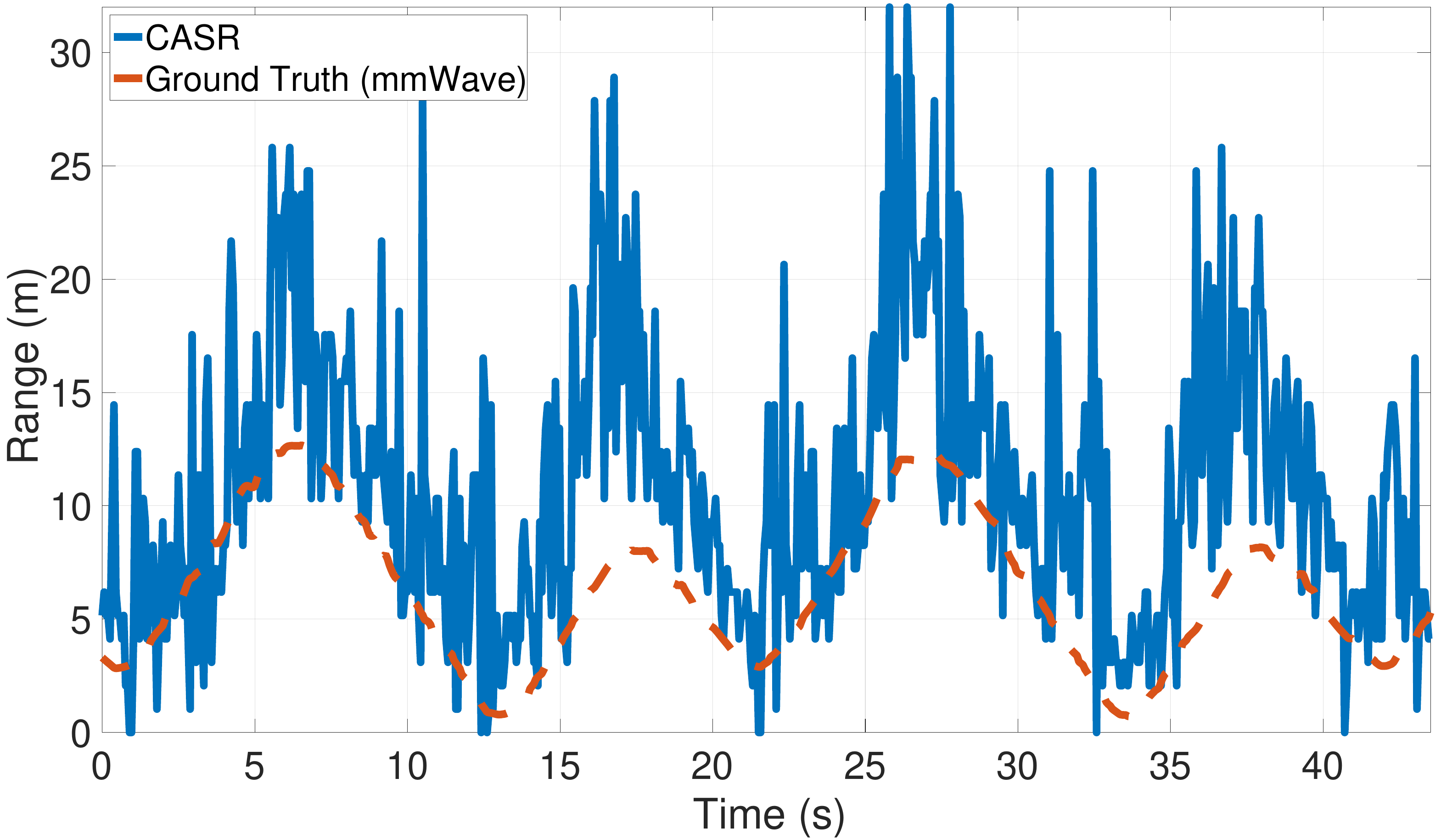}
        \subcaption{CASR (1Tx-2Rx)}
    \end{subfigure}\\
    \begin{subfigure}{\linewidth}
        \includegraphics[width=0.975\linewidth]{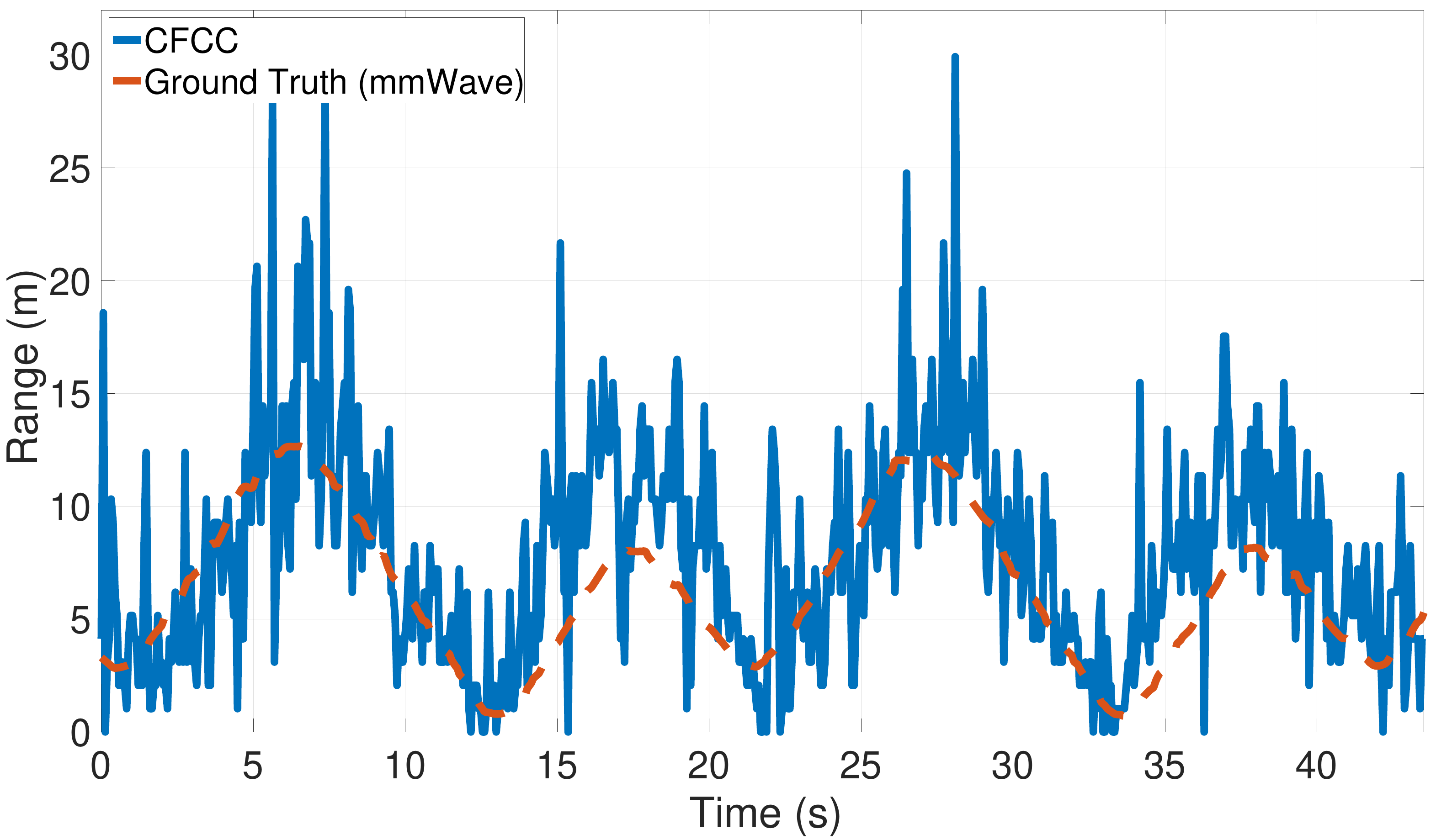}
        \subcaption{CFCC (1Tx-1Rx)}
    \end{subfigure}\\
    \begin{subfigure}{\linewidth}
        \includegraphics[width=0.975\linewidth]{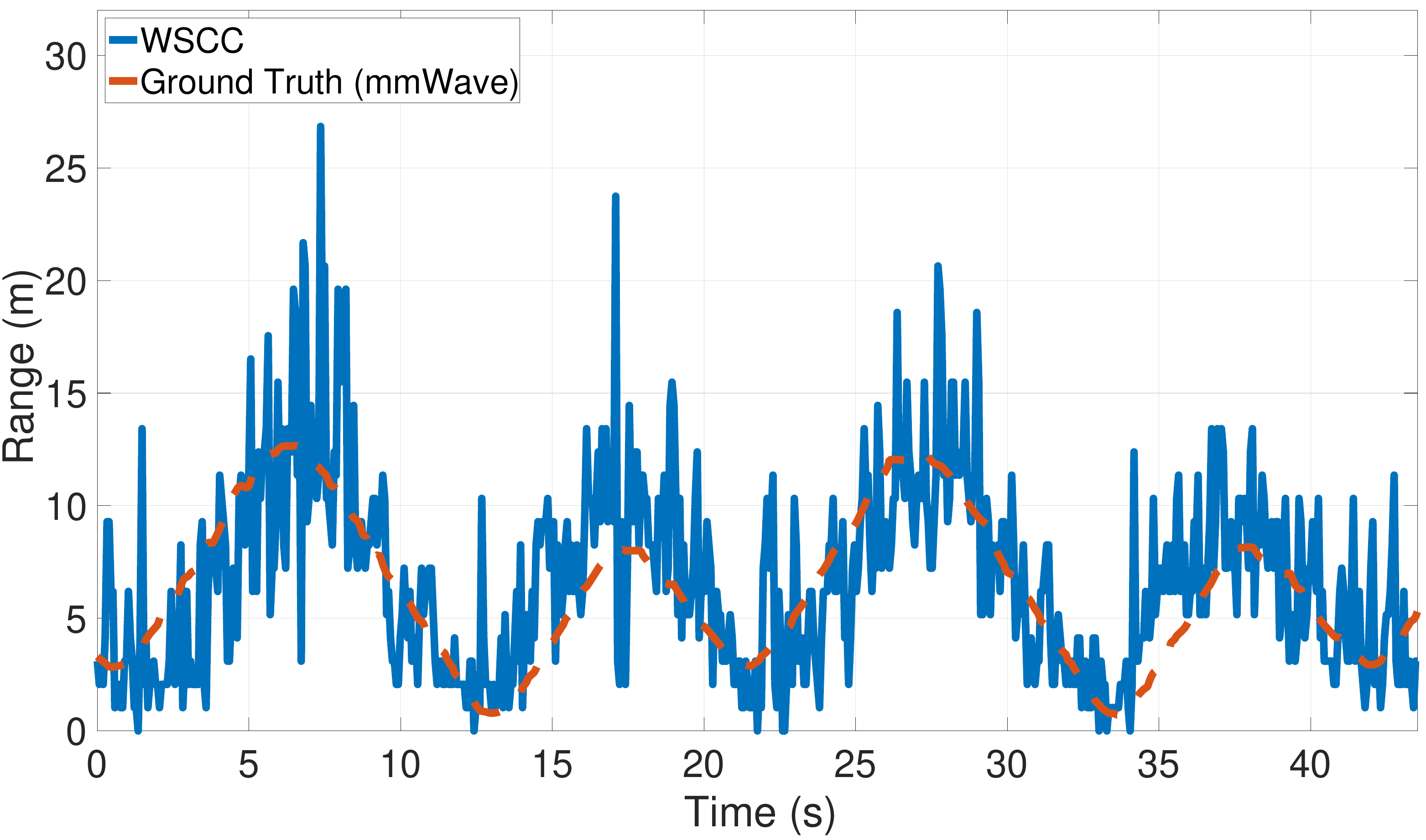}
        \subcaption{\textbf{Our SRCC (1Tx-1Rx)}}
    \end{subfigure}
    \caption{Range}
\label{figure:range_error}
\end{minipage}
%
\begin{minipage}[t]{0.245\linewidth}
\vspace{0pt} 
    \centering
    \begin{subfigure}{\linewidth}
        \includegraphics[width=0.97\linewidth]{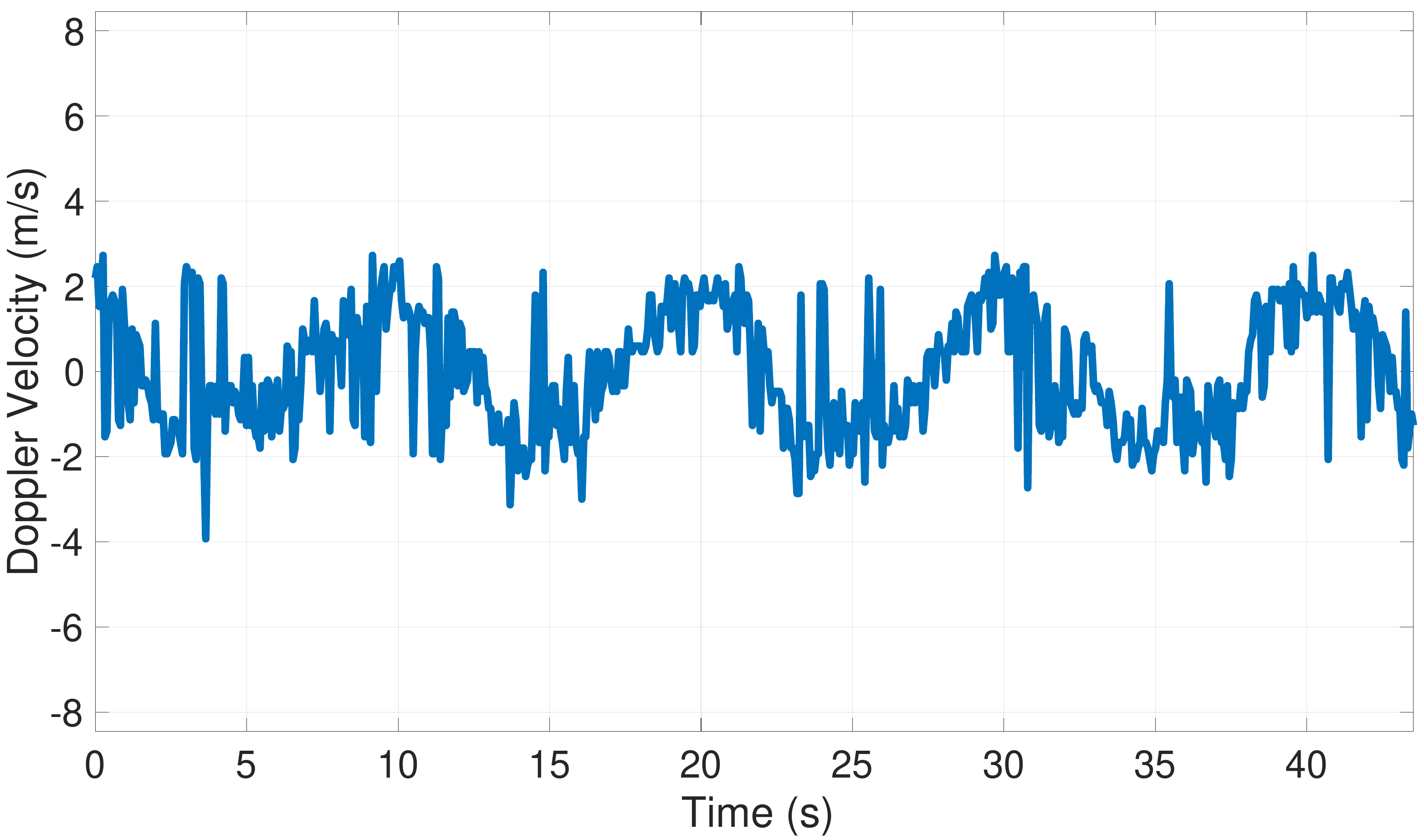}
        \subcaption{CACC (1Tx-2Rx)}
    \end{subfigure}\\
    \begin{subfigure}{\linewidth}
        \includegraphics[width=0.97\linewidth]{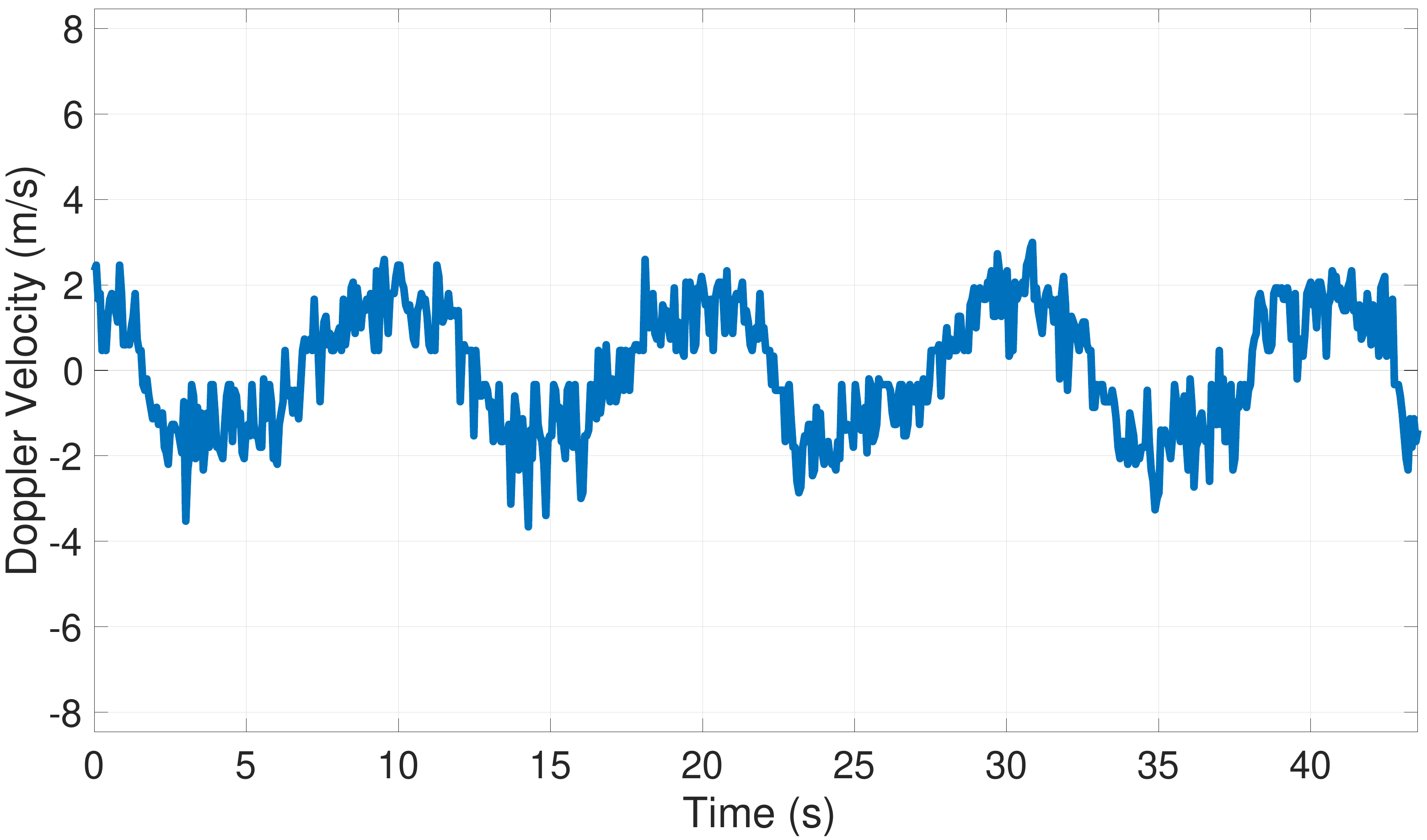}
        \subcaption{DCACC (1Tx-3Rx)}
    \end{subfigure}\\
    \begin{subfigure}{\linewidth}
        \includegraphics[width=0.97\linewidth]{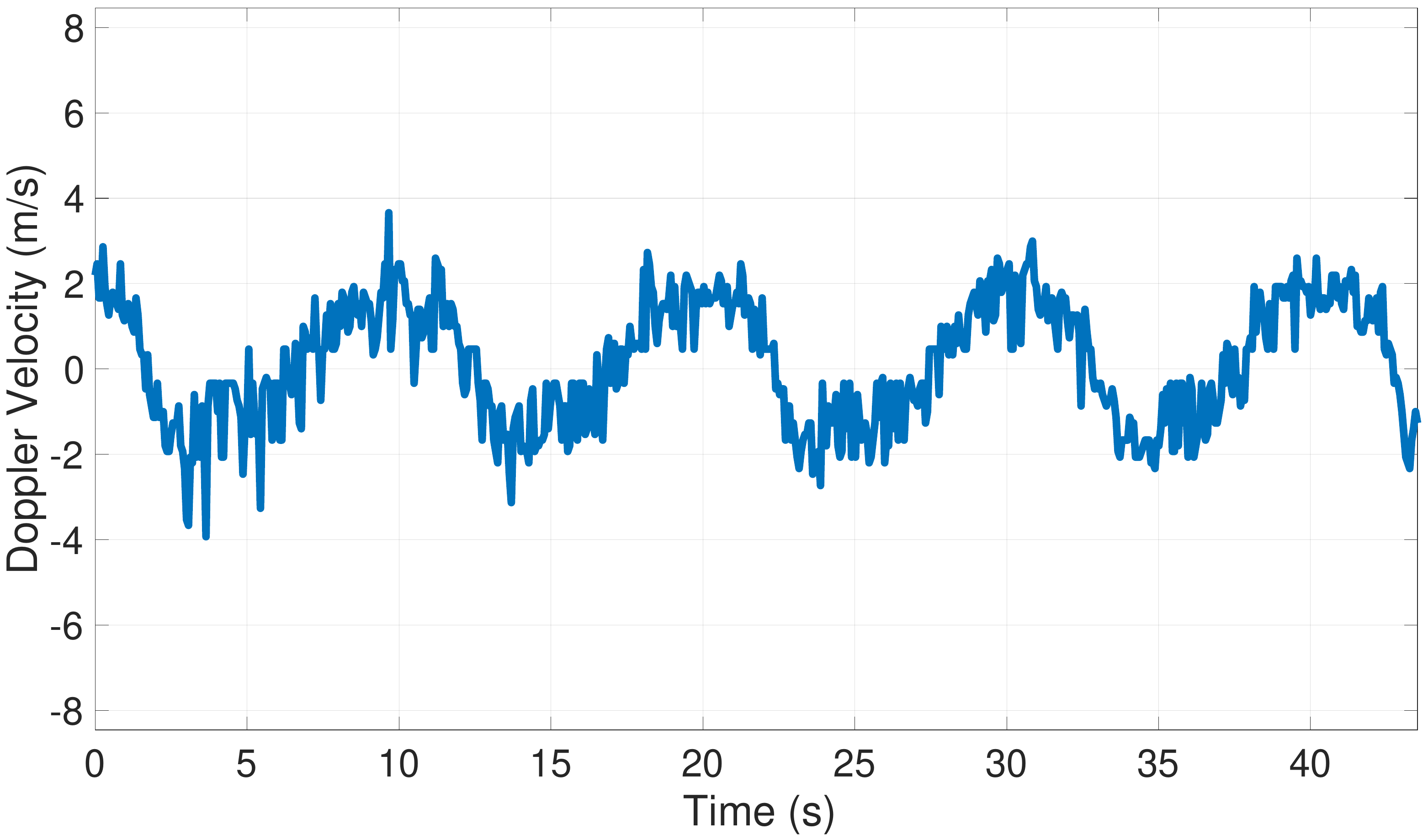}
        \subcaption{CASR (1Tx-2Rx)}
    \end{subfigure}\\
    \begin{subfigure}{\linewidth}
        \includegraphics[width=0.97\linewidth]{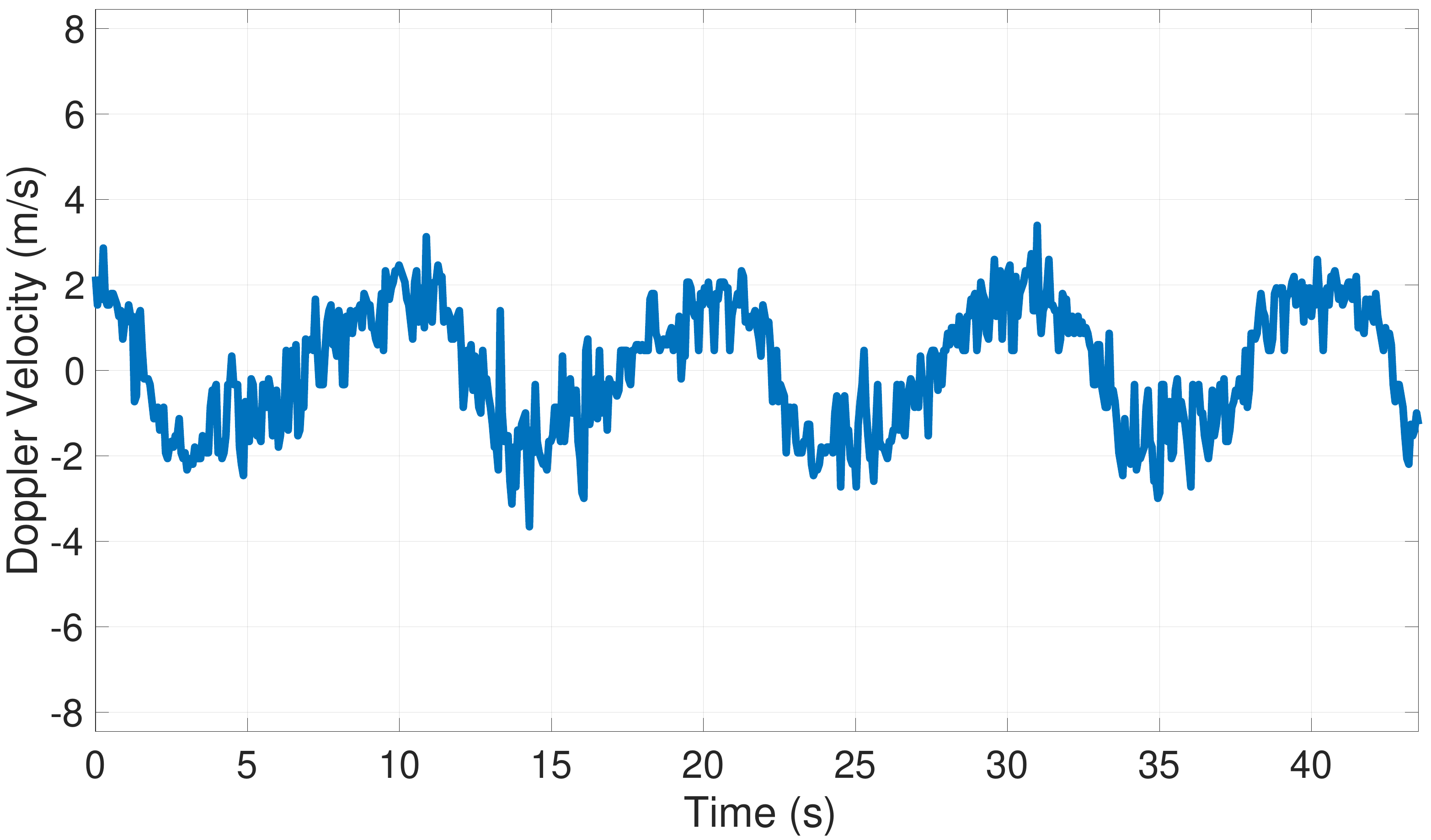}
        \subcaption{CFCC (1Tx-1Rx)}
    \end{subfigure}\\
    \begin{subfigure}{\linewidth}
        \includegraphics[width=0.97\linewidth]{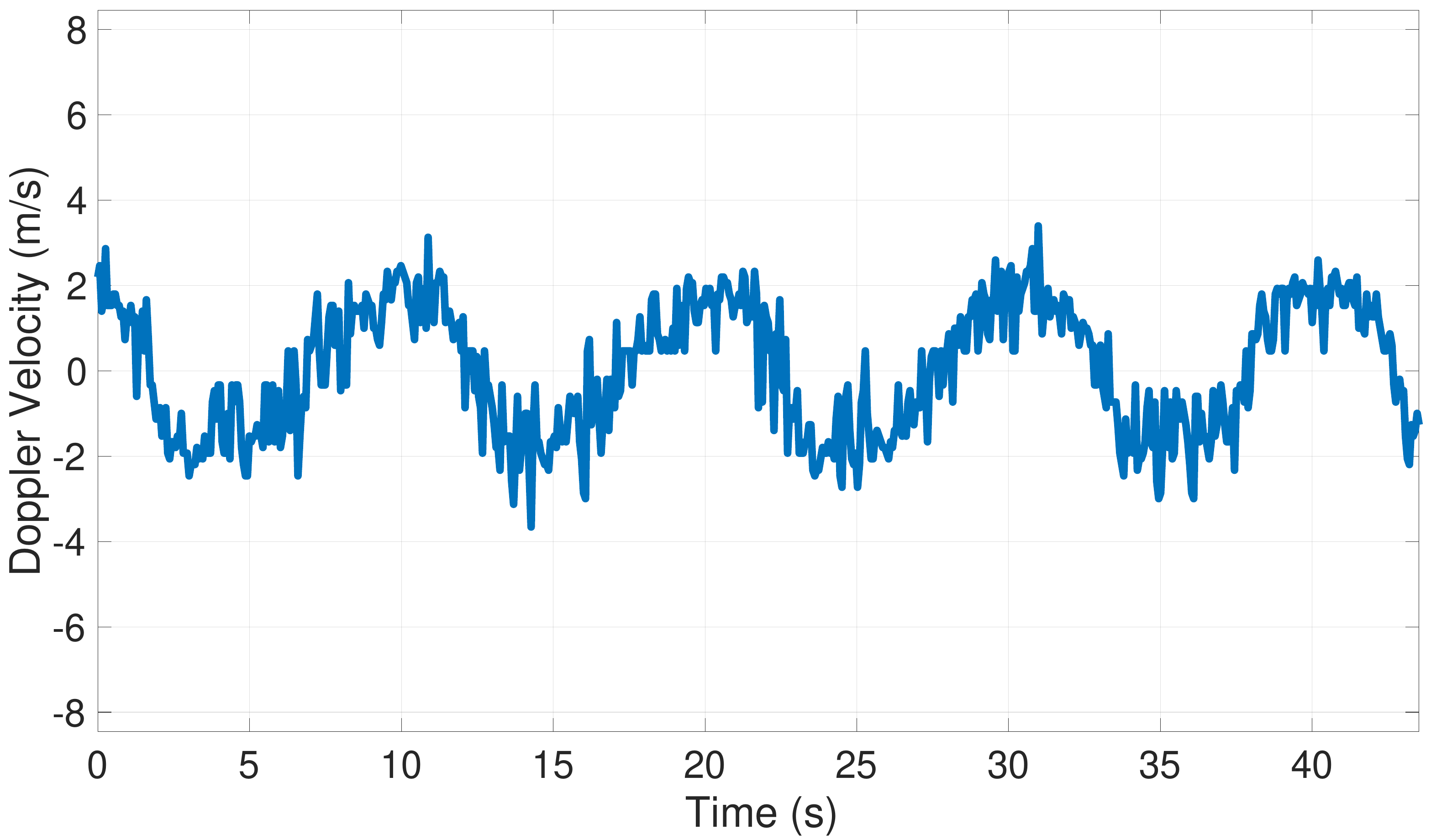}
        \subcaption{\textbf{Our SRCC (1Tx-1Rx)}}
    \end{subfigure}
    \caption{Doppler (w. delay)}
\label{figure:estimated_doppler}
\end{minipage}
%
\begin{minipage}[t]{0.245\linewidth}
\vspace{0pt} 
    \centering
    \begin{subfigure}{\linewidth}
        \includegraphics[width=\linewidth]{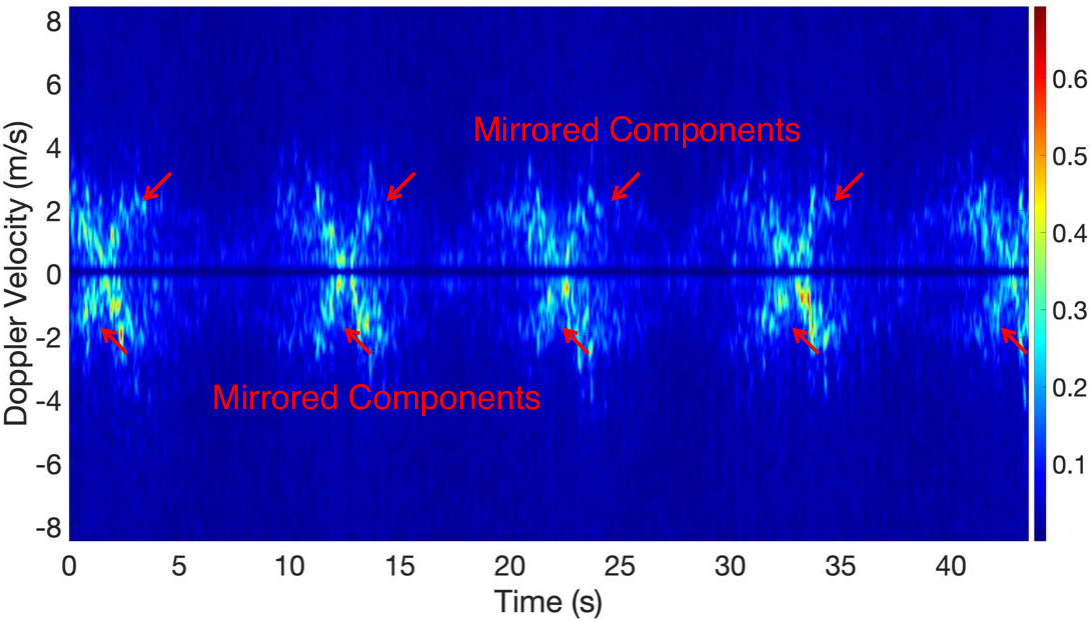}
        \subcaption{CACC (1Tx-2Rx)}
    \end{subfigure}\\
    \begin{subfigure}{\linewidth}
        \includegraphics[width=\linewidth]{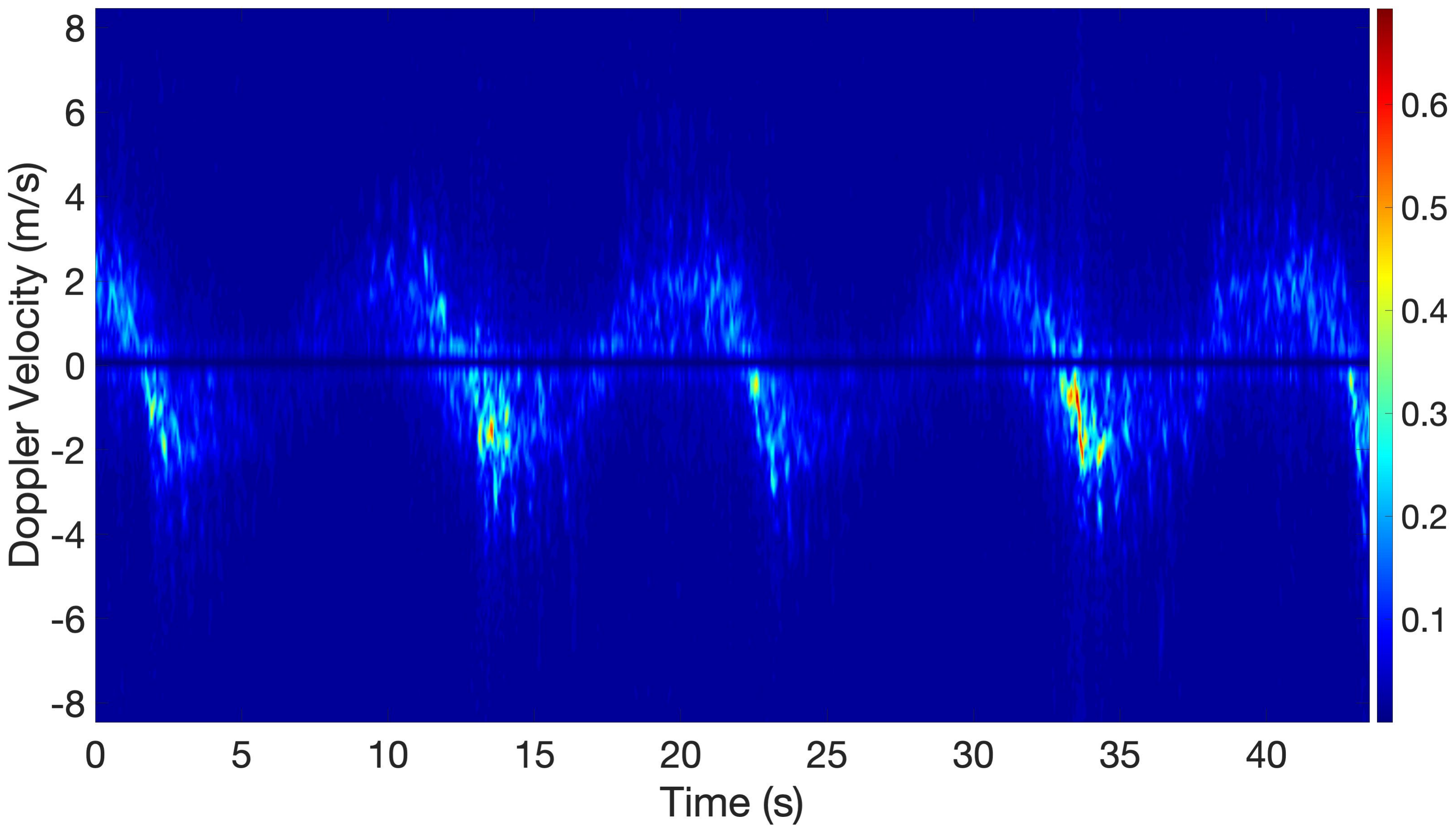}
        \subcaption{DCACC (1Tx-3Rx)}
    \end{subfigure}\\
    \begin{subfigure}{\linewidth}
        \includegraphics[width=\linewidth]{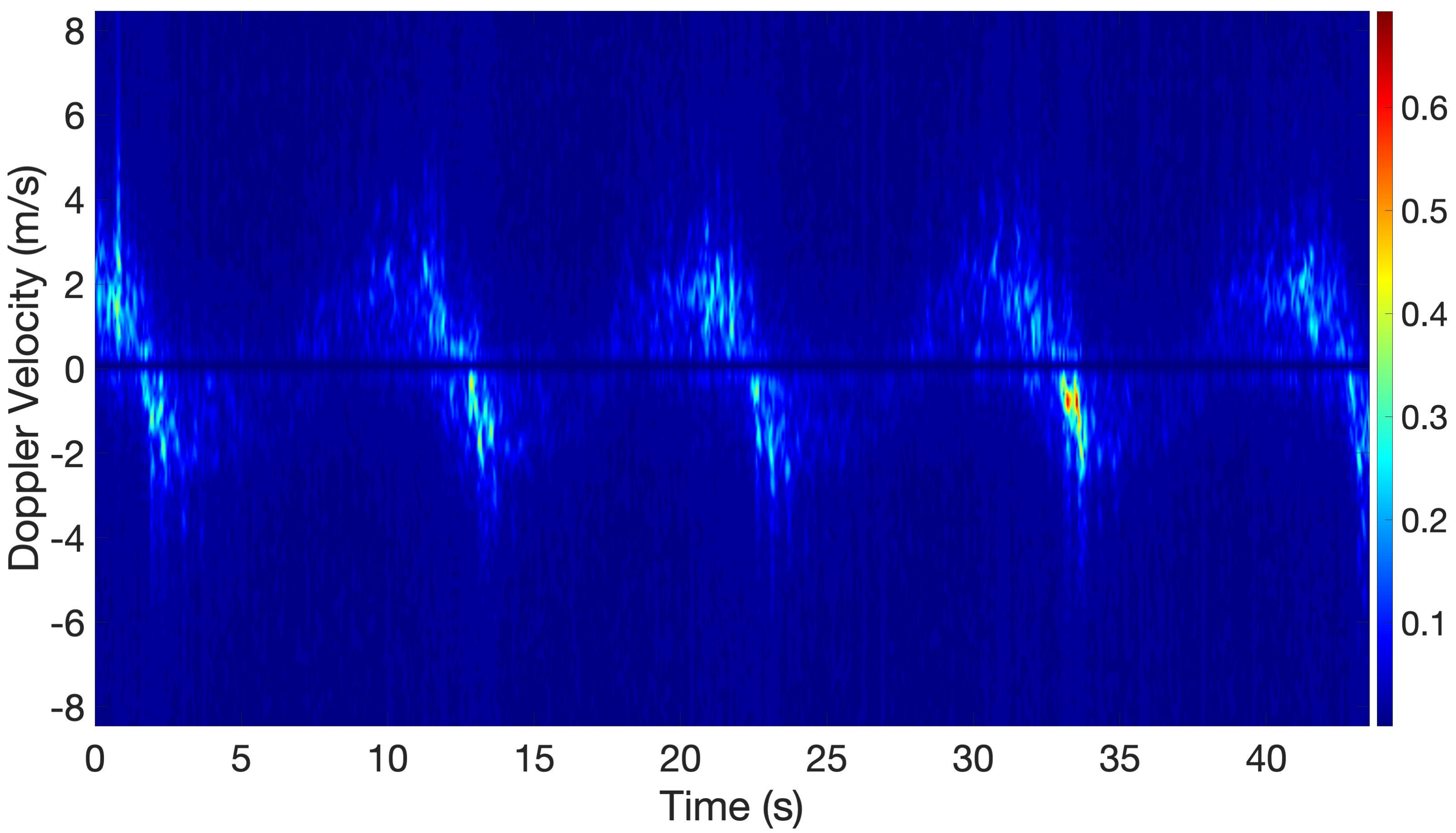}
        \subcaption{CASR (1Tx-2Rx)}
    \end{subfigure}\\
    \begin{subfigure}{\linewidth}
        \includegraphics[width=\linewidth]{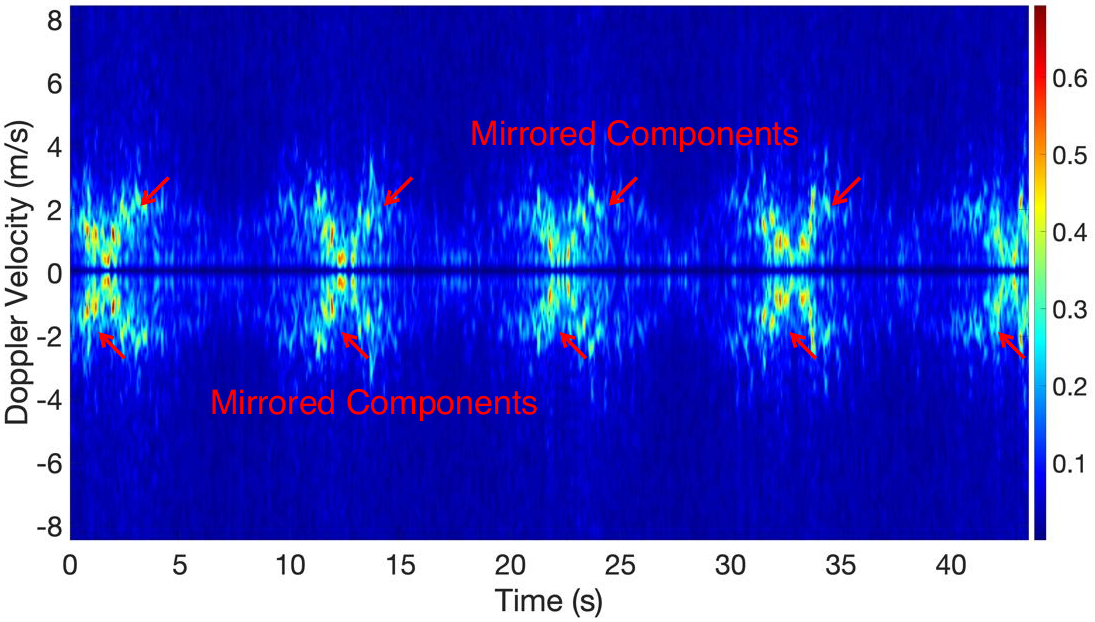}
        \subcaption{CFCC (1Tx-1Rx)}
    \end{subfigure}\\
    \begin{subfigure}{\linewidth}
        \includegraphics[width=\linewidth]{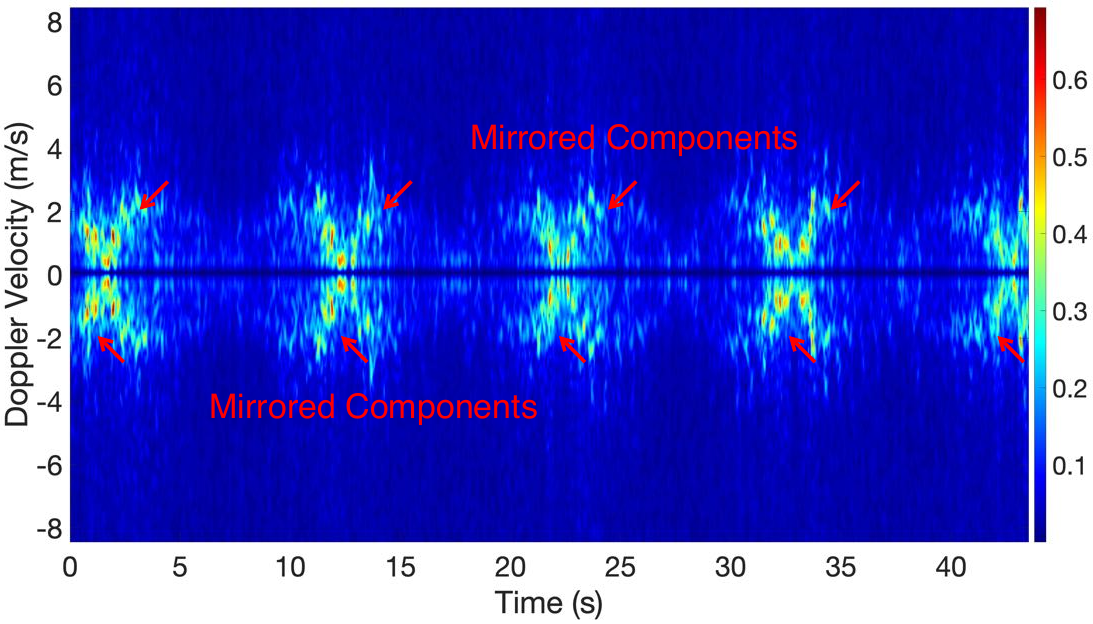}
        \subcaption{\textbf{Our SRCC (1Tx-1Rx)}}
    \end{subfigure}
    \caption{Micro-Doppler (w.o. delay)}
\label{figure:micro_doppler_wo_delay}
\end{minipage}
%
\begin{minipage}[t]{0.245\linewidth}
\vspace{0pt} 
    \centering
    \begin{subfigure}{\linewidth}
        \includegraphics[width=\linewidth]{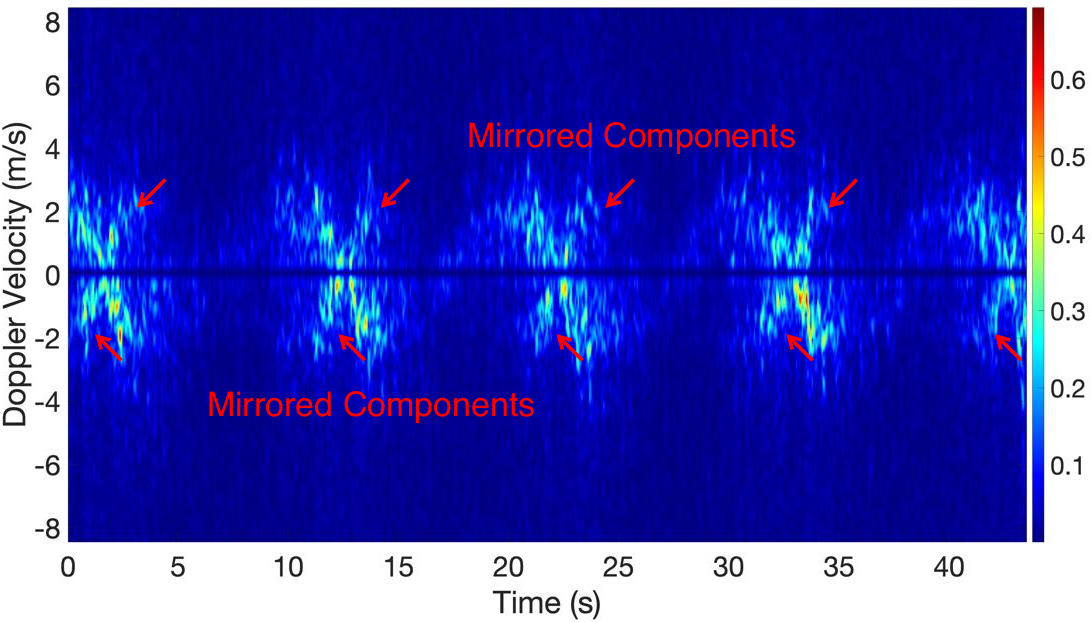}
        \subcaption{CACC (1Tx-2Rx)}
    \end{subfigure}\\
\begin{subfigure}{\linewidth}
        \includegraphics[width=\linewidth]{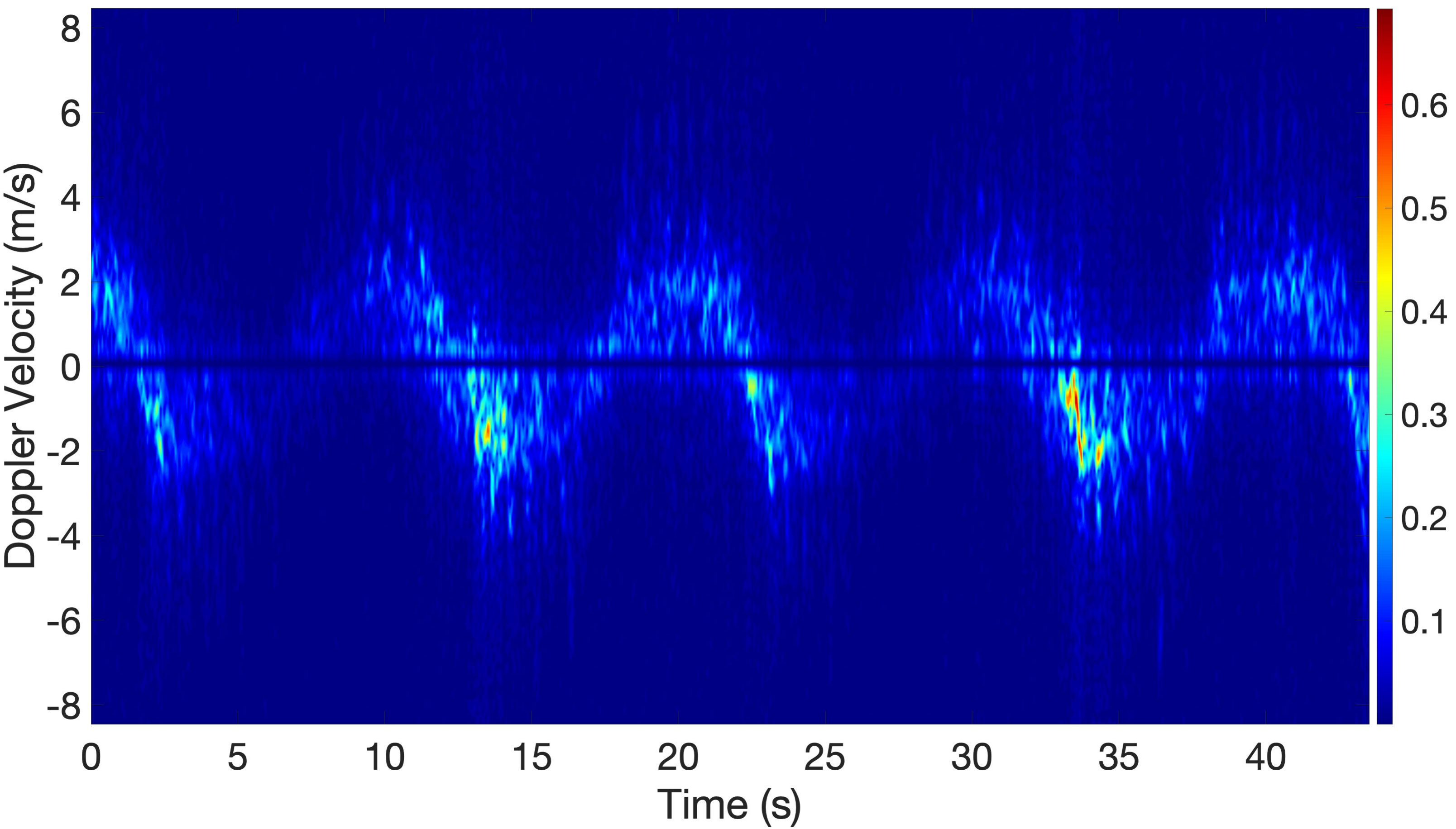}
       \subcaption{DCACC (1Tx-3Rx)}
    \end{subfigure}\\
    \begin{subfigure}{\linewidth}
        \includegraphics[width=\linewidth]{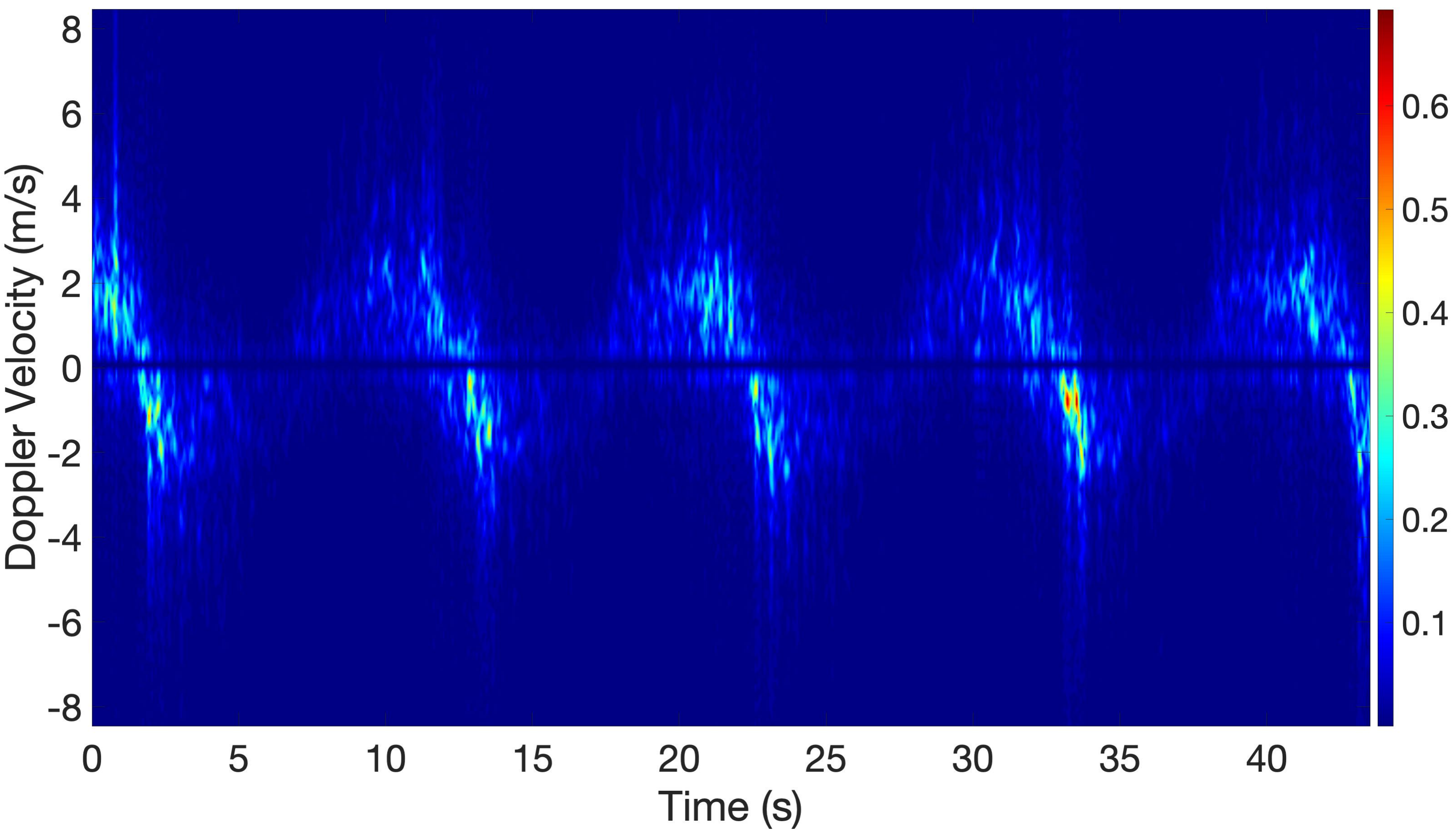}
        \subcaption{CASR (1Tx-2Rx)}
    \end{subfigure}\\
    \begin{subfigure}{\linewidth}
        \includegraphics[width=\linewidth]{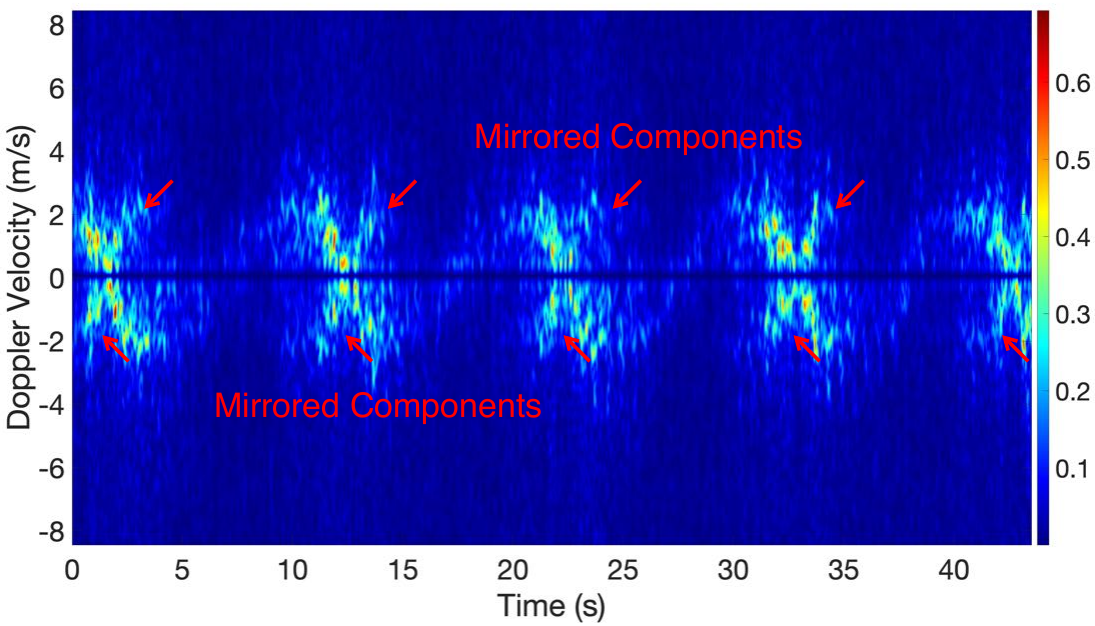}
        \subcaption{CFCC (1Tx-1Rx)}
    \end{subfigure}\\
    \begin{subfigure}{\linewidth}
        \includegraphics[width=\linewidth]{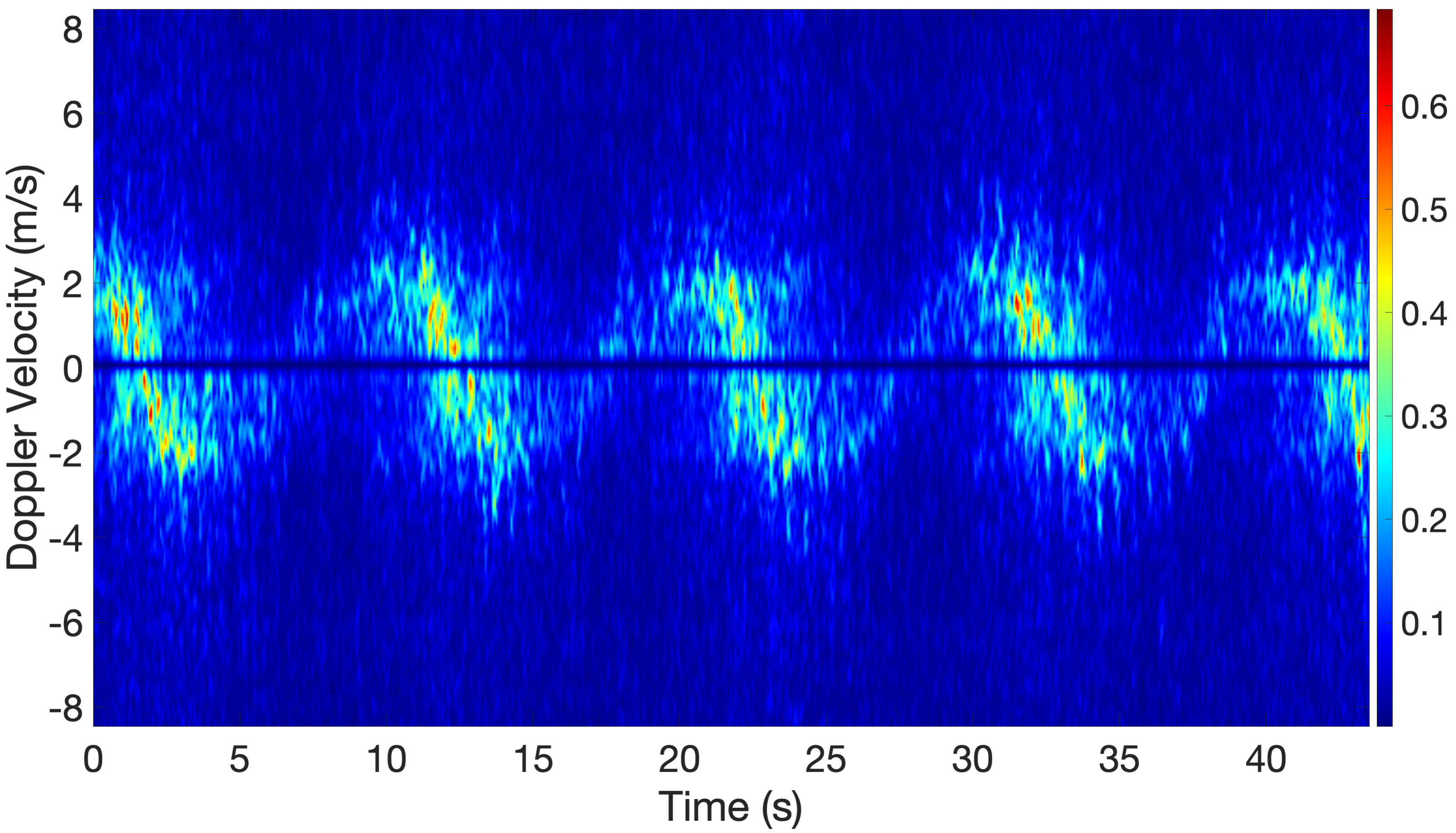}
       \subcaption{\textbf{Our SRCC (1Tx-1Rx)}}
    \end{subfigure}
    \caption{Micro-Doppler (w. delay)}
\label{figure:micro_doppler_w_delay}
\end{minipage}
\vspace{-1em}
\end{figure*}

\subsection{Delay-Doppler Feature Extraction Quality}
\begin{table}[t]
\caption{{Range (i.e., delay) estimation error (m) at 50\% and 70\% percentiles}. Our SRCC achieves the lowest range estimation error across all trajectories. \textit{No smoothing or denoising is applied.}}
\centering
\renewcommand{\arraystretch}{1.2}
\begin{tabularx}{\linewidth}{
>{\centering\arraybackslash}m{0.8cm}  
>{\centering\arraybackslash}m{1.0cm}  
*{4}{>{\centering\arraybackslash}X}   
>{\centering\arraybackslash}m{1.2cm}  
}
\toprule
\textbf{Percentile} & \textbf{Track} & \textbf{CACC} & \textbf{DCACC} & \textbf{CASR} & \textbf{CFCC} & \textbf{\mbox{Our SRCC}} \\
\midrule
\multirow{3}{*}{50\%} 
& Ellipse   & 2.57\,m & 2.40\,m & \textit{3.83\,m} & 2.37\,m & \textbf{2.16\,m} \\
& Linear    & 2.84\,m & 2.25\,m & \textit{4.61\,m} & 2.21\,m & \textbf{2.02\,m} \\
& Rectangle & 2.24\,m & 2.36\,m & \textit{3.55\,m} & 2.42\,m & \textbf{1.98\,m} \\
\midrule
\multirow{3}{*}{70\%} 
& Ellipse   & 4.10\,m & 3.82\,m & \textit{6.30\,m} & 3.82\,m & \textbf{3.42\,m} \\
& Linear    & 4.67\,m & 3.42\,m & \textit{6.83\,m} & 3.66\,m & \textbf{3.22\,m} \\
& Rectangle & 3.84\,m & 3.59\,m & \textit{6.02\,m} & 4.12\,m & \textbf{3.27\,m} \\
\bottomrule
\end{tabularx}
\label{table:range_error}
\vspace{-1em}
\end{table}

\subsubsection{Delay}
Fig. \ref{fig:range_cdf} and Table. \ref{table:range_error} report the raw range estimation error (computed by converting delay to distance using the speed of light for intuitive interpretation) at the 50\% and 70\% percentiles across three trajectories. Fig. \ref{figure:range_error} visualizes the raw range variations for a segment of the elliptical trajectory. Our WiDFS 3.0 with the 1Tx-1Rx SRCC method achieves the lowest error, with the median errors of 2.16 m, 2.02 m, and 1.98 m for the ellipse, linear, and rectangle trajectories, respectively. Even at the 70\% percentile, our SRCC maintains the lowest estimation errors with 3.42 m, 3.22 m, and 3.27 m, respectively. In contrast, the 1Tx-2Rx CASR exhibits the highest estimation errors due to nonlinear distortions across subcarriers. The 1Tx-2Rx CACC suffers from limited mirror suppression. The 1Tx-3Rx DCACC and 1Tx-1Rx CFCC have similar levels of accuracy by using linear transformations without introducing nonlinear artifacts. Overall, our beamforming-based pipeline can achieve better noise suppression.

\subsubsection{Doppler}
Fig. \ref{figure:estimated_doppler} illustrates the extracted Doppler profiles for a single moving target. Except for the 1Tx-2Rx CACC, other methods show similar Doppler values. The CACC suffers from strong Doppler mirroring, where the power at $+f^D$ and $-f^D$ is nearly symmetric, making it challenging to distinguish the correct Doppler bin. Additionally, the antenna pair selection for CACC makes it highly sensitive: when the chosen antenna pair has large signal energy asymmetry, doppler mirror suppression can be effective; however, in cases with small energy differences, the mirror problem become severe. Even with delay-domain filtering, the suppression capability remains limited due to the inherently narrow bandwidth. In contrast, 1Tx-3Rx DCACC and 1Tx-2Rx CASR utilize spatial domain across Rx antennas to achieve more robust mirror suppression. Our 1Tx-1Rx SRCC and CFCC can improve Doppler clarity by incorporating delay-domain filtering.

\subsubsection{Micro-Doppler}
\begin{figure*}
\centering
    \begin{subfigure}[t]{0.245\linewidth}
        \includegraphics[width=\linewidth]{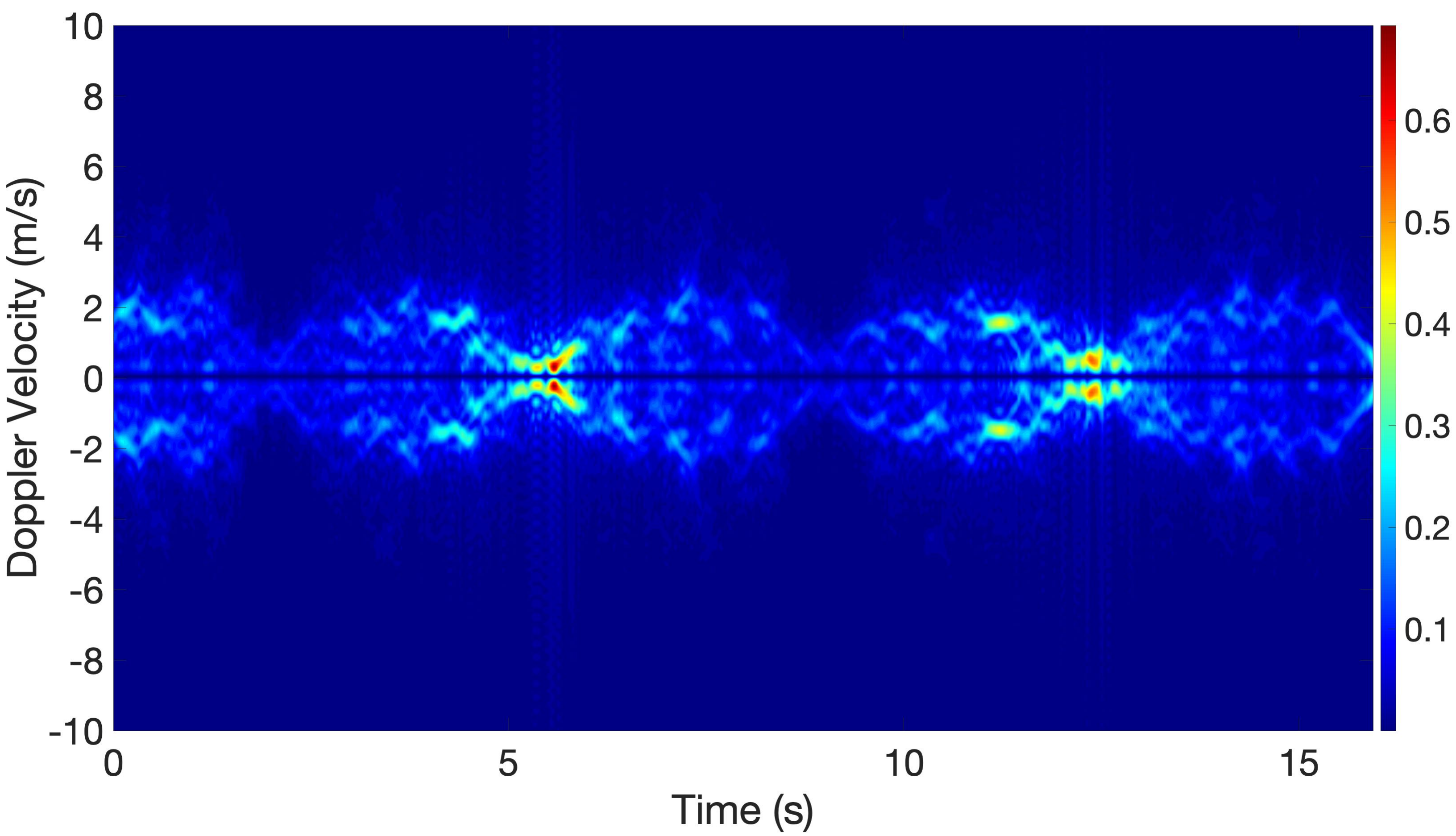}
        \subcaption{CFCC (w.o. delay)}
    \end{subfigure}
    \begin{subfigure}[t]{0.245\linewidth}
        \includegraphics[width=\linewidth]{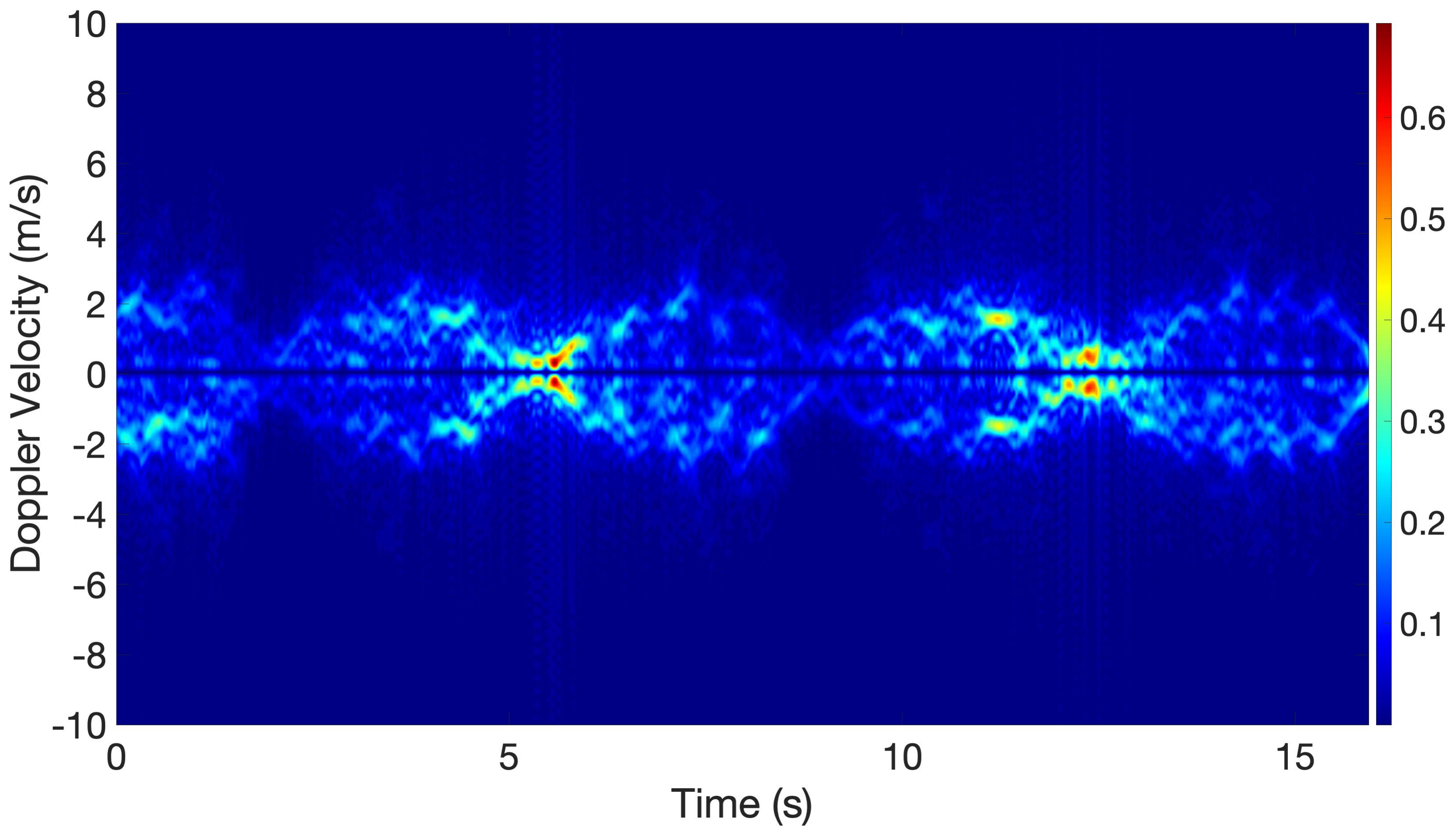}
        \subcaption{CFCC (w. delay)}
    \end{subfigure}
\begin{subfigure}[t]{0.245\linewidth}
        \includegraphics[width=\linewidth]{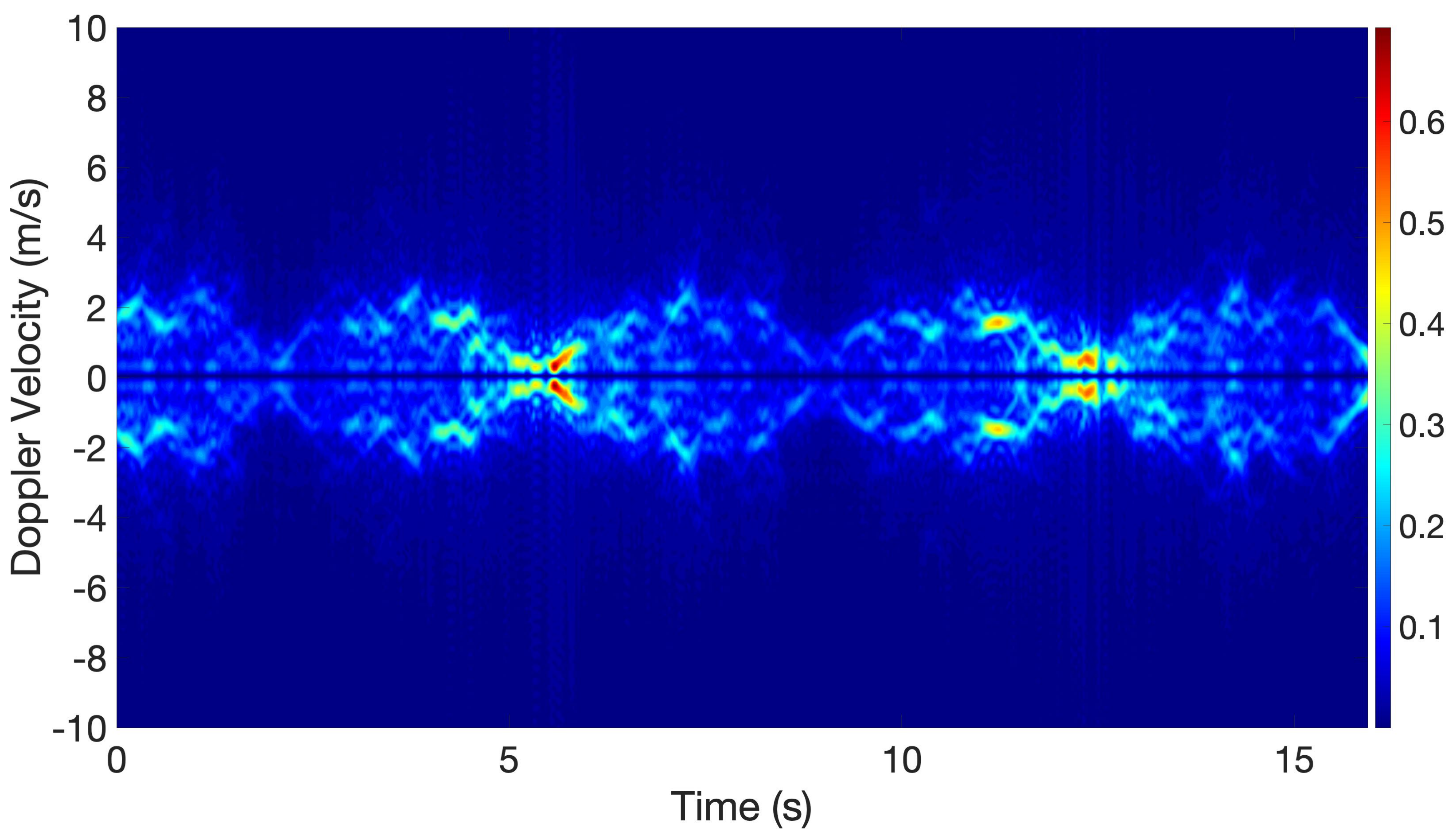}
        \subcaption{Our SRCC (w.o. delay)}
    \end{subfigure}
    \begin{subfigure}[t]{0.245\linewidth}
        \includegraphics[width=\linewidth]{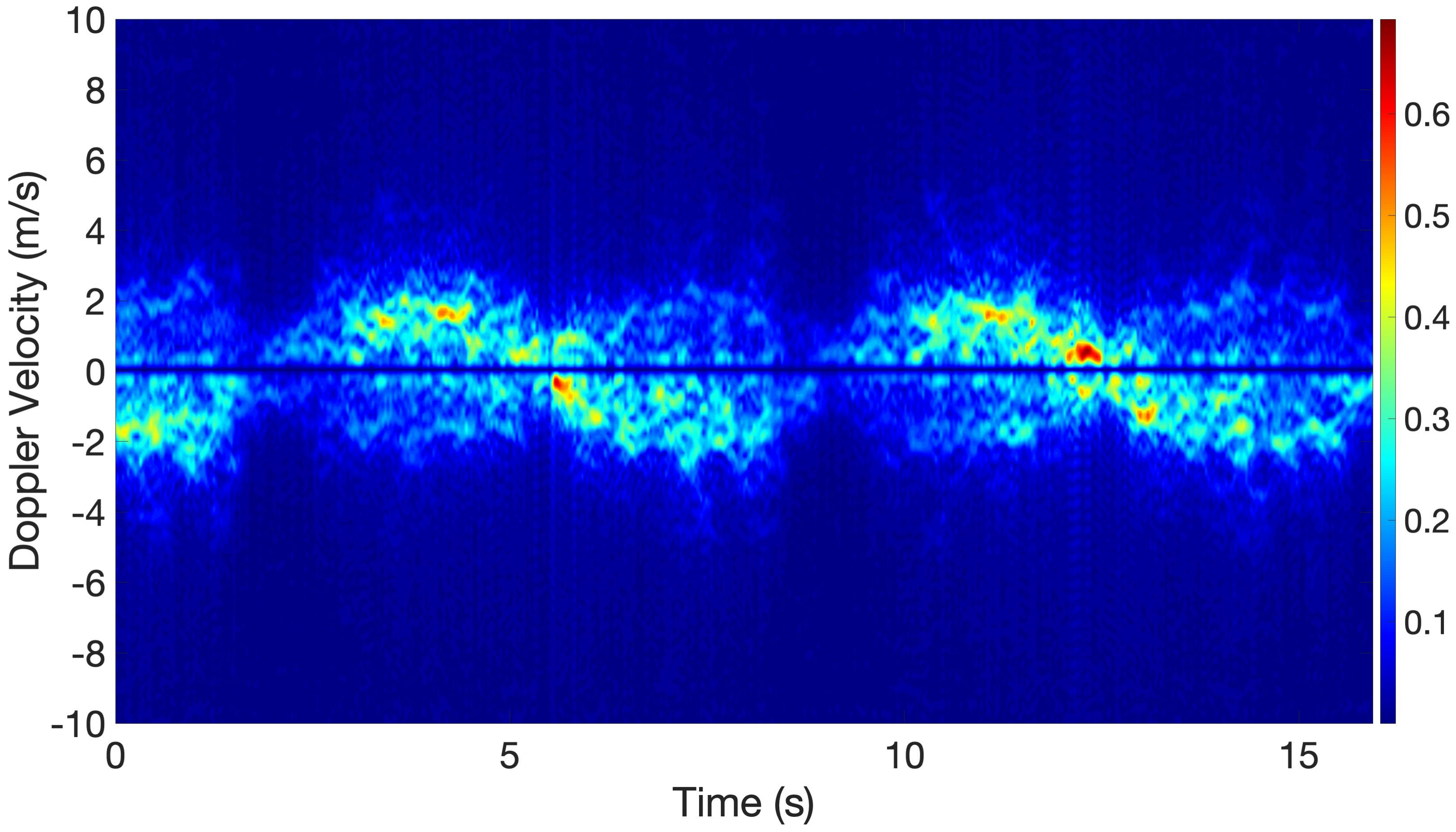}
        \subcaption{\textbf{Our SRCC (w. delay)}}
    \end{subfigure}
\caption{Comparison of 1Tx-1Rx micro-Doppler signatures under 3.1 GHz LTE signals.}
\label{fig:microdoppler_LTE}
\vspace{-1em}
\end{figure*}

Fig. \ref{figure:micro_doppler_wo_delay} and Fig. \ref{figure:micro_doppler_w_delay} present the micro-Doppler signatures extracted without (w.o.) and with (w.) delay-domain filtering, respectively. Without delay filtering, obvious mirrored Doppler components appear in CACC, as well as in 1Tx-1RX CFCC and SRCC. These symmetric artifacts around the zero-Doppler axis severely compromise the interpretation of target dynamics. After applying delay-domain filtering, the three methods exhibit improved suppression of the mirrored components. However, CACC and CFCC still retain residual mirrored energy, whereas our SRCC yields significantly cleaner micro-Doppler signatures. This demonstrates the effectiveness of our delay-aligned beamforming in isolating motion-induced Doppler features and eliminating aliased components. For multi-antenna DCACC and CASR, both leverage spatial domain information to suppress Doppler mirroring, producing clean spectrograms without symmetric distortions. However, CASR suffers from nonlinear distortions in the subcarrier domain, resulting in degraded range estimation. In addition, Fig. \ref{fig:microdoppler_LTE} compares 1Tx-1Rx CFCC and SRCC on LTE data at 3.1 GHz. Compared to 5 GHz WiFi, the LTE signals have a longer wavelength, resulting in smaller Doppler shifts and lower sensing sensitivity. Consistent with the previous WiFi-based results, both methods initially exhibit severe Doppler mirroring when delay filtering is not applied. With delay filtering, CFCC achieves partial suppression, while our SRCC achieves much stronger attenuation of mirrored components, demonstrating robustness across different signals. 

In summary, DCACC effectively suppresses Doppler mirroring and enables accurate range estimation in multi-antenna systems, but AoA estimation requires complex antenna calibration \cite{wang2023single}. In contrast, our SISO SRCC delivers comparable Doppler quality while inherently avoiding inter-antenna phase inconsistencies caused by clock asynchrony and hardware diversity, making it well-suited for compact, low-cost, and calibration-free deployments.

\begin{figure*}
\centering
    \begin{subfigure}[t]{0.325\linewidth}
        \includegraphics[width=\linewidth]{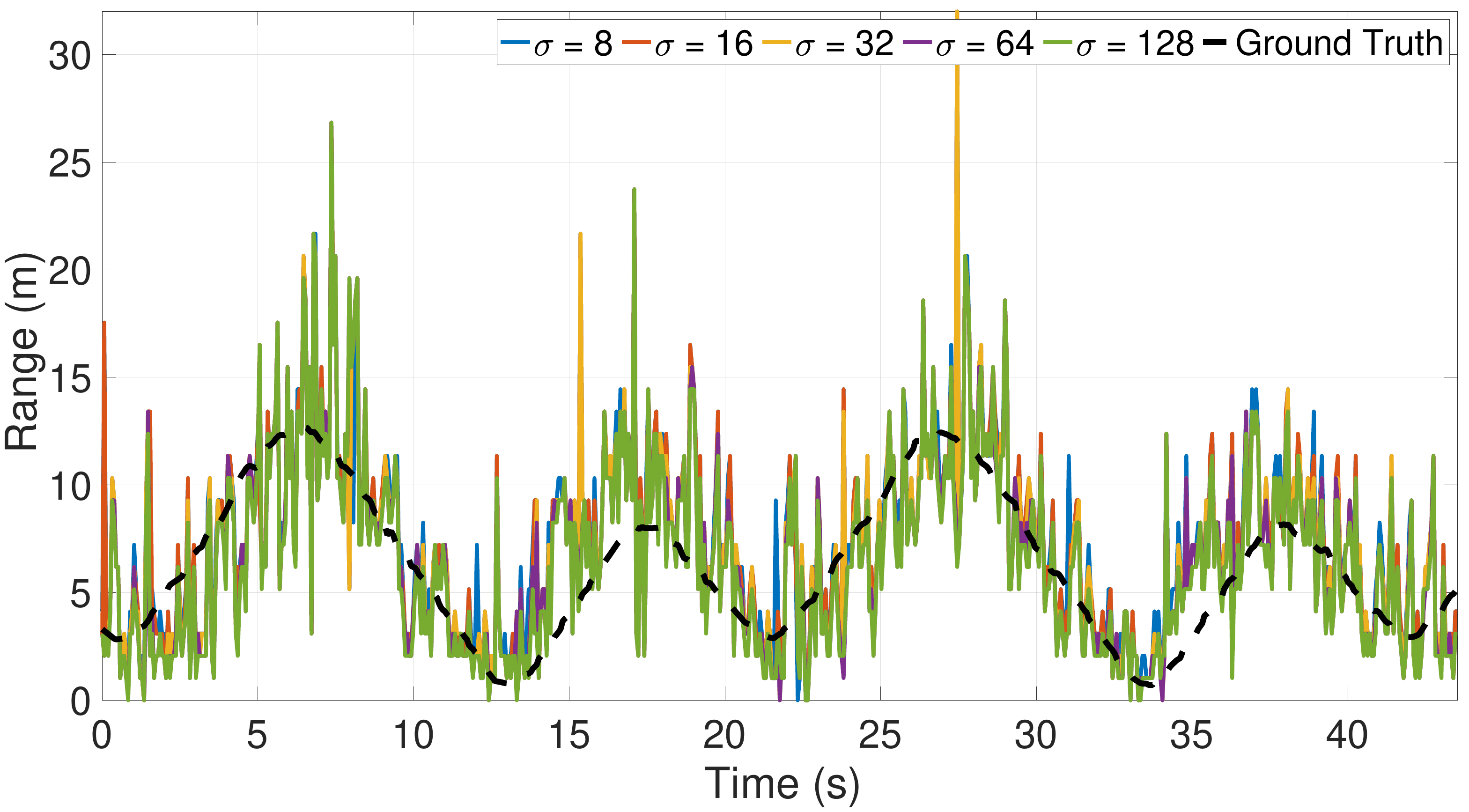}
        \subcaption{Range estimation}
    \end{subfigure}
    \begin{subfigure}[t]{0.325\linewidth}
        \includegraphics[width=\linewidth]{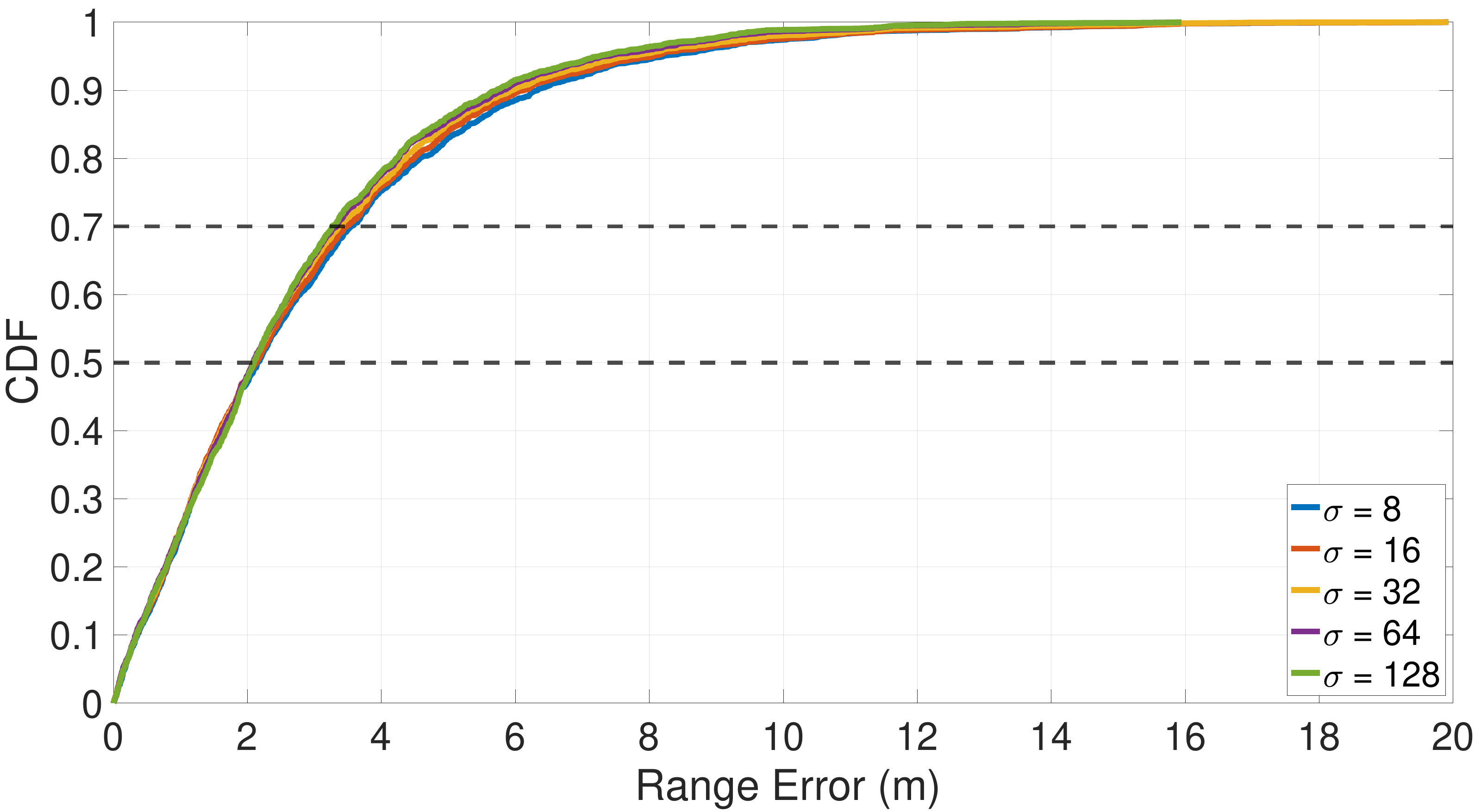}
        \subcaption{CDF of range error}
    \end{subfigure}
\begin{subfigure}[t]{0.325\linewidth}
        \includegraphics[width=\linewidth]{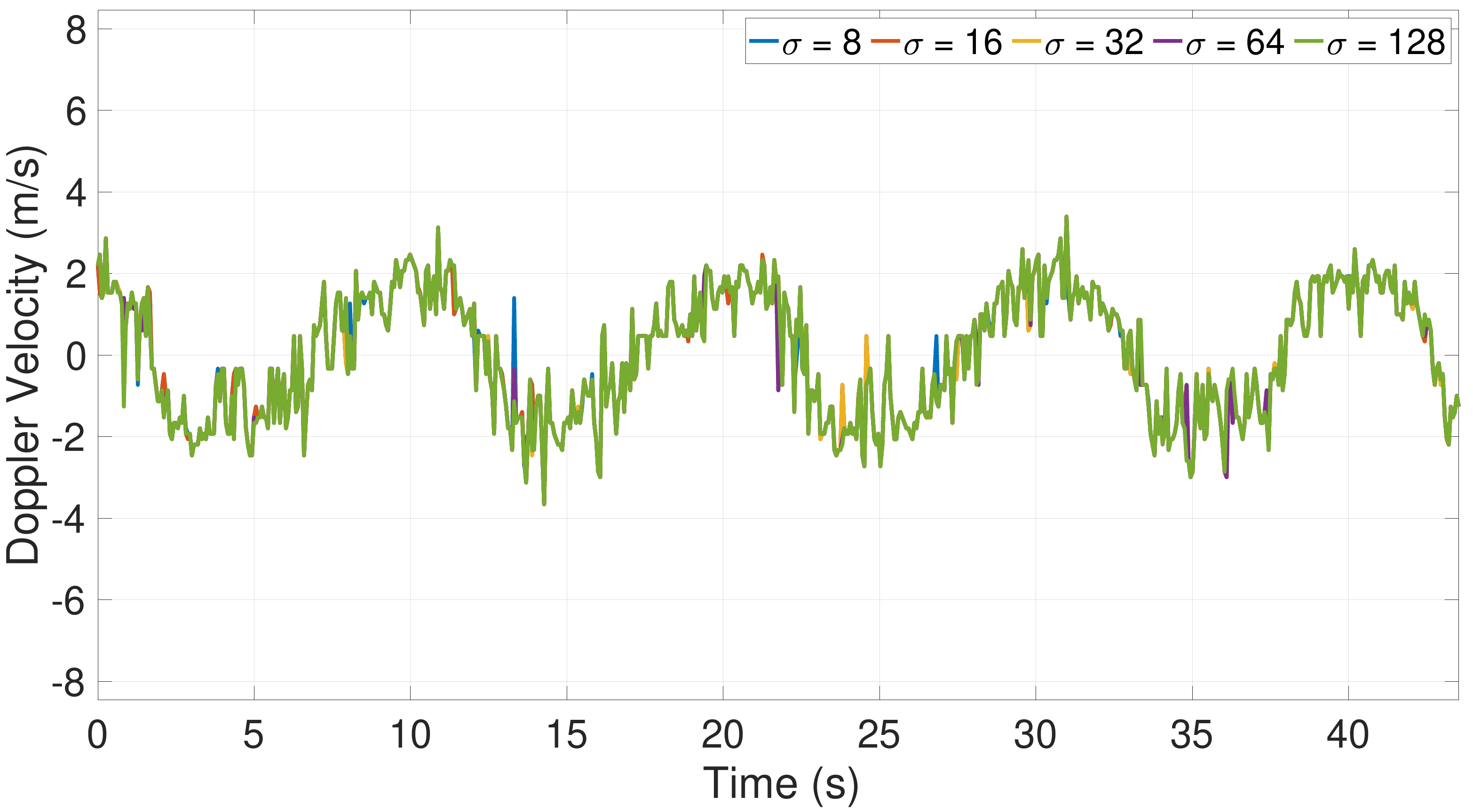}
        \subcaption{ Doppler velocity}
    \end{subfigure}
\caption{Impact of Gaussian window width ($\sigma$)}
\label{fig:window_sigma}
\vspace{-1em}
\end{figure*}

\subsubsection{Impact of Gaussian Window Width}
Fig. \ref{fig:window_sigma} and Table. \ref{tab:window_sigma} show the impact of the Gaussian window width $\sigma$ on delay and Doppler estimation. The smaller $\sigma$ facilitates the isolation of dominant paths. However, insufficient smoothing introduces more noise in the range estimates. Fig. \ref{fig:window_sigma} (a) and (b) shows that increasing $\sigma$ yields a slight improvement in range accuracy. Fig. \ref{fig:window_sigma} (c) shows that Doppler estimation remains relatively stable across a wide range of $\sigma$ values. The larger $\sigma$ combines more adjacent delay components, leading to reduced purity of the constructed signal. Despite this, our delay-domain beamforming effectively suppresses mirror artifacts.

\begin{table}
\caption{{Range estimation errors (m) of our SRCC at the 50\% and 70\% percentiles.} Overall, increasing the Gaussian window width tends to reduce the delay estimation error.}
\centering
\renewcommand{\arraystretch}{1.2}
\begin{tabularx}{\linewidth}{l *{5}{>{\centering\arraybackslash}X}}
\toprule
\textbf{Metric} & $\sigma = 8$ & $\sigma=16$ & $\sigma = 32$ & $\sigma = 64$ & $\sigma = 128$ \\
\midrule
50\% & 2.16\,m & 2.14\,m & {2.08\,m} & \textbf{2.08\,m} & {2.10\,m} \\
70\%  & 3.55\,m & 3.48\,m & 3.37\,m & 3.33\,m & \textbf{3.26\,m} \\
\bottomrule
\end{tabularx}
\label{tab:window_sigma}
\vspace{-1em}
\end{table}

\begin{table}
\centering
\caption{{Feature generation latency on different platforms without any code-level optimization (milliseconds)}.}
\renewcommand{\arraystretch}{1.2}
\setlength{\tabcolsep}{2pt}
\begin{tabularx}{\linewidth}{l c c}
\toprule
\textbf{Platform} & \textbf{Avg. Latency (ms)} & \textbf{Std. Dev. (ms)} \\
\midrule
Raspberry Pi 4B (8GB) & 8.5 & 4.3 \\
MacBook Pro 2019 (Intel i7, 2.6 GHz) & 1.2 & 0.46 \\
\bottomrule
\end{tabularx}
\label{tab:runtime}
\vspace{-1.5 em}
\end{table}

\subsubsection{Feature Generation Overhead}
Table. \ref{tab:runtime} summarizes the average latency on two platforms. On a Raspberry Pi 4B (8GB), the average latency is 8.5 ms with a standard deviation (Std.) of 4.3 ms, without using any optimization such as parallelization or JIT compilation \cite{lam2015numba}. The MacBook Pro (Intel i7, 2.6 GHz) achieves a much lower latency of 1.2 ms. As discussed in Section V. E, the main computational cost arises from MVDR weight estimation across delay bins. These results illustrate that our method supports real-time execution, even on resource-constrained edge devices.

\subsection{Single-Target Sensing Performance}
We evaluat WiDFS 3.0 under single-target settings, focusing on the impact of network architecture, input features, data augmentation, and generalization, all based on our interpretable and unambiguous micro-Doppler inputs.

\begin{table}
\centering
\footnotesize
\caption{Performance comparison of Dataset 1. \textit{Acc.}: Accuracy, \textit{Prec.}: Macro Precision, \textit{Rec.}: Macro Recall, \textit{F1}: Macro F1-score.}
\label{tab:interaction_single_performance}
\renewcommand{\arraystretch}{1.2}
\begin{tabularx}{\linewidth}{c|c|c|c|c|c}
\toprule
\textbf{Feature} & \textbf{Model} & \textbf{Acc.} & \textbf{Prec.} & \textbf{Rec.} & \textbf{F1} \\
\midrule
\multirow{7}{*}{\makecell{Our SRCC \\ (1Tx--1Rx)}}  
& CNN & 0.780 & 0.779 & 0.780 & 0.779 \\
& MLP & 0.796 & 0.794 & 0.794 & 0.792 \\
& SqueezeNet & 0.879 & 0.876 & 0.879 & 0.877 \\
& ShuffleNetV2 & 0.920 & 0.918 & 0.919 & 0.919 \\
& MobileNetV2 & 0.933 & 0.931 & 0.932 & 0.931 \\
& ResNet18 & 0.934 & 0.933 & 0.933 & 0.933 \\
& \textbf{MobileViT-XXS} & \textbf{0.939} & \textbf{0.938} & \textbf{0.938} & \textbf{0.938} \\
\midrule
\makecell{DCACC \\ (1Tx--3Rx)} & \textbf{MobileViT-XXS} & \textbf{0.991} & \textbf{0.991} & \textbf{0.991} & \textbf{0.991} \\
\midrule
\makecell{BVP \\ (Multi-Receiver)} & MobileViT-XXS & 0.850 & 0.849 & 0.849 & 0.849 \\
\bottomrule
\end{tabularx}
\vspace{-1em}
\end{table}

\begin{table}
\centering
\caption{Detailed classification report of each activity class (Dataset 1).}
\label{tab:interaction_class_report}
\renewcommand{\arraystretch}{1.3}
\begin{tabular}{c|c|c|c|c}
\toprule
\textbf{Class} & \textbf{Precision} & \textbf{Recall} & \textbf{F1-score} & \textbf{Test Samples (\#)} \\
\midrule
Clap          & 0.938 & 0.946 & 0.942 & 7510 \\
Draw-O        & 0.932 & 0.941 & 0.936 & 7965 \\
\textbf{Draw-Zigzag} & \textbf{0.969} & \textbf{0.976} & \textbf{0.973} & \textbf{10215} \\
Push\&Pull    & 0.937 & 0.937 & 0.937 & 7740 \\
Slide         & 0.922 & 0.898 & 0.910 & 8774 \\
Sweep         & 0.931 & 0.932 & 0.931 & 7514 \\
\bottomrule
\end{tabular}
\vspace{-1em}
\end{table}

\begin{figure*}[ht]
\centering
    \begin{subfigure}[t]{0.325\linewidth}
        \includegraphics[width=\linewidth]{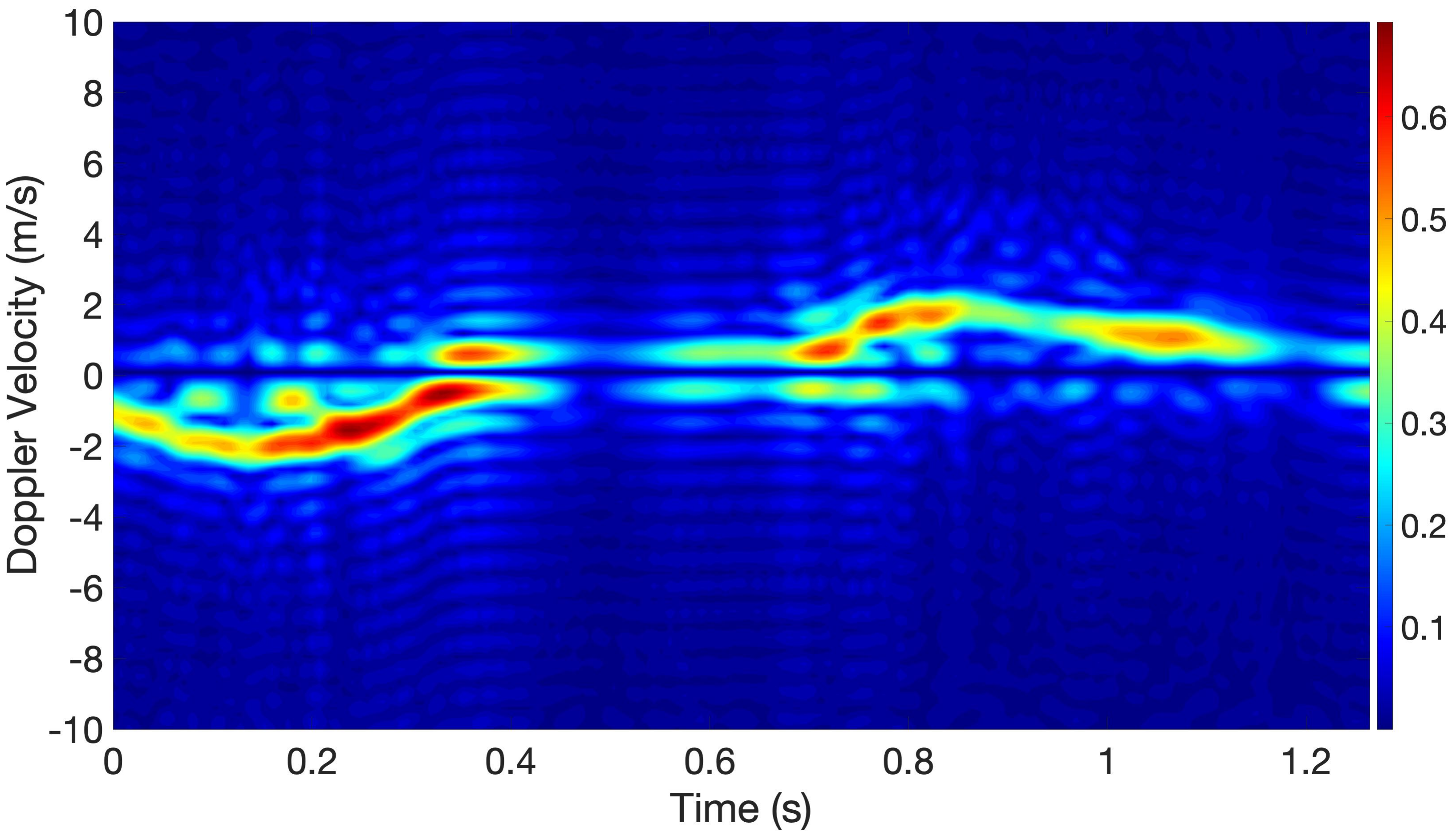}
        \subcaption{DCACC (Push\&Pull)}
    \end{subfigure}
    \begin{subfigure}[t]{0.325\linewidth}
        \includegraphics[width=\linewidth]{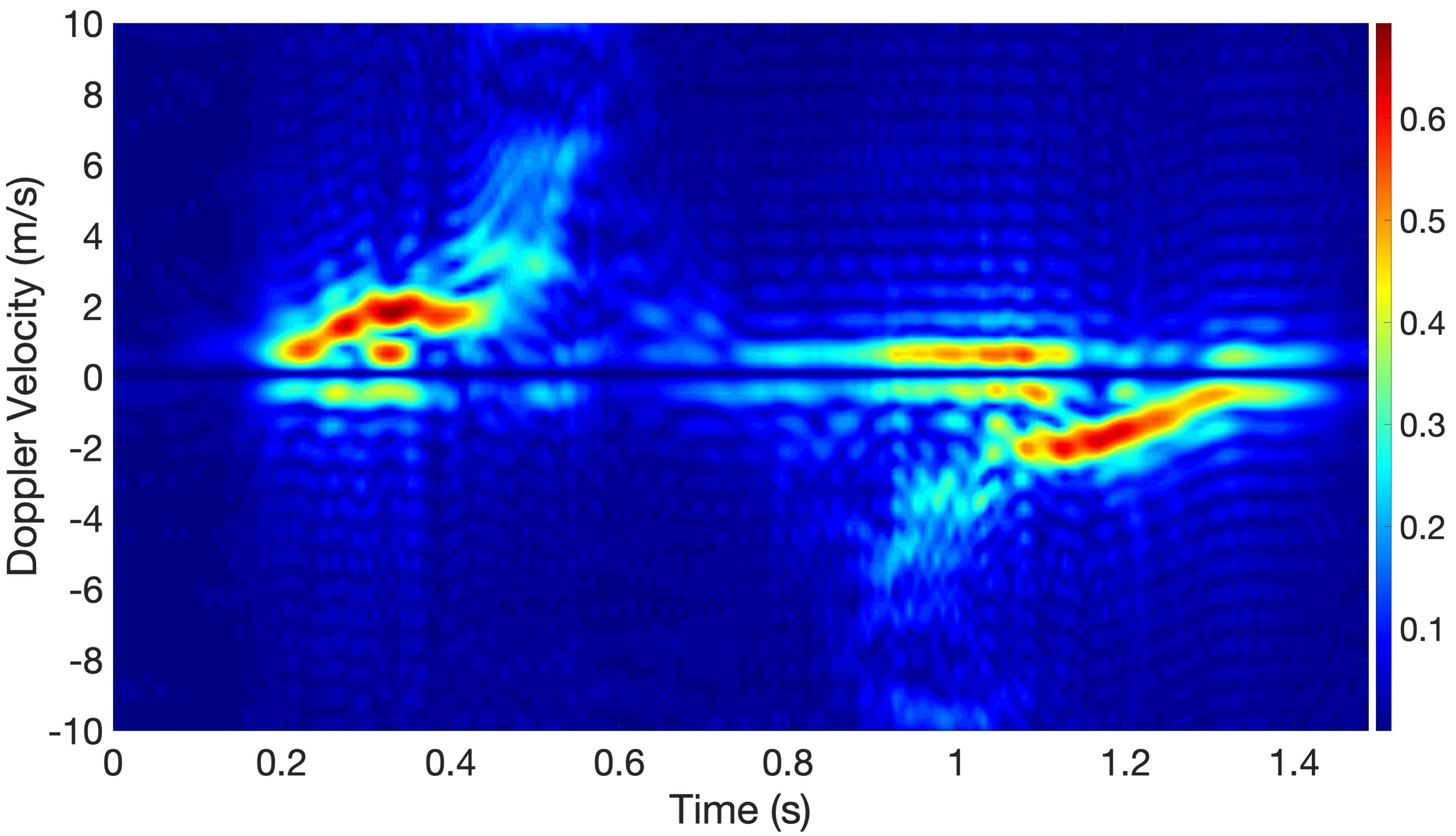}
        \subcaption{DCACC (Sweep)}
    \end{subfigure}
\begin{subfigure}[t]{0.325\linewidth}
        \includegraphics[width=\linewidth]{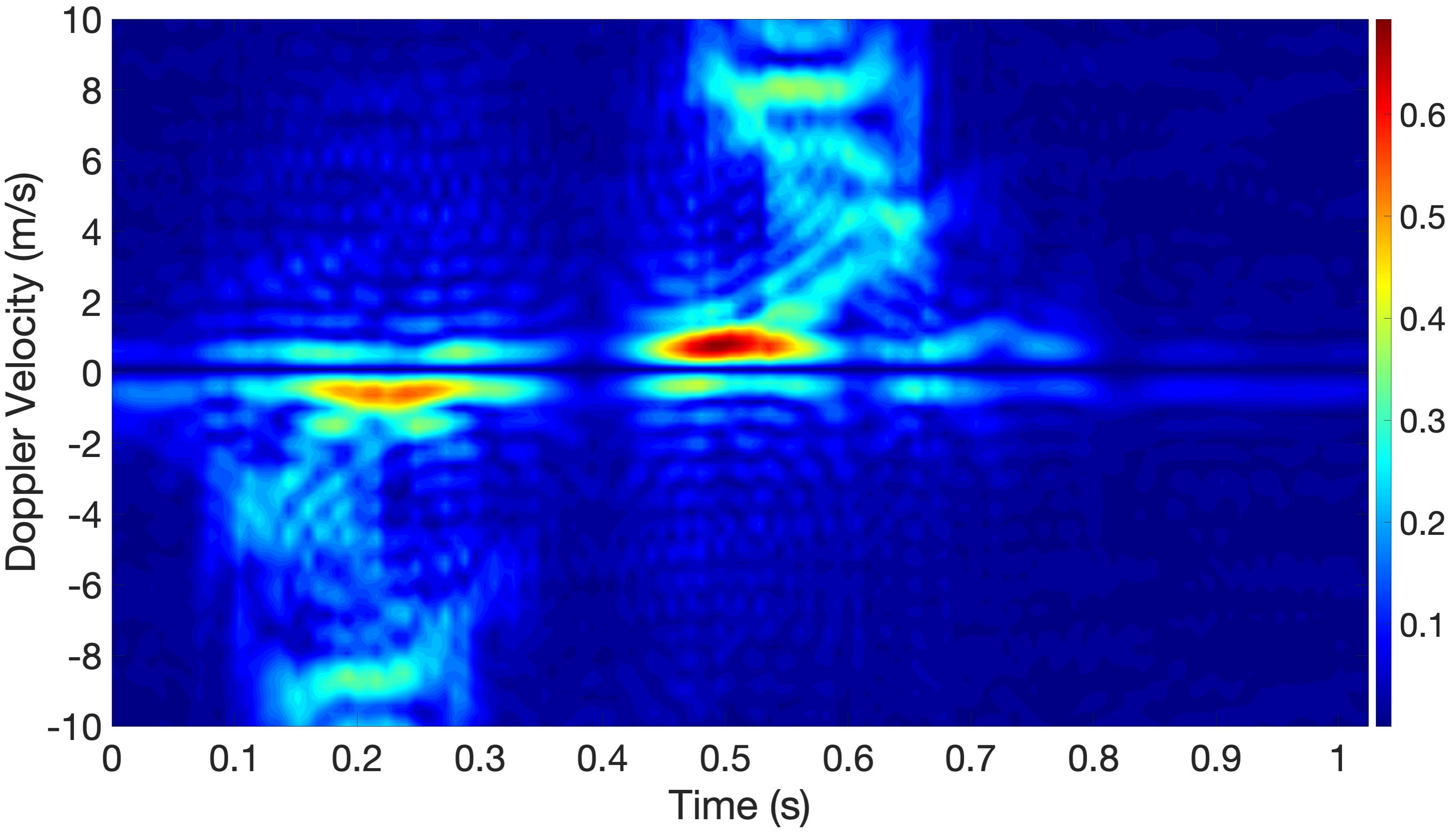}
        \subcaption{DCACC (Clap)}
    \end{subfigure}\\
    \begin{subfigure}[t]{0.325\linewidth}
        \includegraphics[width=\linewidth]{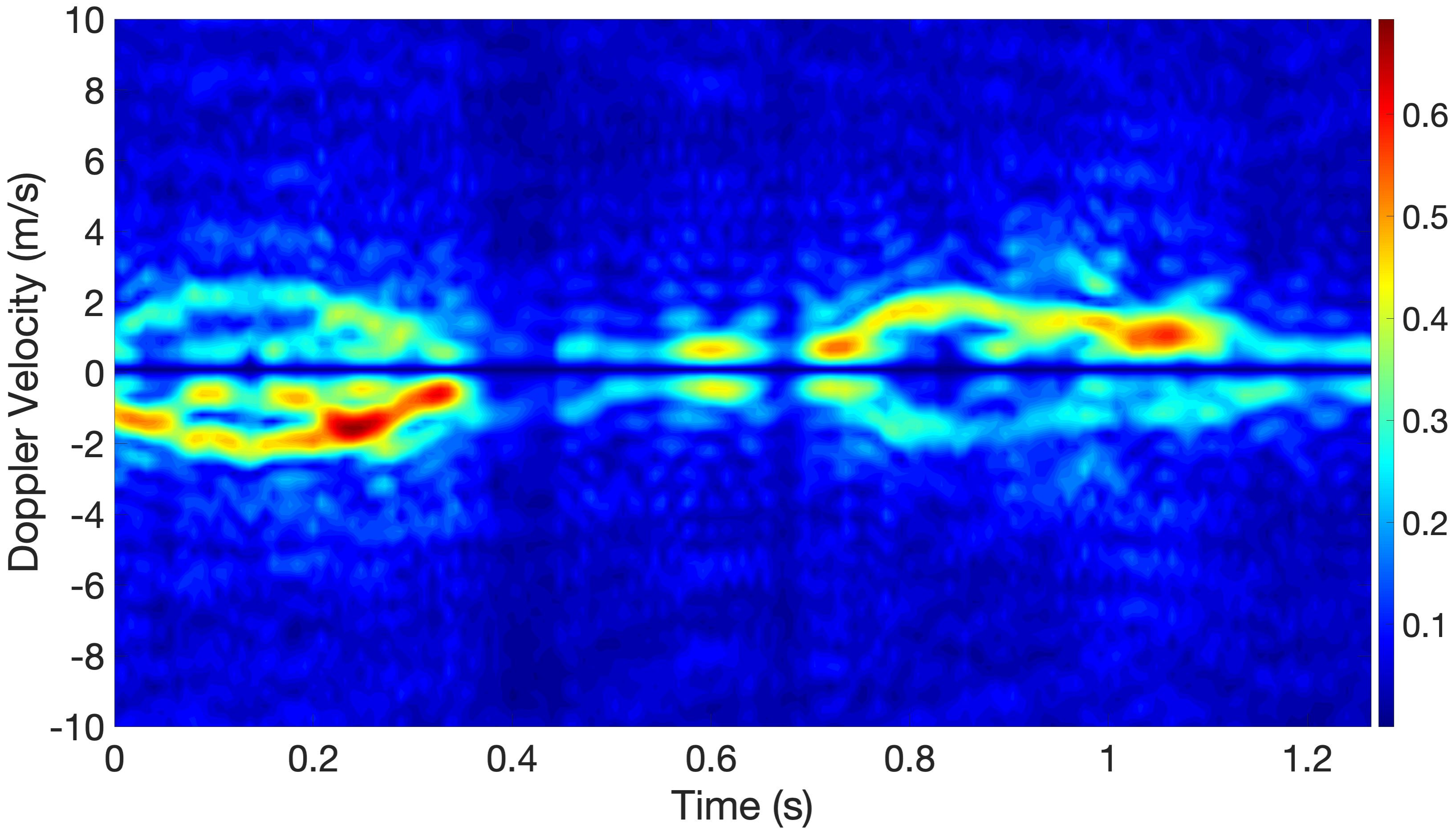}
        \subcaption{\textbf{Our SRCC (Push\&Pull)} }
    \end{subfigure}
    \begin{subfigure}[t]{0.325\linewidth}
        \includegraphics[width=\linewidth]{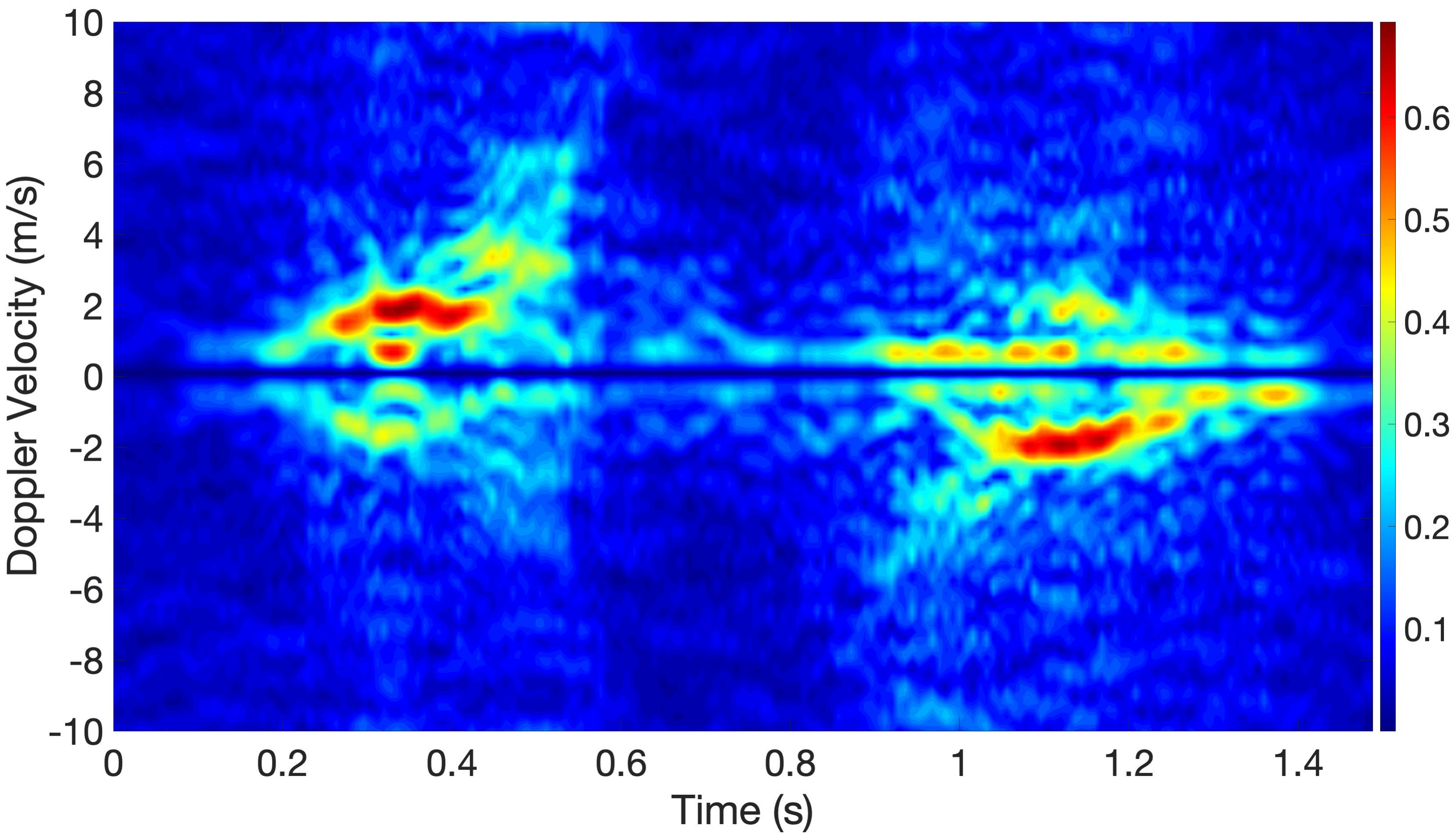}
        \subcaption{\textbf{Our SRCC (Sweep)} }
    \end{subfigure}
\begin{subfigure}[t]{0.325\linewidth}
        \includegraphics[width=\linewidth]{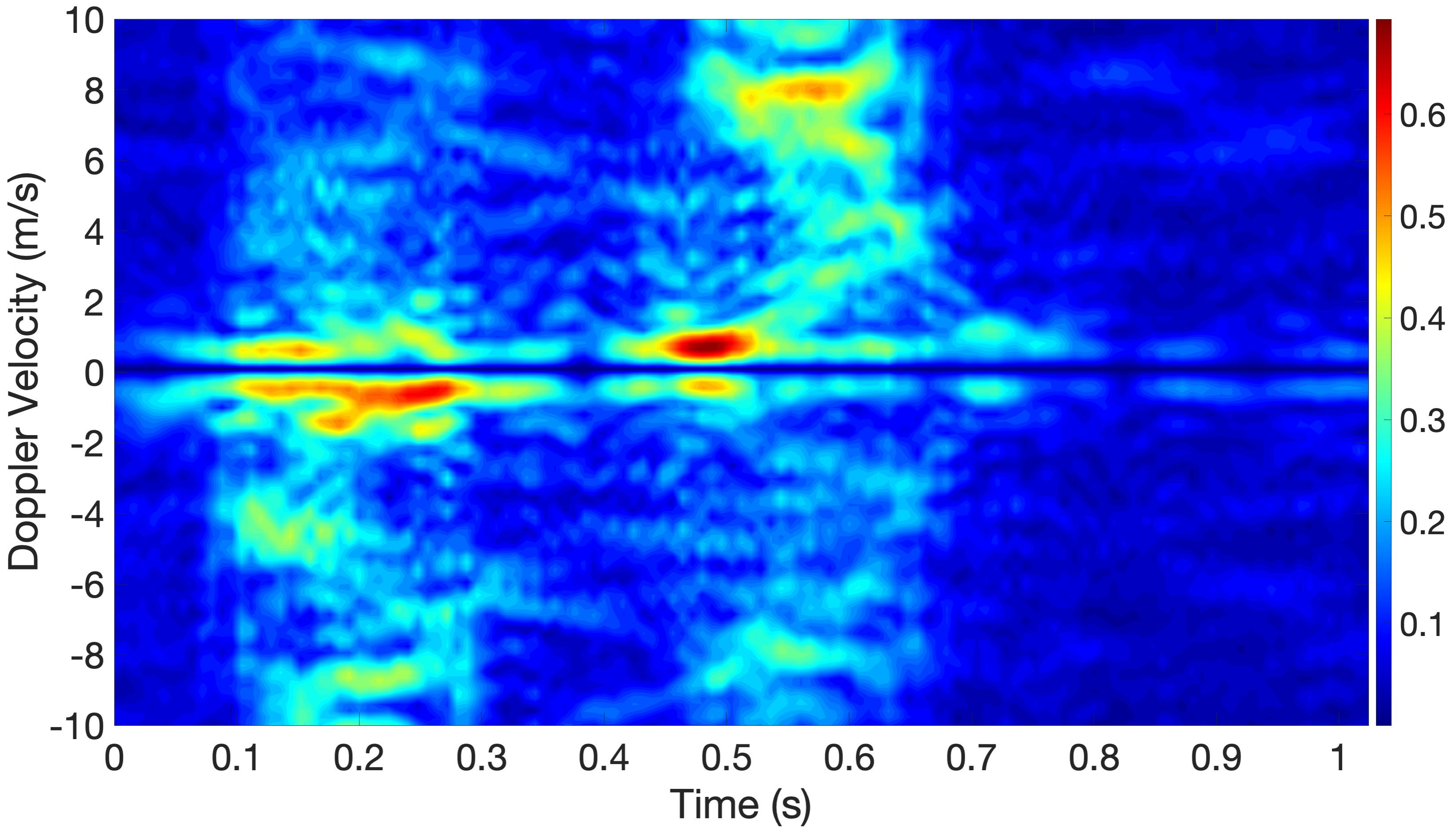}
        \subcaption{\textbf{Our SRCC (Clap)} }
    \end{subfigure}\\
\begin{subfigure}[t]{0.325\linewidth}
        \includegraphics[width=\linewidth]{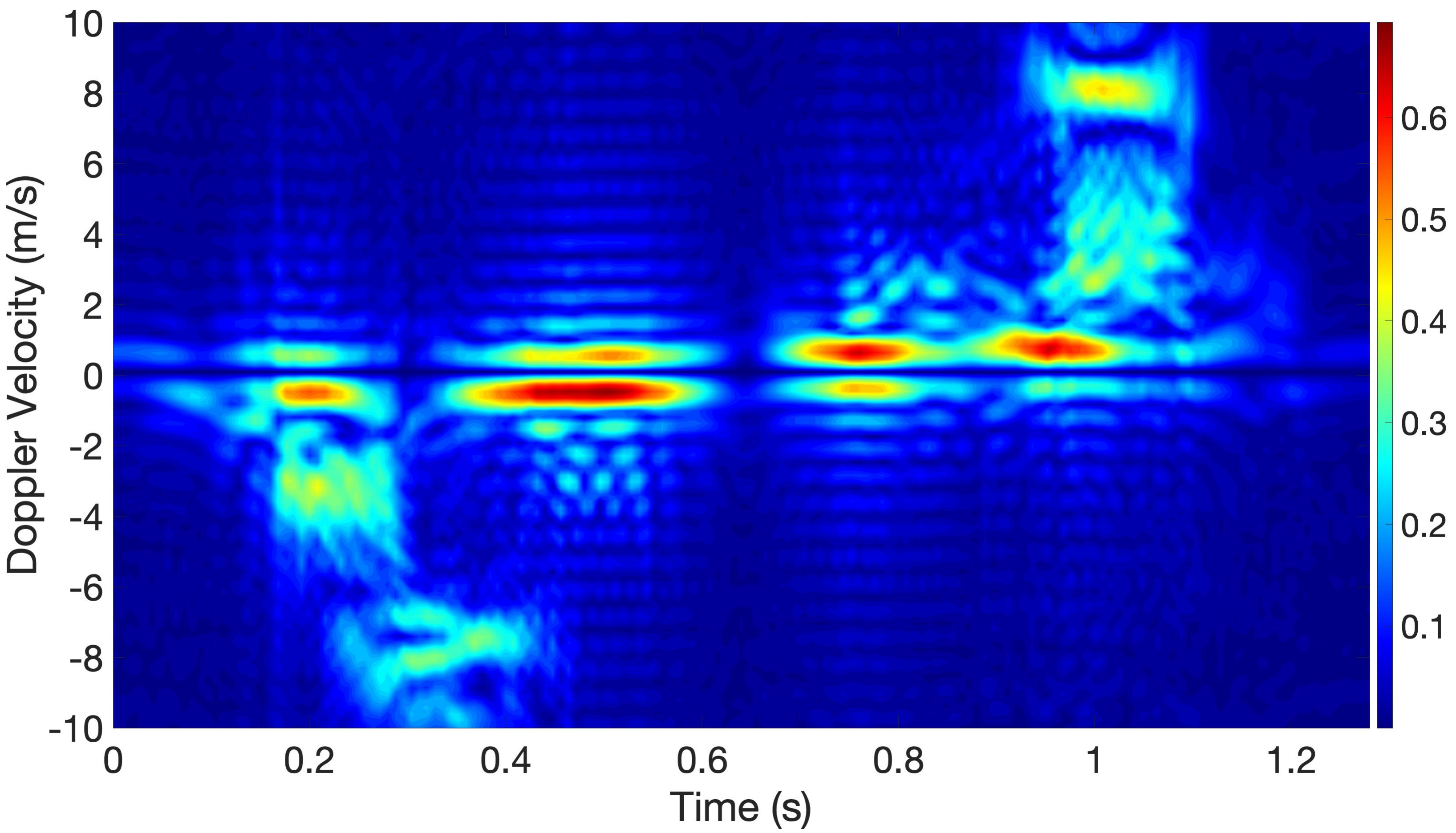}
        \subcaption{DCACC (Slide)}
    \end{subfigure}
    \begin{subfigure}[t]{0.325\linewidth}
        \includegraphics[width=\linewidth]{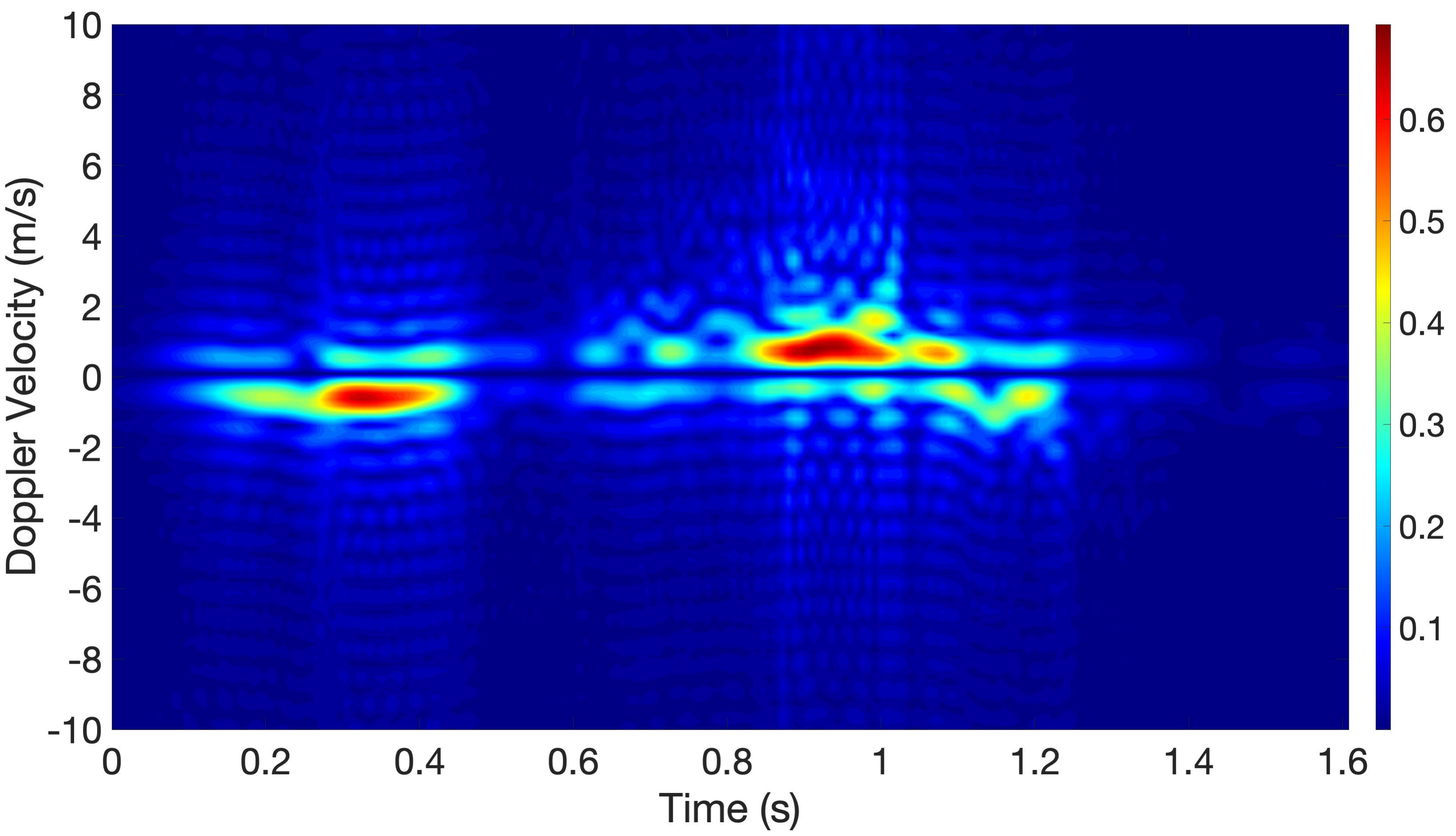}
        \subcaption{DCACC (Draw-O, Horizontal)}
    \end{subfigure}
\begin{subfigure}[t]{0.325\linewidth}
        \includegraphics[width=\linewidth]{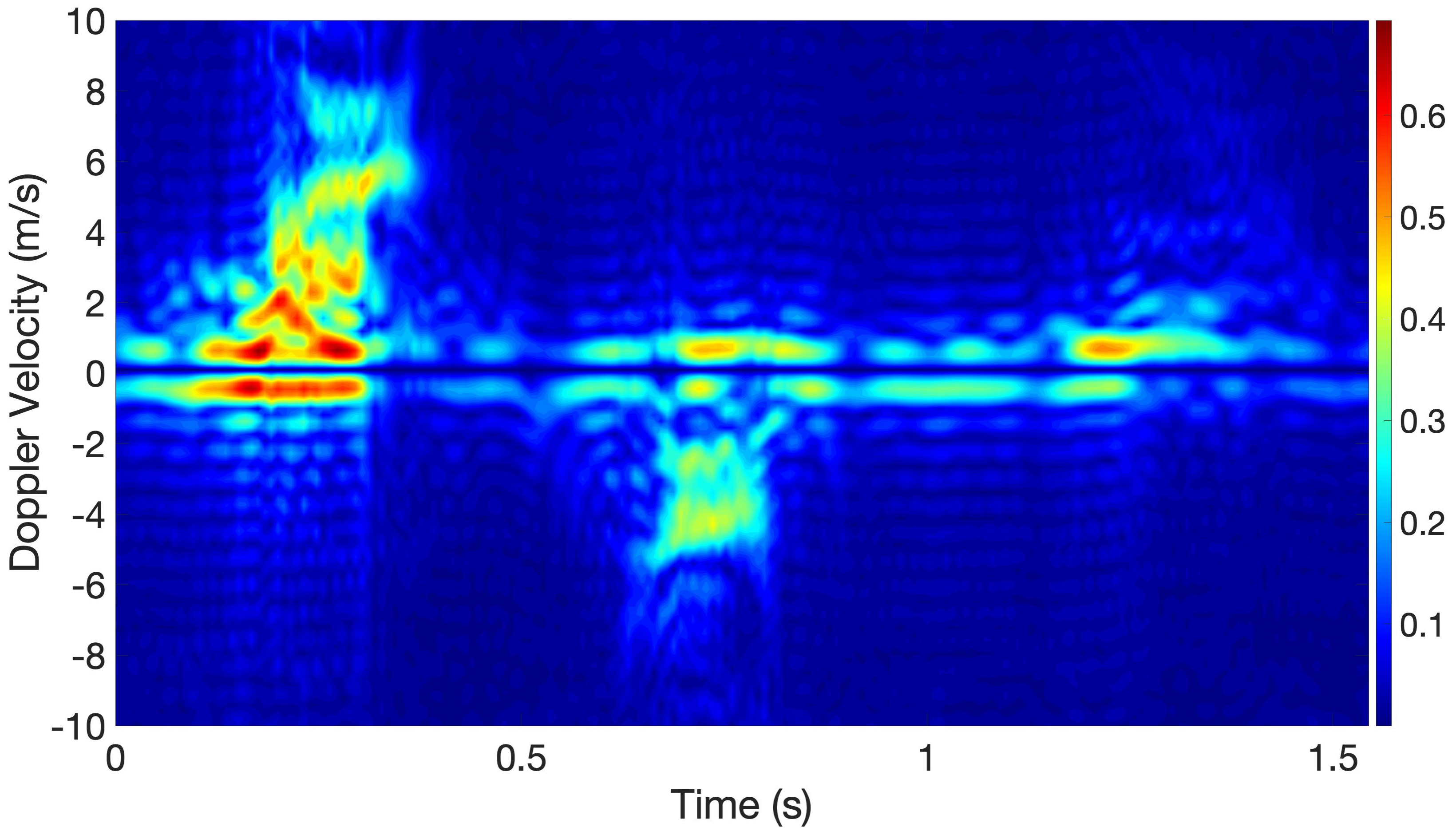}
        \subcaption{DCACC (Draw-Zigzag, Horizontal)}
    \end{subfigure}\\
    \begin{subfigure}[t]{0.325\linewidth}
        \includegraphics[width=\linewidth]{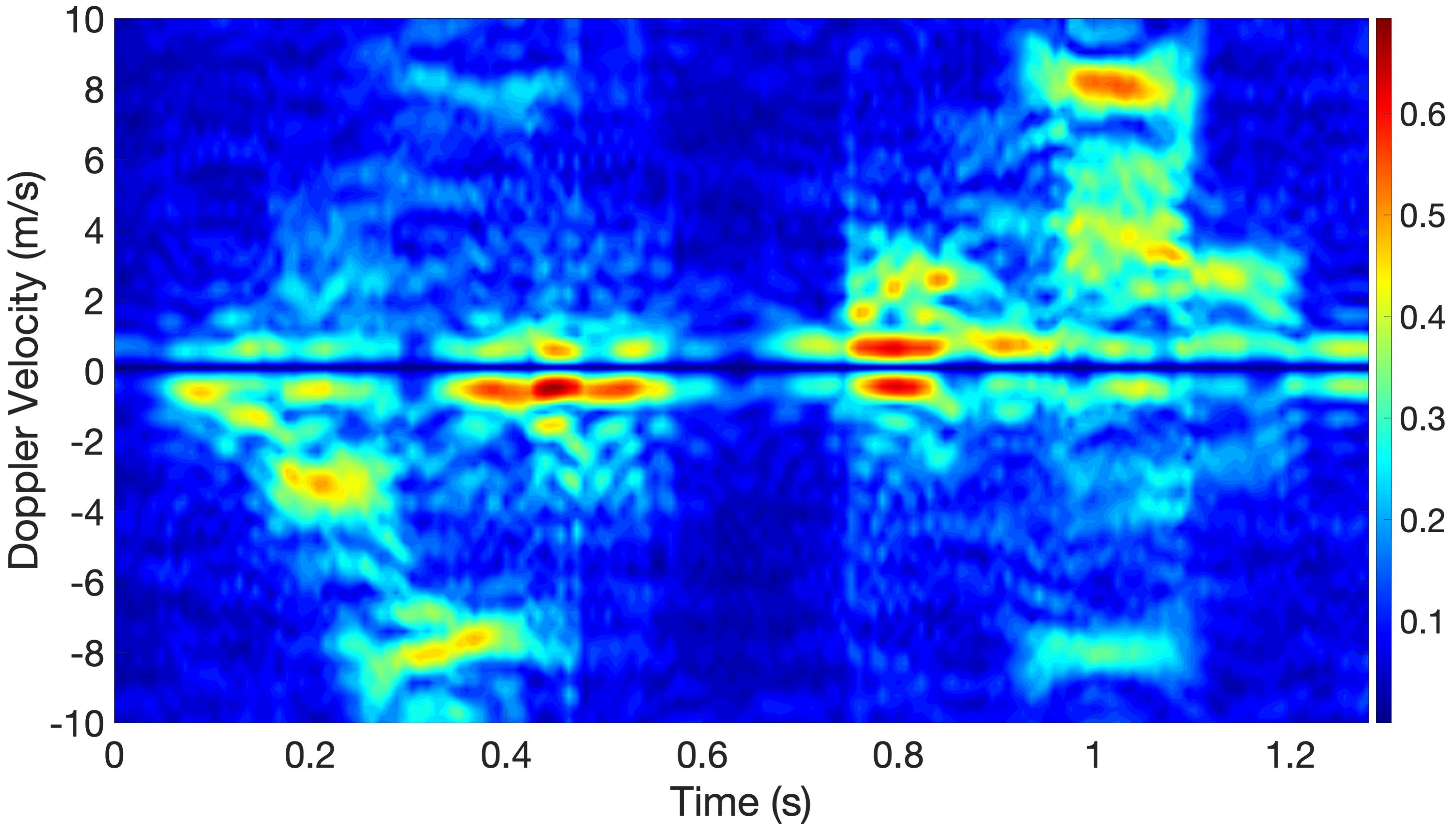}
        \subcaption{\textbf{Our SRCC (Slide)} }
    \end{subfigure}
    \begin{subfigure}[t]{0.325\linewidth}
        \includegraphics[width=\linewidth]{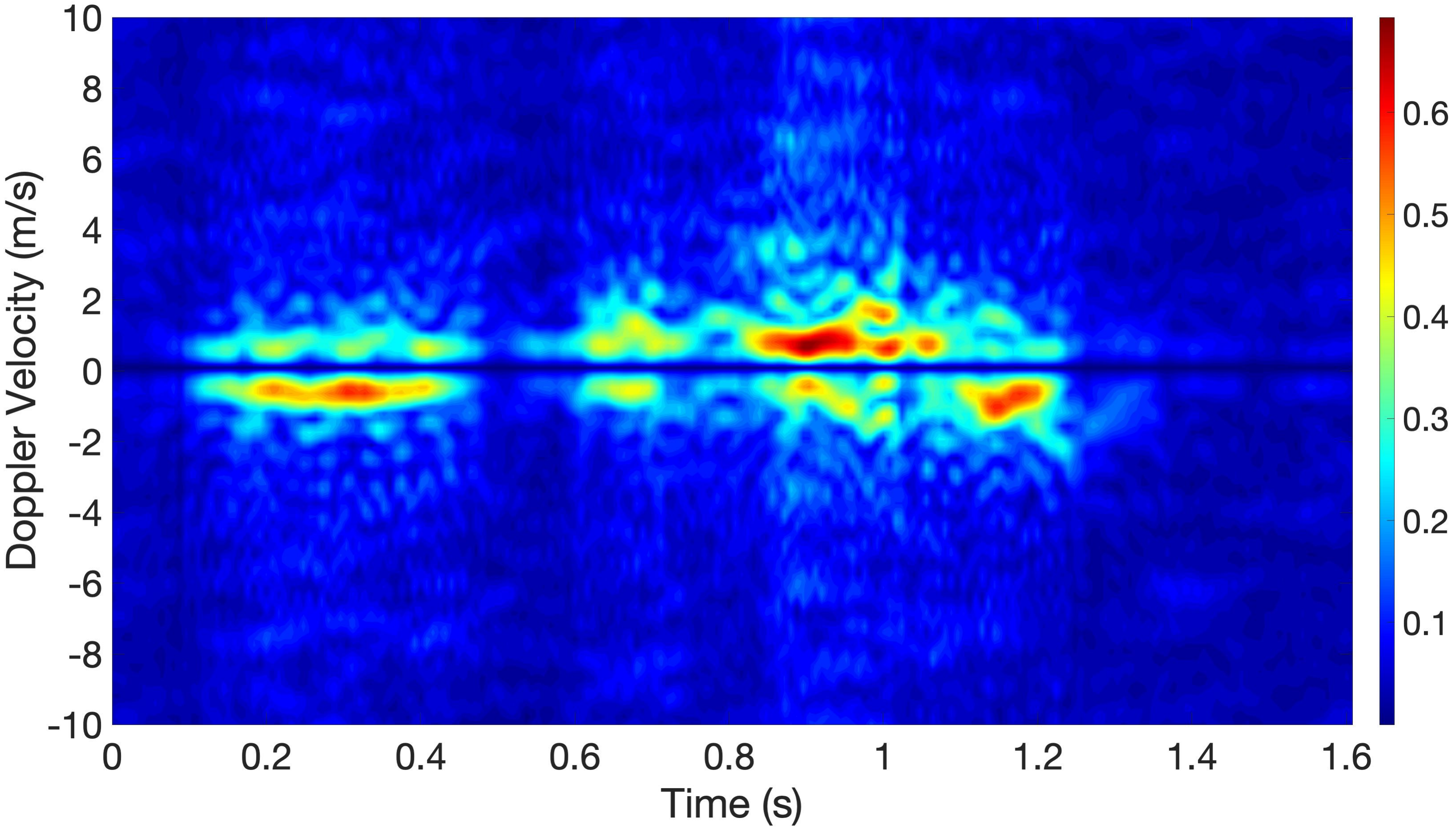}
        \subcaption{\textbf{Our SRCC (Draw-O, Horizontal)} }
    \end{subfigure}
\begin{subfigure}[t]{0.325\linewidth}
        \includegraphics[width=\linewidth]{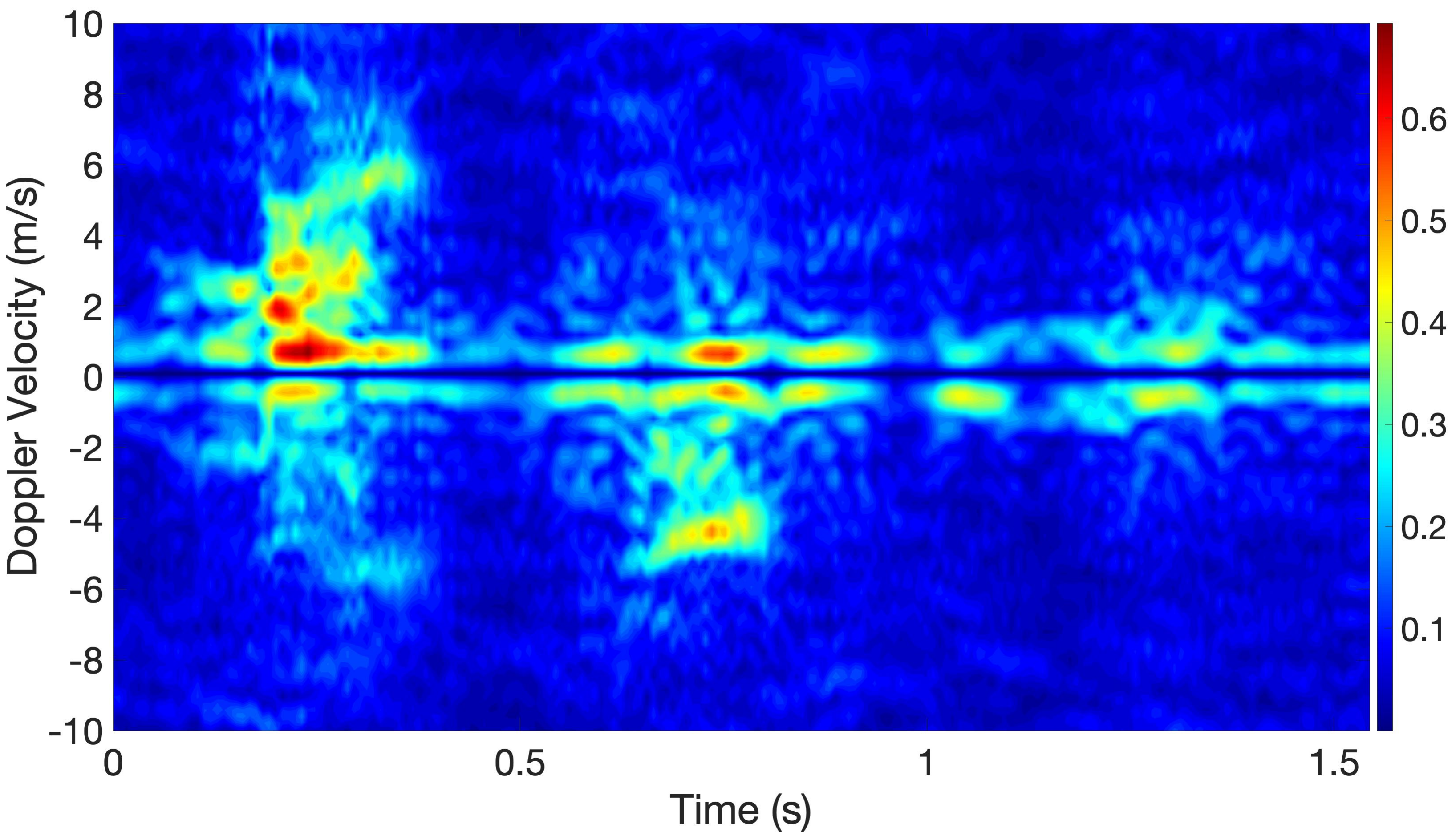}
        \subcaption{\textbf{Our SRCC (Draw-Zigzag, Horizontal)} }
    \end{subfigure}
\caption{Micro-Doppler signatures for six gestures (Dataset 1) using 1Tx-3Rx DCACC (top) and our 1Tx-1Rx SRCC method (bottom), showing effective suppression of mirror Doppler components in SISO configurations.}
\label{fig:microdoppler_examples}
\vspace{-1em}
\end{figure*}

\subsubsection{Overall Performance across Lightweight Models}
We next present detailed evaluations on two datasets of Widar 3.0, including Dataset 1 for human-computer interaction activities, and Dataset 2 for digit gesture recognition. To ensure a fair comparison, all models are trained and evaluated on each dataset using a 70\%-30\% train-test split, with uniform sampling across different gestures. Each receiver captures CSI from three Rx antennas, and in each training epoch, we randomly select the micro-Doppler feature from one of the antennas. In addition, we visualize the extracted micro-Doppler features for six gestures, shown in Fig.\ref{fig:microdoppler_examples}. Different hand movements induce distinct Doppler signatures, enabling the model to capture different motion dynamics for classification. The 1Tx-3Rx DCACC method, which effectively removes Doppler mirroring artifacts, is used as a reference for comparison. Our 1Tx-1Rx SRCC exhibit significant suppression of mirror Doppler components.

\begin{figure*}
\centering
    \begin{subfigure}[t]{0.495\linewidth}
\centering
        \includegraphics[width=0.85\linewidth]{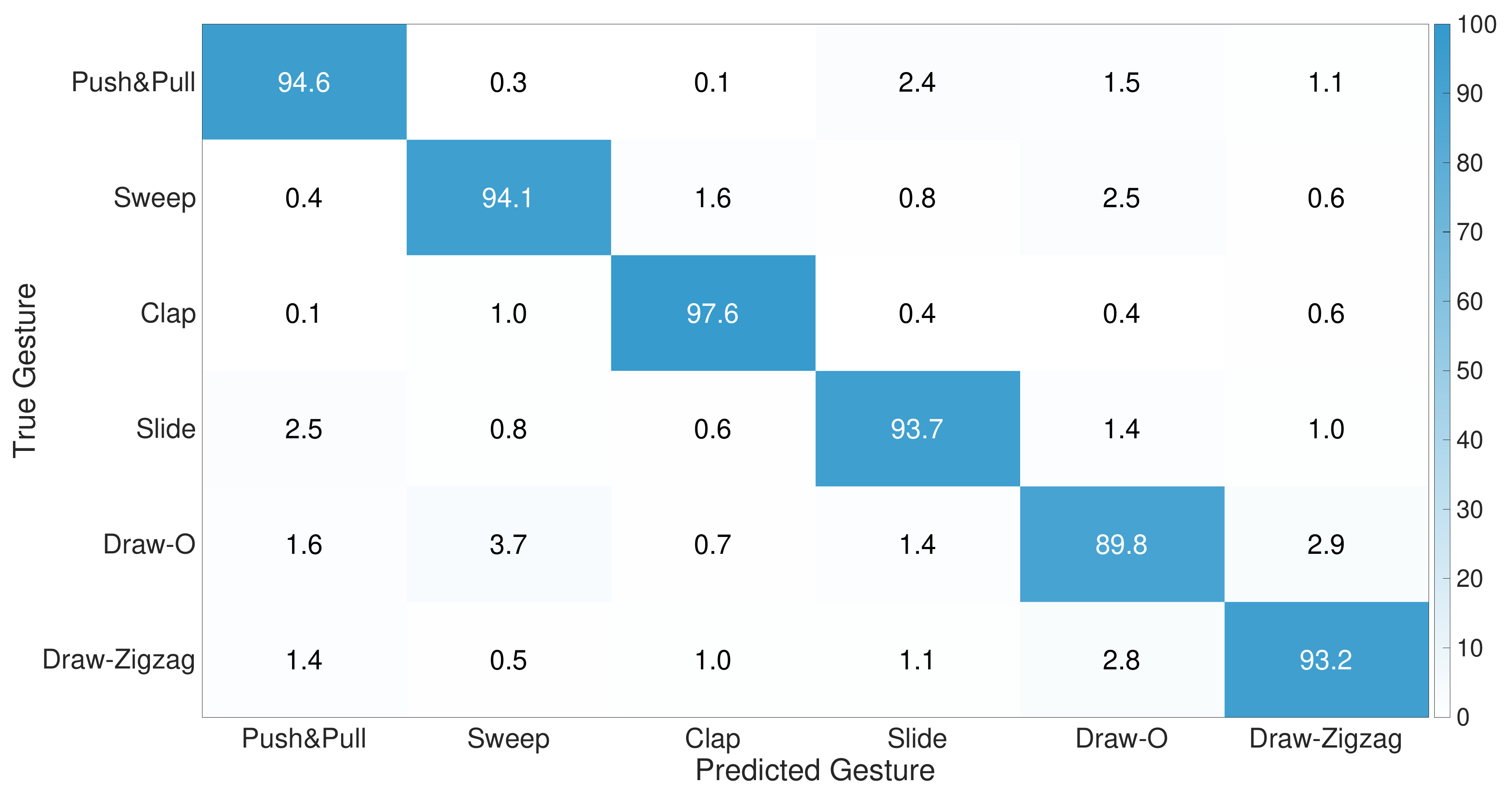}
        \subcaption{Our SRCC (1Tx-1Rx)}
    \end{subfigure}
    \begin{subfigure}[t]{0.495\linewidth}
\centering
        \includegraphics[width=0.85\linewidth]{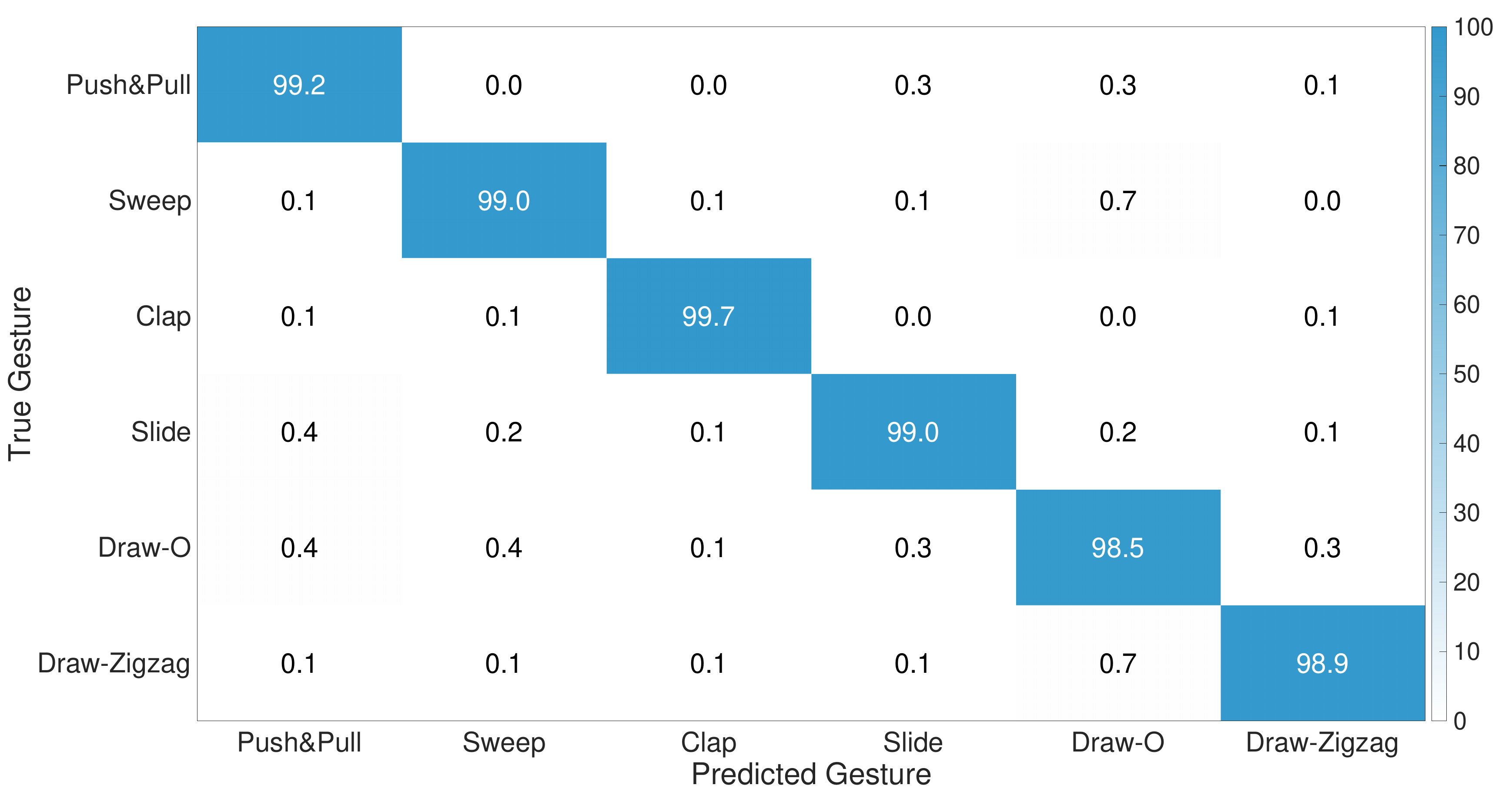}
        \subcaption{DCACC (1Tx-3Rx)}
    \end{subfigure}
\caption{Confusion matrix of human-computer interaction gestures (Dataset 1).}
\label{fig:conf_matrix_interaction}
\vspace{-1em}
\end{figure*}

\begin{figure*}
\centering
    \begin{subfigure}[t]{0.495\linewidth}
\centering
        \includegraphics[width=0.85\linewidth]{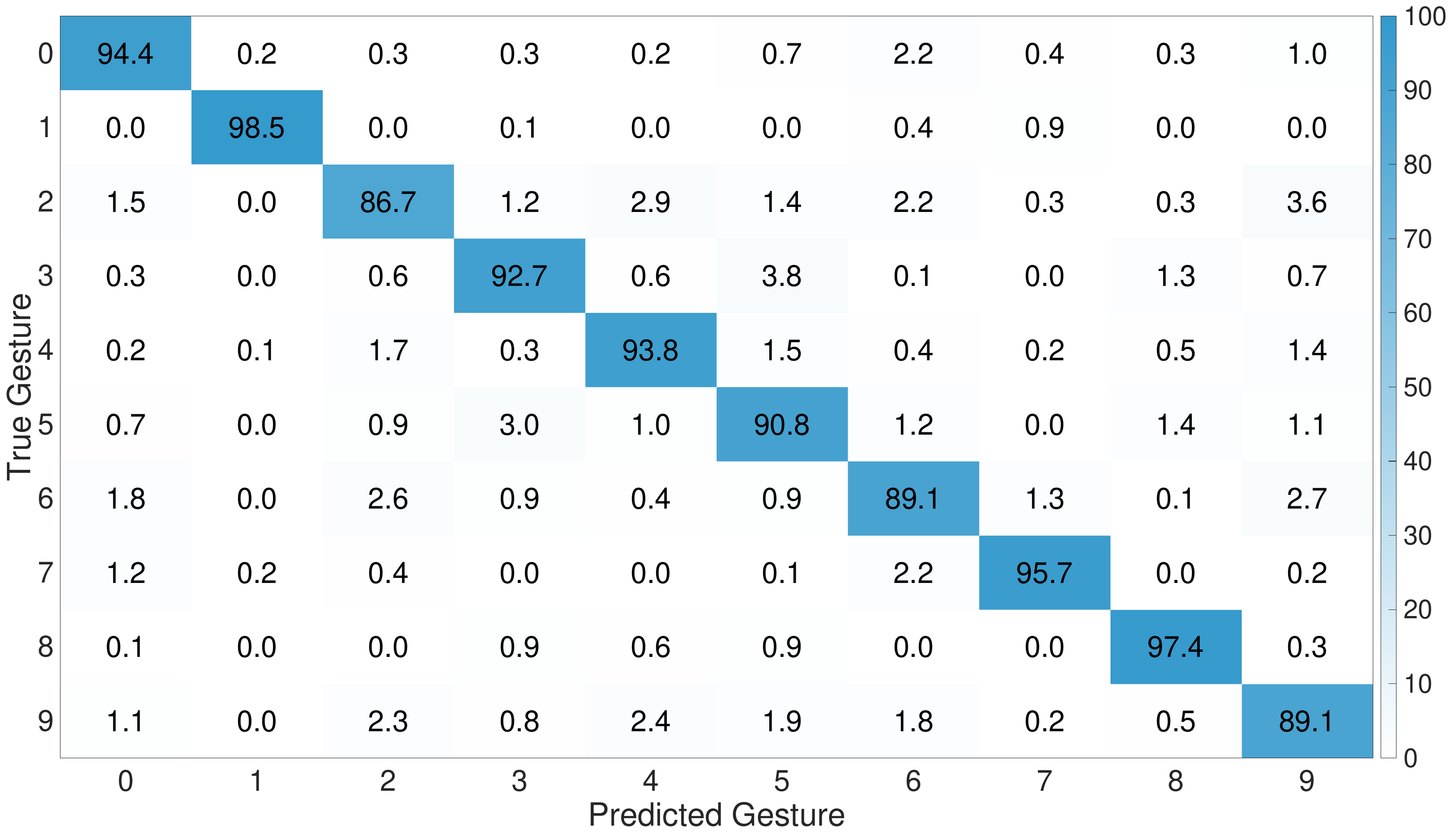}
        \subcaption{Our SRCC (1Tx-1Rx)}
    \end{subfigure}
    \begin{subfigure}[t]{0.495\linewidth}
\centering
        \includegraphics[width=0.85\linewidth]{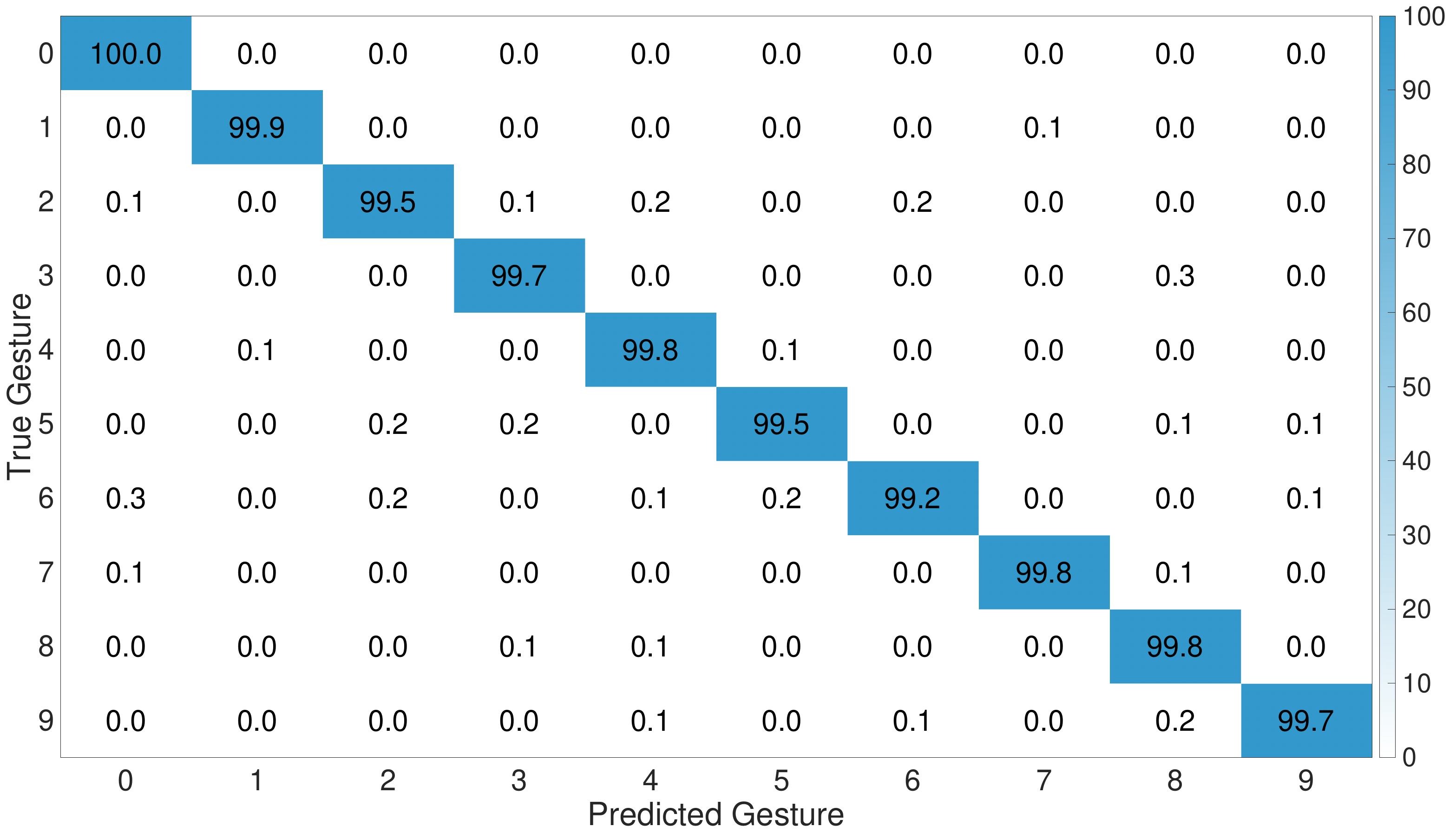}
        \subcaption{DCACC (1Tx-3Rx)}
    \end{subfigure}
\caption{Confusion matrix of numeric drawing gestures (Dataset 2).}
\label{fig:conf_matrix_digits}
\vspace{-1em}
\end{figure*}

\begin{figure*}[t]
\centering
\begin{minipage}[t]{0.328\linewidth}
\vspace{0pt} 
\centering
\includegraphics[width=\linewidth]{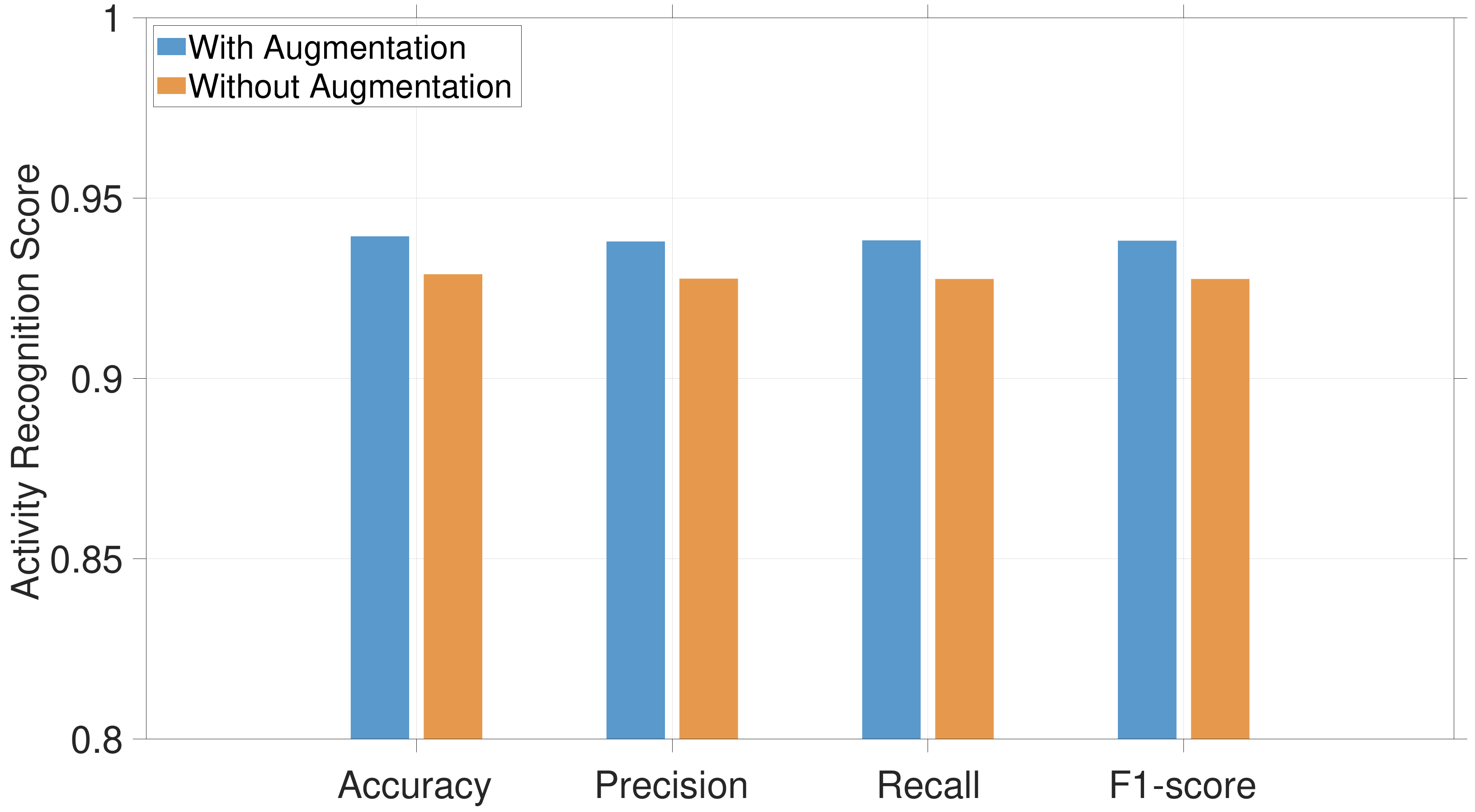}
\caption{Impact of data augmentation}
\label{figure:data_augment}
\end{minipage}
\begin{minipage}[t]{0.328\linewidth}
\vspace{0pt} 
\centering
\includegraphics[width=\linewidth]{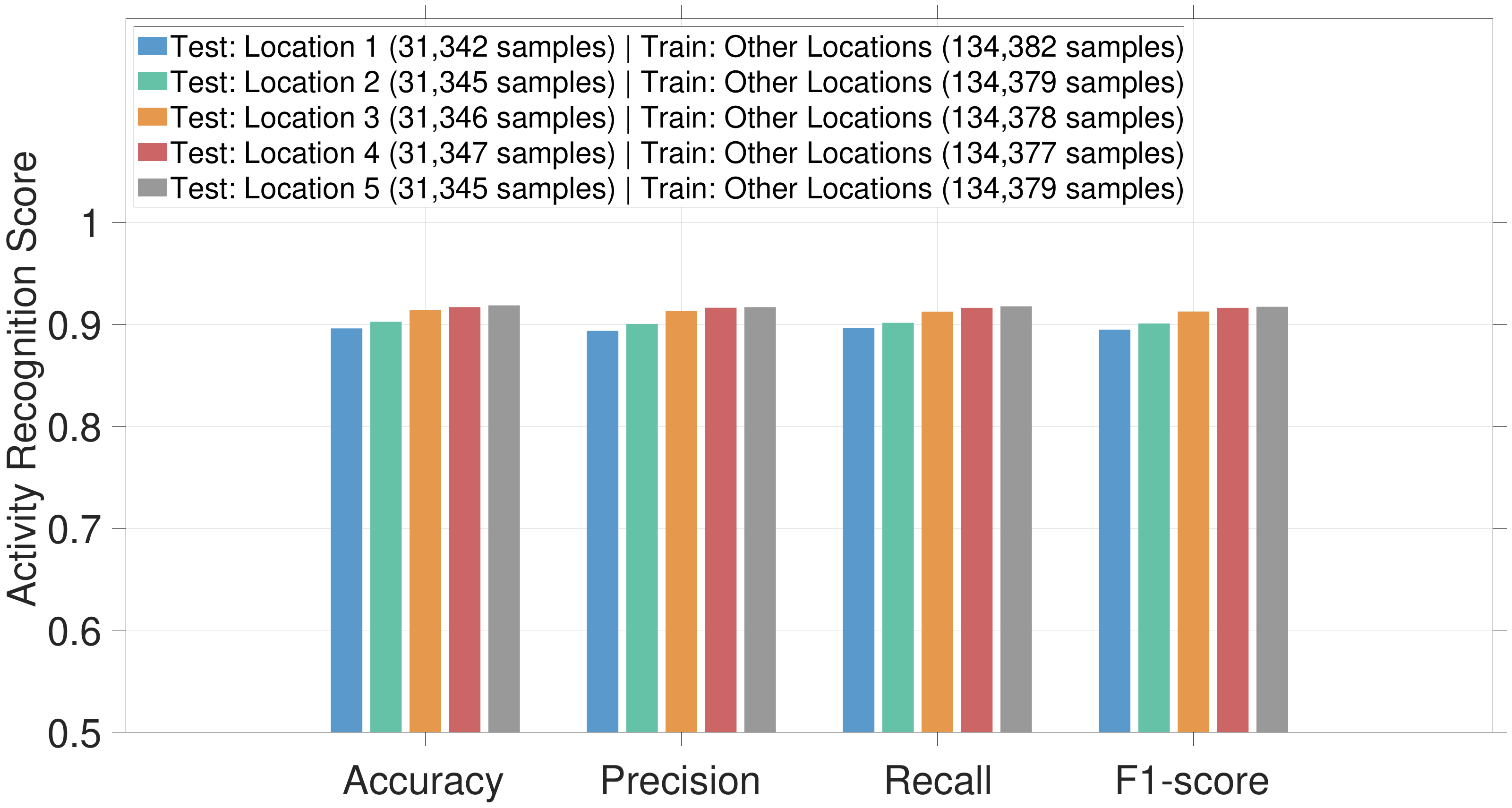}
\caption{Impact of location}
\label{figure:single_location}
\end{minipage}
\begin{minipage}[t]{0.328\linewidth}
\vspace{0pt} 
\centering
\includegraphics[width=\linewidth]{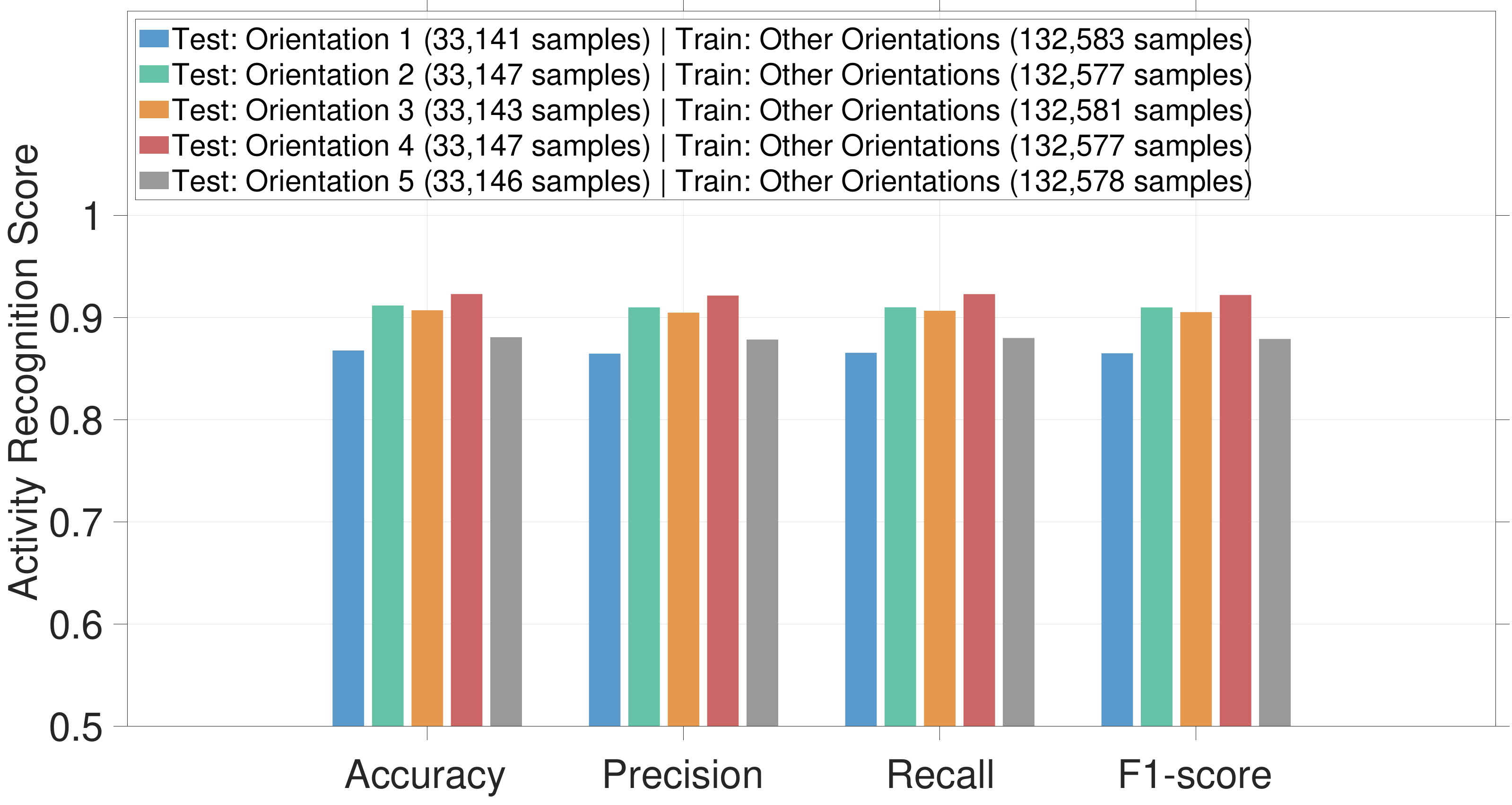}
\caption{Impact of orientation}
\label{figure:single_orientation}
\end{minipage}
\begin{minipage}[t]{0.328\linewidth}
\centering
\includegraphics[width=\linewidth]{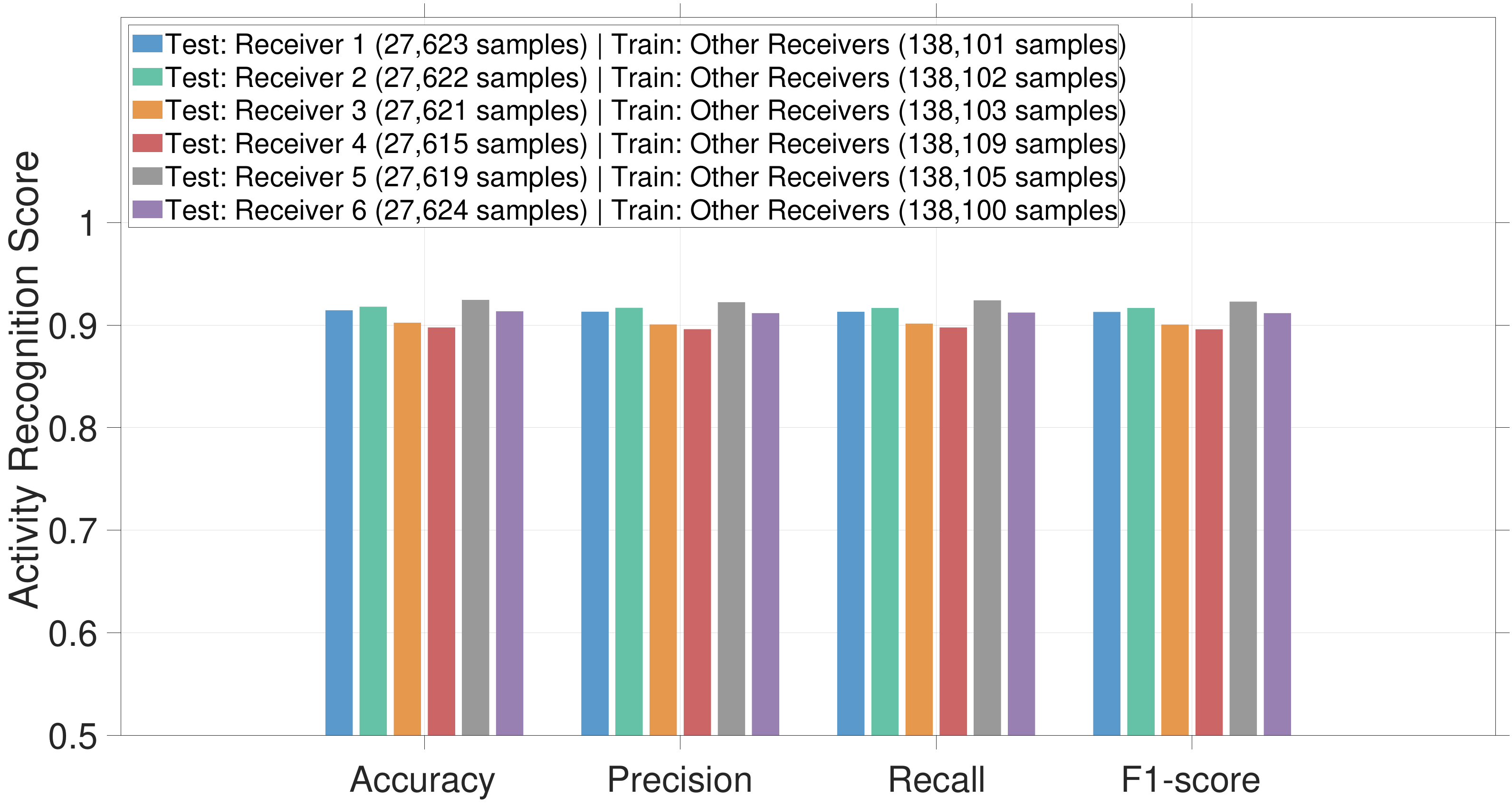}
\caption{Impact of receiver}
\label{figure:single_receiver}
\end{minipage}
\begin{minipage}[t]{0.328\linewidth}
\centering
\includegraphics[width=\linewidth]{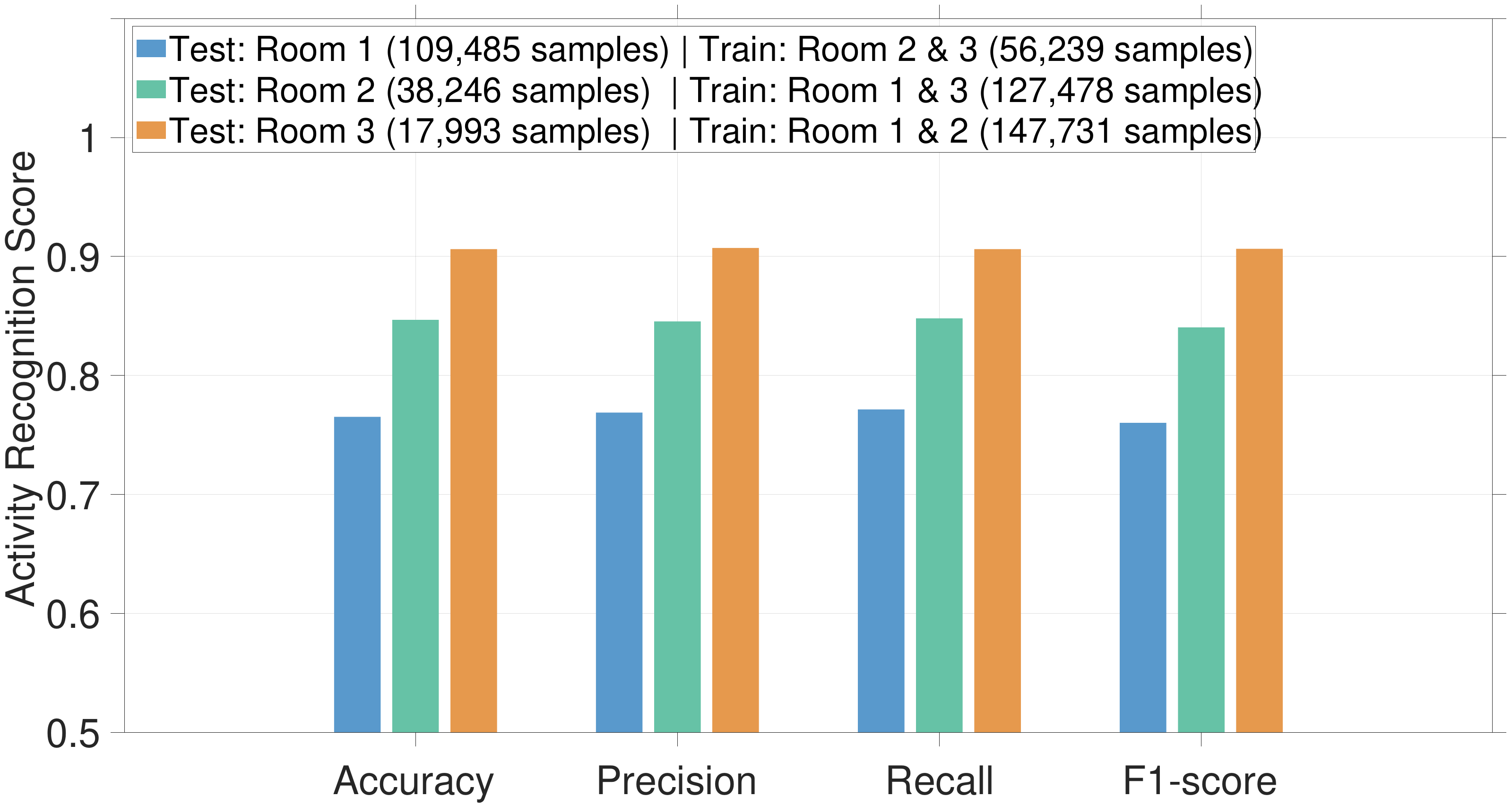}
\caption{Impact of environment}
\label{figure:single_environment}
\end{minipage}
\begin{minipage}[t]{0.328\linewidth}
\centering
\includegraphics[width=\linewidth]{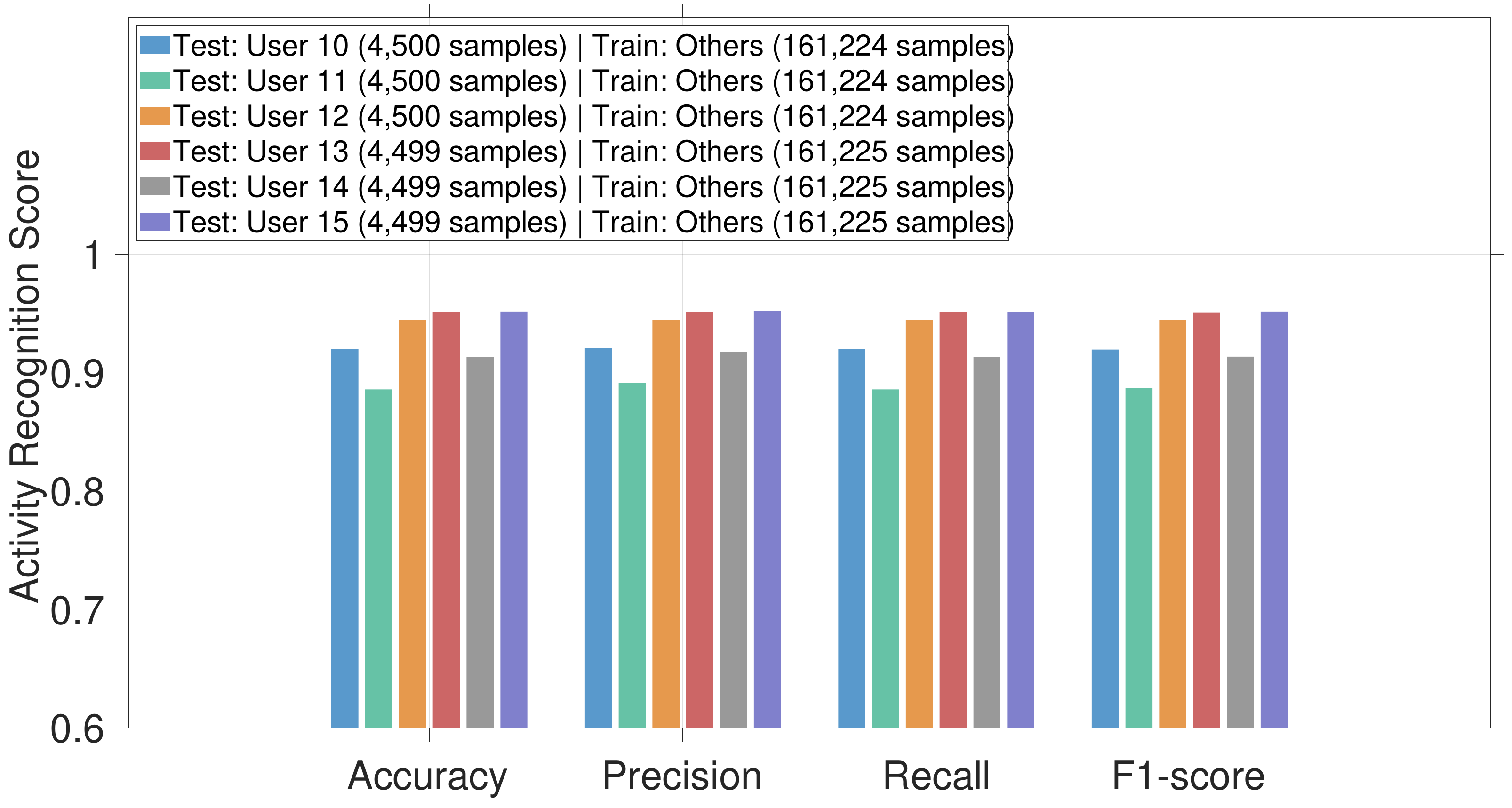}
\caption{Impact of user}
\label{figure:single_user}
\end{minipage}
\vspace{-1em}
\end{figure*}

\begin{itemize}
\item \textit{Dataset 1.} Table. \ref{tab:interaction_single_performance} shows that our 1Tx-1Rx SRCC achieves strong and consistent performance across different lightweight backbones, including MobileNetV2, ShuffleNetV2, SqueezeNet, ResNet18, and MobileViT-XXS. Among them, MobileViT-XXS achieves the best macro F1-score (0.938) and accuracy (0.939), slightly outperforming ResNet18, MobileNetV2, and ShuffleNetV2. However, the MLP and CNN models show limited performance. The MLP processes features independently and lacks temporal inductive bias, limiting its ability to model dynamic gestures. The shallow CNN is constrained by local receptive fields, making it difficult to capture long-range dependencies. In contrast, the attention-based MobileViT model enables global context modeling and spatial preservation, leading to better recognition performance. Table. \ref{tab:interaction_class_report} further details the per-class classification performance on six human-computer interaction gestures. All classes achieve high precision and recall, with the Draw-Zigzag gesture yielding the highest F1-score of 0.973. This may be partially attributed to its relatively higher number of training samples.

\item \textit{Dataset 2.} As summarized in Table. \ref{tab:digit_single_performance} and Table \ref{tab:digit_class_report}, our WiDFS 3.0 still maintains high performance, achieving a macro F1-score of 0.928. We can see that digits like Draw-1, Draw-7 and Draw-8 achieve F1-scores above 0.96, while more ambiguous digits like Draw-2 and Draw-6 yield slightly lower scores. Overall, these results validate the robustness of WiDFS 3.0. The interpretable micro-Doppler features extracted from SISO configurations maintain strong motion discriminability, enabling lightweight models to achieve high performance .

\begin{table}
\centering
\caption{Performance comparison of DataSet 2.}
\label{tab:digit_single_performance}
\renewcommand{\arraystretch}{2.0} 
\resizebox{\columnwidth}{!}{
\begin{tabular}{c|c|c|c|c|c}
\toprule
\makecell{\textbf{Input Feature}} &
\textbf{Model} &
\textbf{Accuracy} &
\textbf{Precision} &
\textbf{Recall} &
\textbf{F1-score} \\
\midrule
\makecell{Our SRCC \\ (1Tx--1Rx)}    
  & \multirow{3}{*}{MobileViT-XXS} 
  & \textbf{0.928} & \textbf{0.928} & \textbf{0.928} & \textbf{0.928} \\
\cline{1-1} \cline{3-6}
\makecell{DCACC \\ (1Tx--3Rx)}   
  &  & \textbf{0.997} & \textbf{0.997} & \textbf{0.997} & \textbf{0.997} \\
\cline{1-1} \cline{3-6}
\makecell{BVP \\ (Multi-Receiver)} 
  &  & 0.859 & 0.861 & 0.859 & 0.859 \\
\bottomrule
\end{tabular}
}
\vspace{-1em}
\end{table}

\item \textit{Performance Comparison.} Table. \ref{tab:interaction_single_performance} and Table. \ref{tab:digit_single_performance} report the recognition performance using DCACC-based features, achieving macro F1-scores of 0.991 and 0.997 for Dataset 1 and 2, respectively. The corresponding confusion matrices are shown in Fig. \ref{fig:conf_matrix_interaction} (a) and Fig. \ref{fig:conf_matrix_digits} (a). The super high recognition accuracy underscore the effectiveness of unambiguous micro-Doppler features in capturing motion dynamics with high discriminability. Our method achieves comparable sensing performance while effectively suppressing Doppler mirroring. However, due to the limitations of single-antenna configurations, the mirror components cannot be entirely eliminated. However, the BVP feature yields lower accuracy on both datasets, primarily due to its reliance on multi-receiver Doppler aggregation without explicit ambiguity suppression, using raw CACC for random phase removal. This makes it more susceptible to mirrored components and inter-receiver calibration errors.
\begin{table}
\centering
\caption{Detailed classification report of each digit gesture class (DataSet 2).}
\label{tab:digit_class_report}
\renewcommand{\arraystretch}{1.3}
\begin{tabular}{c|c|c|c|c}
\toprule
\textbf{Class} & \textbf{Precision} & \textbf{Recall} & \textbf{F1-score} & \textbf{Test Samples (\#)} \\
\midrule
Draw-0  & 0.932 & 0.944 & 0.938 & 1170 \\
\textbf{Draw-1}  & \textbf{0.996} & \textbf{0.985} & \textbf{0.991} & \textbf{1170} \\
Draw-2  & 0.907 & 0.867 & 0.886 & 1170 \\
Draw-3  & 0.926 & 0.927 & 0.926 & 1171 \\
Draw-4  & 0.920 & 0.938 & 0.929 & 1170 \\
Draw-5  & 0.892 & 0.908 & 0.900 & 1169 \\
Draw-6  & 0.894 & 0.891 & 0.893 & 1170 \\
Draw-7  & 0.967 & 0.957 & 0.962 & 1170 \\
Draw-8  & 0.957 & 0.974 & 0.965 & 1170 \\
Draw-9  & 0.891 & 0.891 & 0.891 & 1170 \\
\bottomrule
\end{tabular}
\vspace{-1em}
\end{table}

\item \textit{Data Augmentation} We use MobileViT-XXS to evaluate the impact of data augmentation due to its highest classification performance described above. Fig. \ref{figure:data_augment} shows the augmentation strategy yields more than 1\% improvement in macro F1-score compared to the case without augmentation. The proposed physics-guided augmentation is beneficial for improving model performance.\end{itemize}

\begin{table*}[ht]
\centering
\caption{Performance comparison of different input features  (Dataset 1) under a strict generalization setting: the model is trained on samples from Location \#1 and Receiver \#1 (5,224 samples), and tested on samples from all other locations and receivers (160,500 samples).}
\label{tab:input_feature_summary}
\renewcommand{\arraystretch}{1.2}
\resizebox{\textwidth}{!}{
\begin{tabular}{c|c|c|cccc|c}
\toprule
\textbf{Model} & \textbf{Input Feature}  & \textbf{Minimum Setup} & \textbf{Accuracy} & \textbf{Macro Precision} & \textbf{Macro Recall} & \textbf{Macro F1-score} & \textbf{Accuracy Gap} \\
\midrule
\multirow{8}{*}{\textbf{MobileViT-XXS}} 
& CSI Amplitude \cite{yang2023sensefi}       & 1Tx--1Rx       & 0.366 & 0.362 & 0.357 & 0.355 & 0.510 \\
& BVP \cite{zhang2021widar3}                 & Multi-Receiver & 0.422 & 0.440 & 0.417 & 0.418 & 0.454 \\
& CACC (w.o. delay) \cite{qian2018widar2}    & 1Tx--2Rx       & 0.601 & 0.601 & 0.600 & 0.594 & 0.275 \\
& CACC (w. delay)     & 1Tx--2Rx       & 0.614 & 0.618 & 0.612 & 0.613 & 0.262 \\
& CASR (w. delay) \cite{feng2021lte, ni2023uplink}     & 1Tx--2Rx       & 0.853 & 0.852 & 0.851 & 0.852 & 0.023 \\
& \textbf{DCACC (w. delay)} \cite{wang2023single, 10737138}    & \textbf{1Tx--3Rx}       & \textbf{0.876} & \textbf{0.875} & \textbf{0.873} & \textbf{0.873} & \textbf{/} \\
& CFCC (w. delay) \cite{11079818}     & 1Tx--1Rx       & 0.565 & 0.565 & 0.561 & 0.558 & 0.311 \\
& \textbf{Our SRCC (w. delay)}    & \textbf{1Tx--1Rx}       & \textbf{0.767} & \textbf{0.764} & \textbf{0.765} & \textbf{0.763} & \textbf{0.109} \\
\bottomrule
\end{tabular}
}
\vspace{-1em}
\end{table*}

\subsubsection{Generalization Performance}
We evaluate the generalization of our feature by testing on samples from one target subset while training on all others.

\begin{itemize}
\item \textit{Impact of location:} We train the model on data from four target positions and testing it on the remaining one. For example, when Location \#1 is used for testing, the samples from the other locations (\#2-\#8) are used for training, covering various face orientations, receivers, and room. Here, Locations \#6, \#7, and \#8 contain significantly fewer samples (around 3,000 samples only), while Locations \#1-\#5 have more balanced sample sizes. For fair comparison, we conduct the experiments on Locations \#1-\#5. As shown in Fig. \ref{figure:single_location}, the sample distribution across the five locations is relatively uniform, achieving macro F1-scores ranging from 0.917 to 0.918. The results show the robustness of our SISO micro-Doppler feature in handling spatial shifts of the target's position.

\item \textit{Impact of orientation:} We evaluate the model's generalization under different body orientations. As shown in Fig. \ref{figure:single_orientation}, the model consistently achieves similar sensing accuracy across all five orientations, with F1-scores above 0.88. The experiment result shows the SISO micro-Doppler feature can also maintain strong orientation invariance, enabling reliable activity recognition even under significant changes in a user's facing direction.

\item \textit{Impact of receiver:} This experiment is to assess sensitivity to receiver placement. As shown in Fig. \ref{figure:single_receiver}, our scheme maintains consistently high performance across all receivers, with F1-scores exceeding 0.91 in each case, which can demonstrate the model's strong generalization to varying transceiver positions. 

\item \textit{Impact of environment:} As shown in Fig. \ref{figure:single_environment}, the model achieves the lowest F1-score of 0.74 when tested on Room 1, while the highest F1-score of 0.93 is observed when testing on Room 3. Here, the number of training samples in Room 1 is approximately half that of the other environments, leading to a data imbalance that contributes to the reduced performance. As the training data increases, the test accuracy tends to improve accordingly. In our feature extraction pipeline, static clutter removal is employed to reduce the sensitivity of the extracted micro-Doppler features to environmental differences. 

\item \textit{Impact of User:} To evaluate the generalization ability across different users, we select a subset of six users (User \#10 to \#15) since each user almost contributes an equal number of training and testing samples, thereby eliminating the impact of data imbalance. As shown in Fig.~\ref{figure:single_user}, our feature consistently achieves high F1-scores (0.88--0.95) across all users, indicating strong generalization to user-specific variations.
\end{itemize}

\subsubsection{Comparison of input features} To examine how different input features affect generalization, we construct a strict train-test split: the model is trained only on samples from Location \#1 and Receiver \#1 (5,224 samples), and tested on samples from all other locations and receivers (160,500 samples). A summary of these comparisons is provided in Table \ref{tab:input_feature_summary}.

\begin{itemize}
\item \textit{Commonly-used Features.} We first compare against three commonly used representations: CSI Amplitude, BVP, and Raw Doppler (directly extracted from CACC without delay filtering). These baselines yield F1-scores of only 0.355, 0.418, and 0.594, respectively. The CSI Amplitude feature is highly sensitive to absolute signal strength, which can vary significantly across hardware setups and environments, resulting in poor generalization. The BVP feature is designed to aggregate Doppler information across multiple receivers to improve motion representation. However, it may be less robust to unseen configurations due to its reliance on consistent receiver geometry. Raw Doppler (CACC w.o. delay) suffers from strong mirror-Doppler ambiguity, which undermines its ability to distinguish motion direction.

\item \textit{Enhanced Features.} We evaluate Doppler-enhanced features, including multi-antenna based CACC, CASR, and DCACC, as well as single-antenna based CFCC and SRCC, all incorporating delay-domain filtering. Among them, 1Tx-3Rx DCACC achieves the highest performance with a macro F1-score of 0.873, benefiting from effective Doppler mirror suppression. CASR also performs well (F1-score: 0.852) using two Rx antennas. In contrast, CACC and CFCC show moderate gains (F1-scores: 0.613 and 0.558, respectively), showing that delay filtering helps reduce mirror artifacts but is not sufficient. Notably, our 1Tx-1Rx SRCC method achieves a strong F1-score of 0.763, outperforming several multi-antenna baselines. Overall, these results demonstrate that Doppler mirror suppression is key to improving generalization. Our SRCC achieves the optimal trade-off between mirror suppression and hardware simplicity, offering strong generalization with only a single antenna.
\end{itemize}

\begin{table}[t]
\centering
\caption{Performance comparison of different models on activity recognition and people counting.}
\label{tab:performance}
\renewcommand{\arraystretch}{1.2}
\setlength{\tabcolsep}{3.5pt}
\scriptsize
\begin{tabular}{c|c|c|c|c|c}
\toprule
\multirow{2}{*}{\textbf{Method}} & \multirow{2}{*}{\textbf{Model}} & \multicolumn{3}{c|}{\textbf{Activity Recognition}} & \textbf{People Counting} \\
\cmidrule(lr){3-5} \cmidrule(lr){6-6}
& & \textbf{Precision} & \textbf{Recall} & \textbf{F1-score} & \textbf{Accuracy} \\
\midrule
\multirow{7}{*}{\centering \makecell{Our SRCC \\ (1Tx--1Rx)}} 
& MLP             & 0.720 & 0.405 & 0.507 & 0.493 \\
& CNN             & 0.665 & 0.400 & 0.491 & 0.584 \\
& MobileNetV2     & 0.687 & 0.606 & 0.642 & 0.569 \\
& ShuffleNetV2    & 0.714 & 0.577 & 0.636 & 0.571 \\
& SqueezeNet      & 0.700 & 0.562 & 0.622 & 0.551 \\
& ResNet18        & 0.706 & 0.590 & 0.641 & 0.574 \\
& \textbf{MobileViT-XXS} & \textbf{0.705} & \textbf{0.621} & \textbf{0.659} & \textbf{0.629} \\
\midrule
\makecell{DCACC \\ (1Tx--3Rx)} 
& \textbf{MobileViT-XXS} & \textbf{0.712} & \textbf{0.620} & \textbf{0.662} & \textbf{0.618} \\
\midrule
\makecell{CASR \\ (1Tx--2Rx)} 
& MobileViT-XXS   & 0.665 & 0.101 & 0.174 & 0.383 \\
\bottomrule
\end{tabular}
\end{table}

\begin{table}[t]
\centering
\caption{Per-class performance on activity recognition.}
\label{tab:perclass}
\renewcommand{\arraystretch}{1.2}
\begin{tabular}{l|c|c|c|c}
\toprule
\textbf{Activity} & \textbf{Precision} & \textbf{Recall} & \textbf{F1-score} & \textbf{Test Samples (\#)} \\
\midrule
Jumping         & 0.718 & 0.653 & 0.684 & 490 \\
Lying Down      & 0.724 & 0.689 & 0.706 & 460 \\
Nothing         & 0.648 & 0.489 & 0.558 & 519 \\
Picking Up      & 0.701 & 0.554 & 0.619 & 487 \\
Rotation        & 0.651 & 0.674 & 0.662 & 481 \\
Sitting Down    & 0.655 & 0.564 & 0.607 & 482 \\
Standing Up     & 0.632 & 0.579 & 0.604 & 468 \\
\textbf{Walking}         & \textbf{0.927} & \textbf{0.809} & \textbf{0.864} & \textbf{487} \\
Waving          & 0.687 & 0.578 & 0.628 & 483 \\
\bottomrule
\end{tabular}
\end{table}

\subsection{Multi-Target Sensing Performance}
We evaluate the performance of WiDFS 3.0 in complex multi-target scenarios using the WiMANS dataset. The following experiments use a 3D delay-Doppler-time tensor as input, preserving temporal motion dynamics and spatial separability.

\subsubsection{Overall Performance across Lightweight Models}
For our 1Tx-1Rx SRCC method, we randomly select a feature from one transmitter-receiver antenna pair in each training epoch. Table \ref{tab:performance} presents the results for various lightweight models. We can see that MobileViT-XXS achieves the highest F1-score of 0.659 for activity recognition and the best people counting accuracy of 0.629. MobileNetV2 and ShuffleNetV2 also perform competitively, with F1-scores exceeding 0.63. For comparison, the 1Tx-3Rx DCACC configuration further improves recognition performance, benefiting from enhanced Doppler mirror suppression. It achieves an F1-score of 0.662 in activity recognition and a people counting accuracy of 0.618. In contrast, the 1Tx-2Rx CASR yields a significantly lower F1-score of only 0.174 due to poor recall, highlighting its limited suitability in complex multi-target environments. In addition, per-class activity recognition results are detailed in Table \ref{tab:perclass}. The \textit{Walking} class, which exhibits the largest body movements, achieves the highest F1-score across all classes. Overall, these results demonstrate the effectiveness and robustness of our 1Tx-1Rx SRCC feature. However, multi-target recognition remains inherently challenging due to the limited bandwidth, which constrains delay resolution, and overlapping Doppler signatures from multiple individuals further reduce per-target separability, making accurate activity and people count estimation more difficult.

\begin{table*}[t]
\centering
\caption{Performance comparison of different features under a cross-environment train-test split.}
\label{tab:performance}
\renewcommand{\arraystretch}{1.2}
\small
\begin{tabular}{c|c|c|c|c|c|c}
\toprule
\multirow{2}{*}{\textbf{Model}} & 
\multirow{2}{*}{\textbf{Method}} & 
\multirow{2}{*}{\textbf{Input Feature}} & 
\multicolumn{3}{c|}{\textbf{Activity Recognition}} & 
\textbf{People Counting} \\
\cmidrule(lr){4-6} \cmidrule(lr){7-7}
 & & & Precision & Recall & F1-score & Accuracy \\
\midrule
\multirow{7}{*}{MobileViT-XXS} 
& \multirow{4}{*}{Our SRCC (1Tx--1Rx)}  
    & Amplitude-Time        & 0.571 & 0.413 & 0.473 & 0.385 \\
&  & Delay-Time            & 0.484 & 0.387 & 0.427 & 0.336 \\
&  & Doppler-Time          & 0.618 & 0.490 & 0.543 & 0.373 \\
&  & \textbf{Delay-Doppler-Time} & \textbf{0.654} & \textbf{0.537} & \textbf{0.588} & \textbf{0.385} \\
\cmidrule{2-7}
& DCACC (1Tx--3Rx) & \textbf{Delay-Doppler-Time} & \textbf{0.713} & \textbf{0.508} & \textbf{0.591} & \textbf{0.436} \\
\cmidrule{2-7}
& CASR (1Tx--2Rx)  & Delay-Doppler-Time & 0.646 & 0.094 & 0.162 & 0.309 \\
\bottomrule
\end{tabular}
\vspace{-1em}
\end{table*}

\begin{figure}
\centering
    \begin{subfigure}[t]{\linewidth}
\centering
        \includegraphics[width=0.8\linewidth]{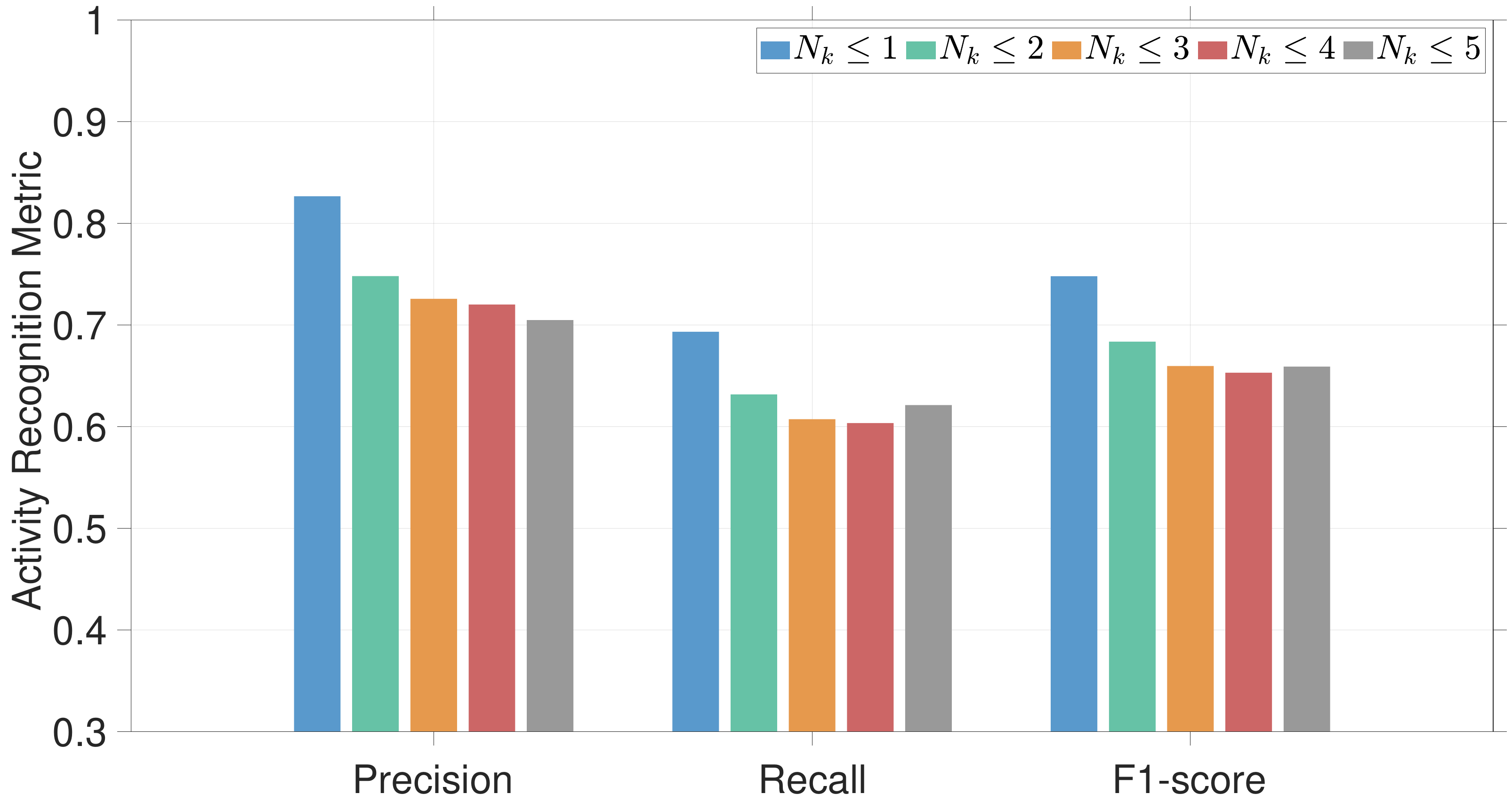}
        \subcaption{Activity recognition}
    \end{subfigure}
\\
    \begin{subfigure}[t]{\linewidth}
\centering
        \includegraphics[width=0.8\linewidth]{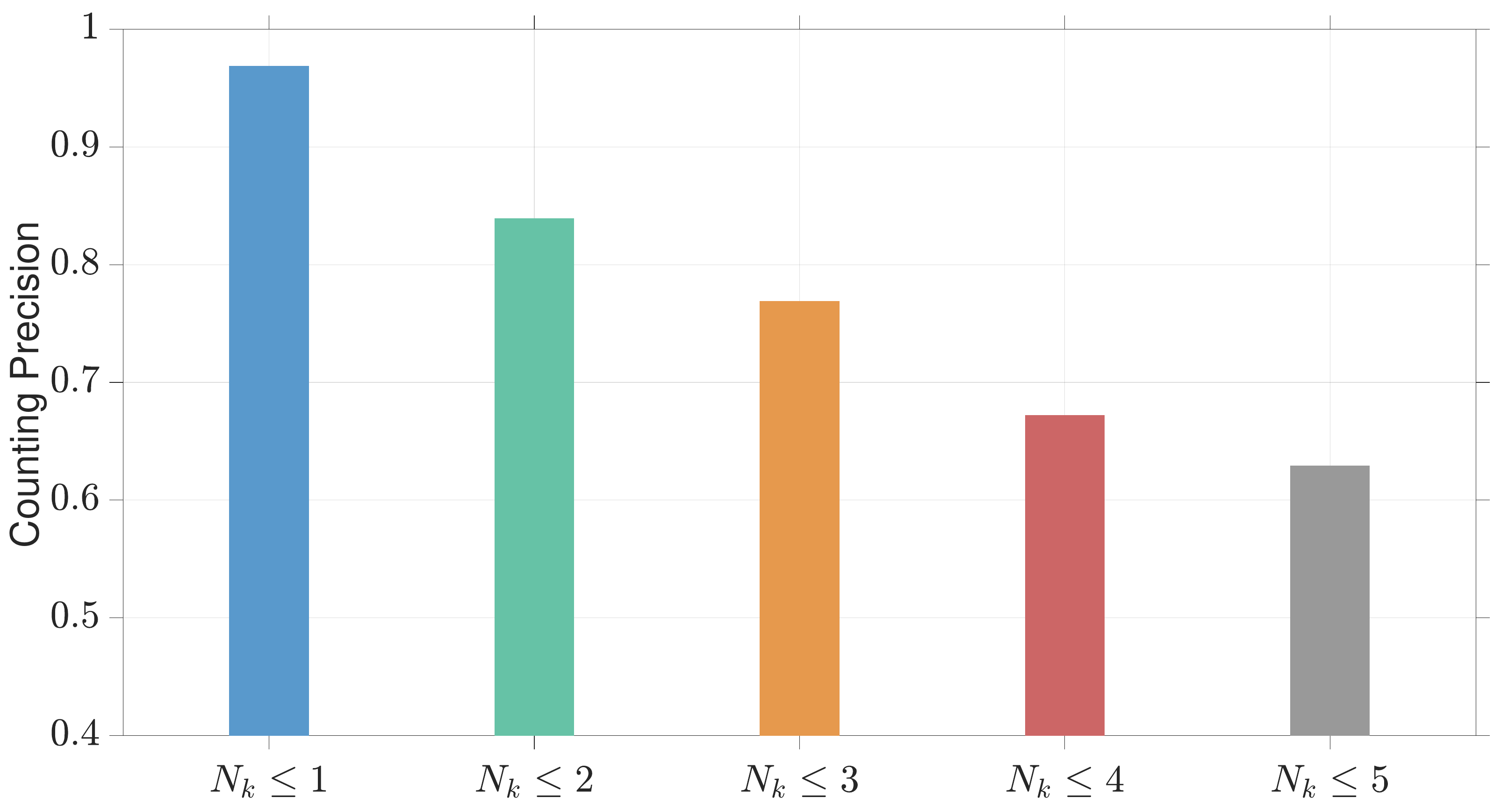}
        \subcaption{People counting}
    \end{subfigure}
\caption{Impact of the number of people on sensing performance.}
\label{fig:num_people}
\vspace{-1em}
\end{figure}

\subsubsection{Input Feature Comparison}
To assess the effectiveness of different input features, we adopt an corss-environment train-test split: the model is trained on data collected in the \textit{classroom} and \textit{empty room}, and tested on the unseen \textit{meeting room} environment. MobileViT-XXS is used for all comparisons. As shown in Table \ref{tab:performance}, the extracted delay-Doppler-time feature based on our 1Tx-1Rx SRCC method achieves an F1-score of 0.588 and a people counting accuracy of 0.385, outperforming all 2D baselines such as Amplitude-Time (F1: 0.473), Delay-Time (F1: 0.427), and Doppler-Time (F1: 0.543). These 2D features suffer from dimensionality reduction: Doppler-Time and Delay-Time projections discard either spatial or spectral information, while Amplitude-Time lacks physically meaningful motion encoding. In contrast, the 3D feature preserves full spatiotemporal information, leading to improved generalization across environments. We also compare with other multi-antenna methods. The 1Tx-3Rx DCACC approach achieves the highest performance, owing to its enhanced suppression of Doppler mirroring through spatial diversity. In contrast, the 1Tx-2Rx CASR method performs poorly (F1: 0.162), as it is primarily designed for single-target scenarios and struggles to generalize in complex multi-target cases. Overall, these results demonstrate the effectiveness of the proposed SRCC method, which achieves a good trade-off between performance and hardware simplicity. Maintaining both delay and Doppler dimensions is crucial for generalization, especially under multi-target conditions.

\subsubsection{Impact of Number of People}
We use the MobileViT-XXS model to evaluate the impact of the number of people. As shown in Fig. \ref{fig:num_people}, both activity recognition and people counting performance tend to degrade as the number of people increases. Specifically, the F1-score for activity recognition drops from 0.748 (for $N_k \leq 1$) to 0.659 (for $N_k \leq 5$), while counting accuracy decreases from 0.969 to 0.629. This degradation is expected due to increased spectral overlap, occlusion, and motion aliasing. With more users, the temporal and spatial separability of delay-Doppler-time patterns diminishes, making it more difficult for the network to distinguish fine-grained motion features.

\section{Conclusion}
This work presents WiDFS 3.0, a practical and lightweight bistatic sensing framework for SISO-based ISAC systems. It can operate with a single antenna at both the transmitter and receiver, while remaining extensible to multi-antenna setups. We propose a SRCC technique for efficient CSI random phase removal and introduce a delay-domain beamforming pipeline that produces a robust 3D delay-Doppler-time representation. This feature preserves key motion cues while suppressing interference and ambiguity, and can be effectively utilized by compact neural networks. Extensive experiments show that WiDFS 3.0 consistently outperforms conventional methods and exhibits strong feature generalization in both single- and multi-target scenarios. Overall, it offers a scalable, cost-efficient, and robust sensing solution that brings ISAC closer to practical deployment.

\appendix 
\section*{CRLB Derivation for CSI Phase Estimation with Windowing}
\label{appendix:crlb}

Assume the reconstructed CSI $\mathcal{CSI}_{i,j}$ is distorted by additive complex Gaussian noise $\mathcal{N}(0, \eta^2)$. We treat the phase $\phi_{i,j} = \angle \mathcal{CSI}_{i,j}$ as the parameter to be analyzed. Under the complex Gaussian noise model, the Fisher Information for $\phi_{i,j}$ is given by:
\begin{equation}
\mathcal{I}(\phi_{i,j}) = \frac{2|\mathcal{CSI}_{i,j}|^2}{\eta^2},
\end{equation}
so the CRLB becomes:
\begin{equation}
\text{Var}(\hat{\phi}_{i,j}) \geq \frac{1}{\mathcal{I}(\phi_{i,j})} = \frac{\eta^2}{2 \|\mathcal{CSI}_{i,j}\|^2}.
\end{equation}

From the energy consistancy of the FFT, we approximate $\|\mathcal{CSI}_{i,j}\|^2$ by
\begin{equation}
\|\mathcal{CSI}_{i,j}\|^2 \propto \left\| \mathcal{G}(\tau - \tau_j^{\text{peak}}) \cdot h_j(\tau) \right\|^2.
\end{equation}
Therefore, the phase estimation variance is bounded by:
\begin{equation}
\text{Var}(\hat{\phi}_{i,j}) \geq \frac{\eta^2}{2 \left\| \mathcal{G}(\tau - \tau_j^{\text{peak}}) \cdot h_j(\tau) \right\|^2}.
\end{equation}



\ifCLASSOPTIONcaptionsoff
  \newpage
\fi



%

\bibliographystyle{IEEEtran}
\bibliography{main.bbl}

\vspace{-30pt}

\begin{IEEEbiography} [{\includegraphics[width=1.3in,height=1.25in,clip,keepaspectratio]{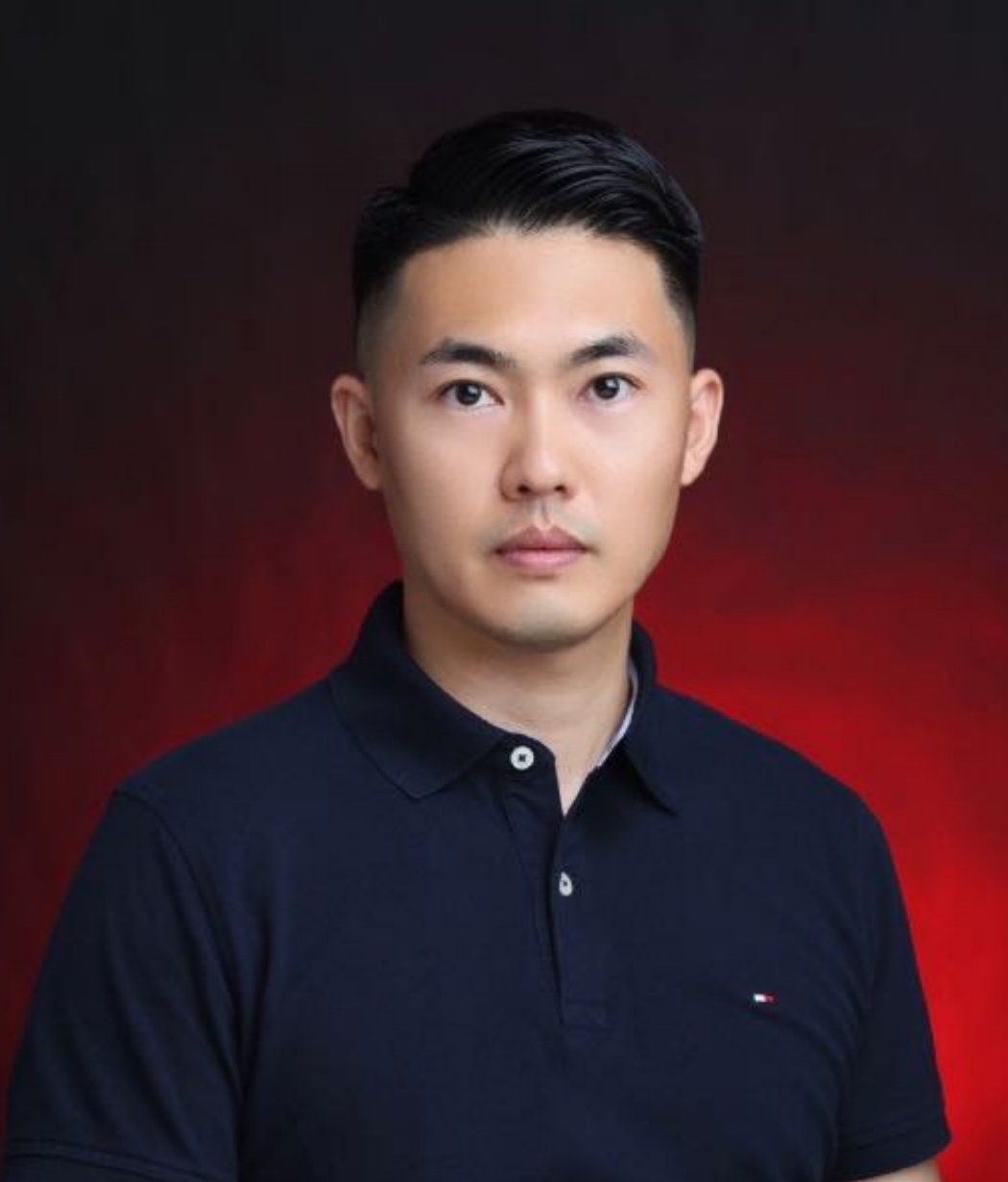}}] {Zhongqin Wang} is presently working as a Postdoctoral Research Fellow in the School of Electrical and Data Engineering at the University of Technology Sydney. He worked as a Lecturer at the School of Information Engineering, Capital Normal University, Beijing China, from 2022 to 2023. He attained his Ph.D. degree from the University of Technology Sydney, Australia, in 2021, and a M.S. degree from Nanjing University of Posts and Telecommunications, Nanjing, China, in 2014. His research interests include Radio Sensing and ISAC.
\end{IEEEbiography}
\vspace{-3em}

\begin{IEEEbiography}
[{\includegraphics[width=1in,height=1.25in,clip,keepaspectratio]{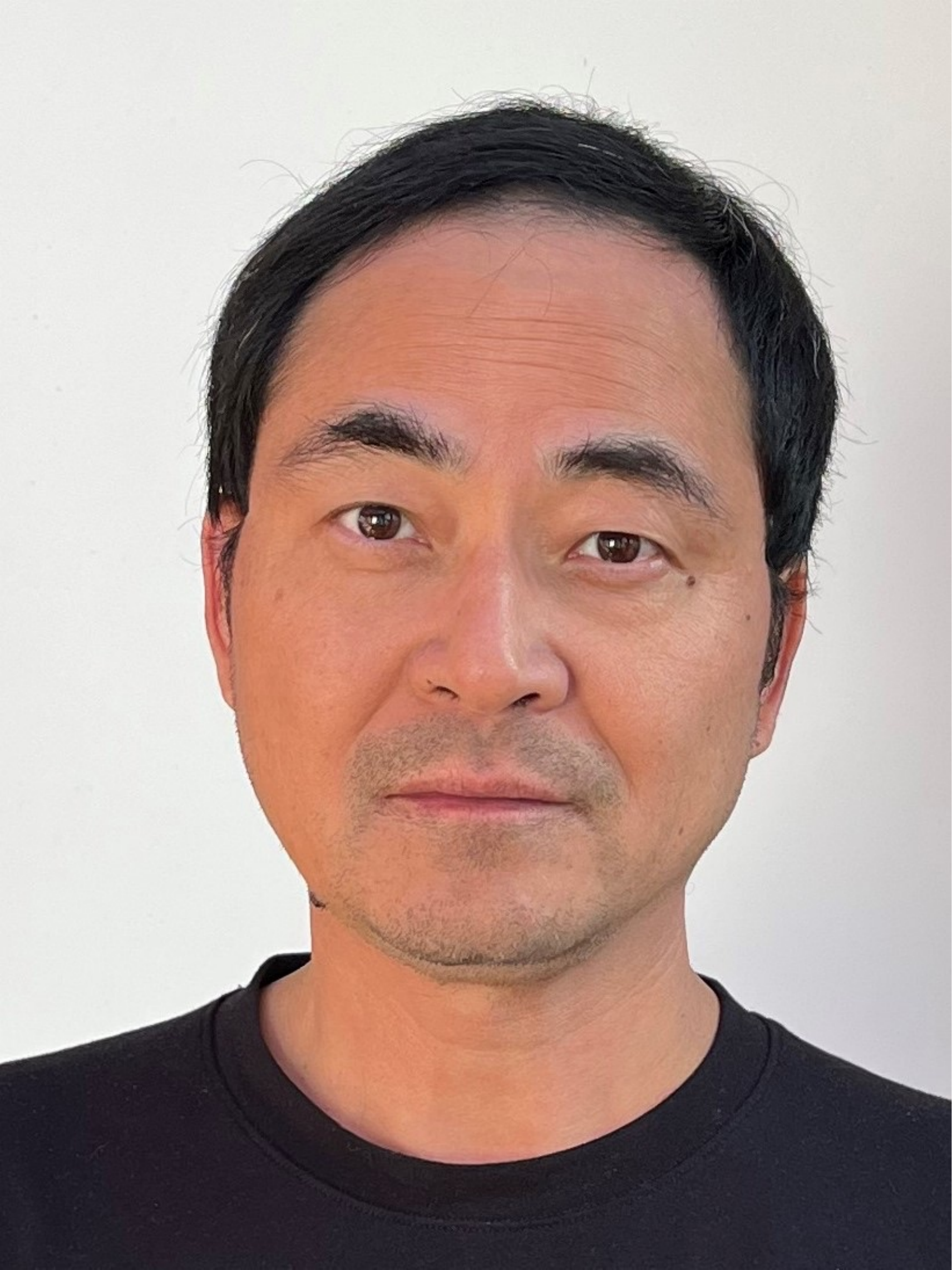}}]{J. Andrew Zhang} (M'04-SM'11) received the B.Sc. degree from Xi'an JiaoTong University, China, in 1996, the M.Sc. degree from Nanjing University of Posts and Telecommunications, China, in 1999, and the Ph.D. degree from the Australian National University, Australia, in 2004.

Currently, Dr. Zhang is a Professor in the School of Electrical and Data Engineering, University of Technology Sydney, Australia. He was a researcher with Data61, CSIRO, Australia from 2010 to 2016, the Networked Systems, NICTA, Australia from 2004 to 2010, and ZTE Corp., Nanjing, China from 1999 to 2001.  Dr. Zhang's research interests are in the area of signal processing for wireless communications and sensing, with a focus on integrated sensing and communications. He has published more than 300 papers in leading journals and conference proceedings, and has won 6 best paper awards for his work, including in IEEE ICC2013. He is a recipient of CSIRO Chair's Medal and the Australian Engineering Innovation Award in 2012 for exceptional research achievements in multi-gigabit wireless communications.
\end{IEEEbiography}
\vspace{-3em}

\begin{IEEEbiography}
[{\includegraphics[width=1.3in,height=1.25in,clip,keepaspectratio]{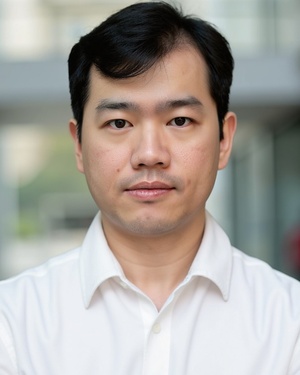}}] {Kai Wu} (Member, IEEE) received the B.E. degree from Xidian University, Xi'an, China, in 2012, and the Ph.D. degree from Xidian University in 2019 and from the University of Technology Sydney (UTS), Sydney, Australia, in 2020. From Nov 2017 to Nov 2018, he was a visiting scholar at DATA61, Commonwealth Scientific and Industrial Research Organisation (CSIRO). 

He is currently a Lecturer with the School of Electrical and Data Engineering (SEDE) and the Global Big Data Technologies Centre (GBDTC) at UTS. He is the system architect of the TPG-UTS Networking Sensing Lab.  
His research interests include space/time/frequency signal processing and its applications in radar and communications and their joint designs. He published an authored book on joint communications and sensing (JCAS), aka integrated sensing and communications (ISAC), in December 2022. 

His UTS Ph.D. degree was awarded "Chancellor's List 2020." His Xidian PhD thesis was awarded the ``Best Ph.D. Thesis Award 2019'' by the Chinese Institute of Electronics. He was awarded the Exemplary Reviewer for IEEE TRANSACTIONS ON COMMUNICATIONS, 2021. He is a Tutorial Speaker of WCNC'20, ICC'20, ISCIT'23, and RadarConf'23, presenting JCAS fundamentals and advancement. He was the TPC and special session (Co-)Chair/Member of numerous international conferences, e.g., ICC'20-23 and ISCIT'23. He is serving as the EiC Assistant for the IEEE ISAC-ETI Newsletter. He is an Associate Editor for IEEE Trans. on Mobile Computing, and has been a Guest Editor for the Special Issues in IEEE Journals. 
\end{IEEEbiography}
\vspace{-3em}

\begin{IEEEbiography} [{\includegraphics[width=1in,height=1.25in,clip,keepaspectratio]{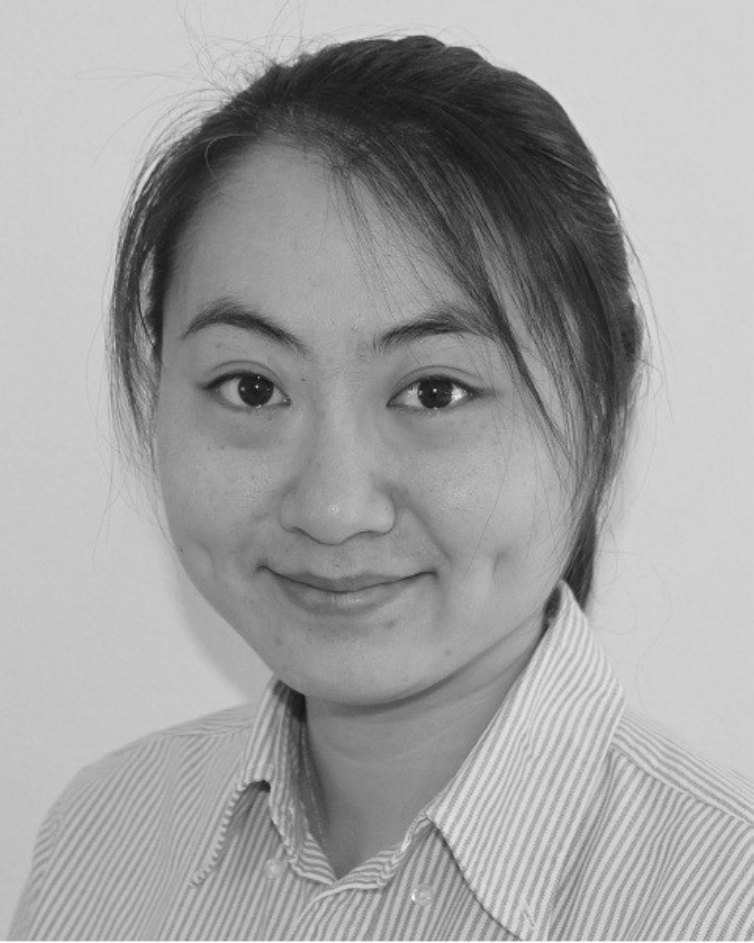}}]{Min Xu} (M'10) is currently a Professor at University of Technology Sydney. She received the B.E. degree from the University of Science and Technology of China, Hefei, China, in 2000, the M.S. degree from National University of Singapore, Singapore, in 2004, and the Ph.D. degree from University of Newcastle, Callaghan NSW, Australia, in 2010. Her research interests include multimedia data analytics, computer vision and machine learning. She has published over 100 research papers in high quality international journals and conferences. She has been invited to be a member of the program committee for many international top conferences, including ACM Multimedia Conference and reviewers for various highly-rated international journals, such as IEEE Transactions on Multimedia, IEEE Transactions on Circuits and Systems for Video Technology and much more. She is an Associate Editor of Journal of Neurocomputing.
\end{IEEEbiography} 
\newpage
\vspace{-3em}

\begin{IEEEbiography} [{\includegraphics[width=1in,height=1.3in, clip,keepaspectratio]{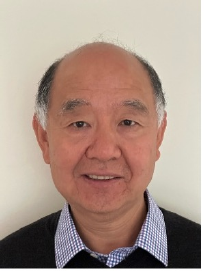}}] Y. Jay Guo (Fellow' 2014) received a Bachelor's Degree and a Master's Degree from Xidian University in 1982 and 1984, respectively, and a Ph.D Degree from Xian Jiaotong University in 1987, all in China. His current research interests include 6G antennas, mm-wave and THz communications and sensing systems as well as big data technologies such as deep machine learning and digital twin. He has published six books and over 800 research papers, and he holds 27 international patents. 

Jay is a Fellow of the Australian Academy of Engineering and Technology, Royal Society of New South Wales and IEEE. He has won a number of the most prestigious Australian national awards. Together with his students and postdocs, he has won numerous best paper awards at international conferences such as IEEE AP-S, EuCAP and ISAP. He was a recipient of the prestigious 2023 IEEE APS Sergei A. Schelkunoff Transactions Paper Prize Award.

Jay is a Distinguished Professor and the founding Director of Global Big Data Technologies Centre (GBDTC) at the University of Technology Sydney (UTS), Australia. He is the founding Technical Director of the New South Wales (NSW) Connectivity Innovation Network (CIN). He is also the Founding Director of the TPG-UTS Network Sensing Lab. Before joining UTS in 2014, Prof Guo served as a Research Director in CSIRO for over nine years. Prior to CSIRO, he held various senior technology leadership positions in Fujitsu, Siemens and NEC in the U.K. 
\end{IEEEbiography}

\end{document}